## *Antiferroelectric Smectic Ordering as a Prelude to the Ferroelectric Nematic:*
## *Introducing the Smectic $Z_A$ Phase*

Xi Chen[1], Vikina Martinez[1], Eva Korblova[2], Guillaume Freychet[3], Mikhail Zhernenkov[3],
Matthew A. Glaser[1], Cheng Wang[4], Chenhui Zhu[4], Leo Radzihovsky[1],
Joseph E. Maclennan[1], David M. Walba[2], Noel A. Clark[1]*

[1]*Department of Physics and Soft Materials Research Center,*
*University of Colorado, Boulder, CO 80309, USA*

[2]*Department of Chemistry and Soft Materials Research Center,*
*University of Colorado, Boulder, CO 80309, USA*

[3]*Brookhaven National Laboratory, National Synchrotron Light Source-II*
*Upton, NY 11973, USA*

[4]*Advanced Light Source, Lawrence Berkeley National Laboratory, Berkeley, CA 94720, USA*

*Abstract*

We have structurally characterized the liquid crystal phase that appears as an intermediate state when a dielectric nematic, having polar disorder of its molecular dipoles, transitions to the almost perfectly polar-ordered ferroelectric nematic. This intermediate phase, which fills a 100-year-old void in the taxonomy of smectics and which we term the "smectic $Z_A$", is antiferroelectric, with the nematic director and polarization oriented parallel to smectic layer planes, and the polarization alternating in sign from layer to layer. The period of this polarization wave (~180 Å) is mesoscopic, corresponding to ~40 molecules side-by-side, indicating that this lamellar structure is collectively stabilized. A Landau free energy, originally formulated to model incommensurate antiferroelectricity in crystals, describes the key features of the nematic–$SmZ_A$–ferroelectric nematic phase sequence.

## INTRODUCTION

Proper ferroelectricity in liquids was predicted in the 1910's by P. Debye [1] and M. Born [2], who applied the Langevin-Weiss model of ferromagnetism to propose a liquid-state phase change in which the ordering transition is a spontaneous polar orientation of molecular electric dipoles. A century later, in 2017, two groups independently reported novel nematic phases in the polar molecules of *Fig. 1A*, the "splay nematic" in the molecule RM734 [3,4,5] and a "ferroelectric-like nematic" phase in the molecule DIO [6]. These nematic phases were subsequently demonstrated to be ferroelectric in RM734 [7] and in DIO [8], and a study of their binary mixtures (*Fig. 1B*) showed that these are the same phase, termed $N_F$ here, a uniaxially symmetric, spatially homogeneous, ferroelectric nematic liquid having >90% polar ordering of its longitudinal molecular dipoles [7,9] (*Fig. 1D*). These mixtures also exhibit, at higher temperature, a nonpolar, paraelectric nematic (N) phase (*Fig 1C*).

RM734 and DIO are members of separate molecular families that represent distinctly different molecular structures, and, at the present time, are the only classes of mesogen to exhibit nematic ferroelectricity. In the course of the remarkable developments around nematic ferroelectricity, molecules with a variety of substitutions around the RM734 and DIO molecular themes have been studied and many of these have also been found to exhibit the $N_F$ phase [8,10].

The initial study of DIO reported the "M2" phase, a second, unidentified liquid crystal phase in a $\Delta T = 15\ °C$ temperature range between the N and $N_F$ phases (*Fig 1B*) [6]. Here we report that this phase exhibits an equilibrium, sinusoidal electron-density modulation of 8.8 nm periodicity. The M2 is, remarkably, a density-modulated, antiferroelectric phase, exhibiting lamellar order with ~17.5 nm repeats, comprising pairs of 8.8 nm-thick layers of opposite ferroelectric polarization (*Fig. 1D*), related by reflection symmetry about a plane normal to the director. As in other layered liquid crystal (LC) phases, these two-dimensional (2D) layers tile three-dimensional (3D) space with fluid spatial periodicity, a structuring step that largely controls the optic and electro-optic (EO) behavior of bulk samples.

In one of the stunning achievements of condensed matter physics, Georges Friedel combined his optical microscopic observation of elliptical and hyperbolic focal conic defect lines in LCs with knowledge of the Dupin cyclides (complex, equally-spaced, 3D-curved surfaces that envelop certain single-parameter families of spheres) to infer the existence of the fluid molecular layer ordering of smectic liquid crystals, without the benefit of x-ray scattering [11]. He termed such phases "smectic", following the Greek, as similar structures were found in concentrated solutions of neat soaps [12]. Friedel's samples also exhibited what he termed "homeotropy", where the preferred orientation in the smectic of the uniaxial optic axis, the local average long molecular axis, given by the director $\boldsymbol{n}$, is normal to the inferred planes. In time, as understanding of the polymorphism of LCs advanced, it became necessary to subdivide the class of smectic-like phases by structure, and Friedel's specific "smectic" geometry, the one having $\theta = 0^\circ$, where $\theta$ is the angle between the layer normal $\boldsymbol{z}$ and optic axis $\boldsymbol{n}$, was appropriately named the "smectic A" [13]. Here, following a path of textural analysis inspired by Friedel, and aided by synchrotron-based SAXS, we report a fluid, layered smectic phase with $\theta = 90^\circ$, the structural antipode to Friedel's smectic A, which we classify as the "smectic Z" [14].

## RESULTS

*X-ray scattering* – DIO was synthesized using standard synthetic schemes (*Fig. S1*). SAXS (*Fig. 2*) and WAXS images (*Figs. S2,S3*), obtained upon cooling DIO from the Iso phase with $q_z$ along the magnetically-aligned nematic director, show the previously observed [6,3], intense, diffuse scattering features at $q_z \sim 0.25$ Å$^{-1}$ and $q_y \sim 1.4$ Å$^{-1}$ from end-to-end and side-by-side molecular positional pair correlations, respectively. In the N phase, at low $q$ there is only low-intensity scattering that varies smoothly with $q_z$. However, upon cooling to the M2 phase, a pair of diffraction spots appear on this background, visible between $T \approx 83^\circ$C and $T \approx 68^\circ$C and with the integrated peak intensity a maximum at $T \sim 76^\circ$C. Scans of intensity through these peaks are shown in *Figs. 2D,S4,S5*. Simultaneously observed is a continuous reorientation, without change of shape, of the entire WAXS/SAXS scattering pattern, through an angle $\delta(T)$, starting at $T \approx 83^\circ$C and reaching $\delta = 12^\circ \pm 1^\circ$ at $T = 75^\circ$, as seen in *Fig. 2B*. The diffraction peaks at T = 75$^\circ$C are at wavevector $q_M = 0.071$Å$^{-1}$ (*Fig. 2D*), indicating an electron-density modulation of wavelength $w_M = 8.8$ nm, with $\boldsymbol{q}_M$ along $\boldsymbol{y}$, normal to the nematic director, the local mean molecular long-axis orientation, $\boldsymbol{n}$, along z. The WAXS peak at $q_y \sim 1.4$ Å$^{-1}$ corresponds to a side-by-side molecular spacing in DIO of $\sim 0.42$ nm in the direction of the modulation, so the period of this modulation is supermolecular, $\sim 20$ molecules wide. The growth of such a periodic structure in a uniaxial, field-aligned nematic should give powder scattering in the ($q_x,q_y$) plane normal to $\boldsymbol{n}$. That only a single pair of diffraction spots appear indicates that there is only a single diffraction ring in this plane intersecting the Ewald sphere (approximately the ($x,z$) plane), indicating that this ordering could be either 1D lamellar smectic-like, or 2D hexagonal columnar. The textural studies and electro-optic experiments presented below show definitively that this is a lamellar phase. Macroscopically, this phase is a fluid, as confirmed by its diffuse scattering features in the WAXS, which differ very little from those of the fluid nematic phase (*Fig. S2*).

As indicated above, the diffraction pattern rotates on cooling, with the lateral diffraction spots remaining in the equatorial plane of the SAXS/WAXS patterns. This implies that the entire sample volume probed by the beam (a 1 mm x 2 $\mu$m x 20 $\mu$m cylinder), the director $\boldsymbol{n}$, and the layering wavevector $\boldsymbol{q}_M$, have all uniformly reoriented together through $\delta(T)$, a rotation of $\boldsymbol{n}$ away from the orientation preferred by the magnetic field. This kind of temperature-dependent director/layer reorientation accompanying layer formation upon cooling of smectic LCs has been observed in magnetically aligned [15] and surface-aligned [16] samples undergoing nematic to smectic A (SmA) to smectic C (SmC) [17,18,19] or N to SmC phase transitions upon cooling [20], in the context of the study of surface-stabilized ferroelectric LCs [21]. This behavior is understood to be a result of fluid smectic layers contracting everywhere upon cooling, in the SmC case due to an increasing tilt of the molecular long axes away from the smectic layer normal with decreasing temperature. The result of layers of thickness $w_M(T)$ growing into a bulk sample with an aligned director and then homogeneously shrinking in thickness is sketched in *Fig.2B*. A local reorientation of the layers through the angle $\delta(T) = \cos^{-1}[w(T)/w_M]$, under conditions of low compressive stress of the layering, preserves the original pitch, $w_M$, of the layering as it was formed at the transition. In a large (1 mm-diameter) x-ray capillary, the pattern of such reorientation is complex, and different on each cooling run. On a larger scale, defects, such as parabolic focal conics [22] or the planar breaks of a chevron structure sketched in *Fig. 2C*, mediate changes in the sign of $\delta(T)$. In the experiment shown in *Fig. 2*, the scattering volume was accidentally filled with a single such reorienting domain. In thin cells with alignment layers, layer shrinkage is typically accommodated by the chevron layer structure and its characteristic defects, discussed below.

These observations and, in particular, the simple appearance of the scattering pattern in *Fig. 2B*, make it clear that the M2 is a form of fluid, lamellar phase with layers defined by a sinusoidal density modulation in which the modulation wavevector $q_M$ is normal to the mean molecular long-axis $n$, constraining $n$ to be parallel to the $(x,z)$ plane of the layers. Such a structure is orthorhombic biaxial, with a principal symmetry/optic axis triad $(l, q_M, n)$, where $l$, the layer normal, is the unit vector mutually normal to both $q_M$ and $n$. Related biaxial lamellar phases have been achieved only in amphiphilic systems, such as the "LAM_N" phase in bola-amphiphiles [23], and the hybrid DNA/cationic liposome lamellar phases [24] which employ strong nanophase segregation to isolate anisotropic 2D nematic-like molecular monolayers within a lamellar phase. We will show that in the M2 phase each layer is structurally and electrically polar, with alternating polarity from layer to layer, an antiferroelectric condition shared by some biaxial phases of bentcore molecules [25].

Finding similar behavior in a molecule as simple as DIO is truly remarkable, so we classify this newly identified phase to be among the thermotropic smectics of rod-shaped mesogens, as sketched in *Fig. 2F*, naming it the "smectic Z", such that the smectic A, the phase having $q_M$ parallel to $n$ ($\theta = 0°$), and the smectic Z, the phase having $q_M$ normal to $n$ ($\theta = 90°$) represent the lower and upper limits of tilt angles, $\theta$. We may also consider that the M2 could be classified as a kind of nematic, in analogy with the twist-bend nematic ($N_{TB}$) phase. The $N_{TB}$ phase is charac-terized by a spontaneous, heliconical precession of the nematic director, a periodic modulation that is helically ($C_\infty$) glide symmetric and therefore has no accompanying electron density variation or non-resonant x-ray scattering: different positions along the helix are relatively rotated but otherwise physically equivalent, making the $N_{TB}$ a kind of nematic, with a modulation that requires the use of resonant scattering in order to be observable with x-rays. However, in the case of a finite amplitude, periodic modulation of the director orientation involving splay and/or bend, as we report below and is sketched in **Fig. 2E**, for example, there is no equivalent translational symmetry, as planes with maximum and zero director distortion, or with a particular polarization magnitude, are inequivalent and therefore have different electron density. Such structures are therefore smectic. In the literature, there are many proposals for, and examples of, modulated nematic director fields based on structures that have director splay and bend [26], including the splay nematic model proposed for RM734 [4,5], but all such phases are inhomogeneous with respect to density and are therefore smectic.

The observation of a single pair of diffraction spots (with no harmonics) suggests that, as in many smectics, the density modulation of the layering may be nearly sinusoidal. However, this notion should be considered with caution, as **Figs. 2,S4,S5** show that the fundamental peak scattering intensity at wavevector $q_M$ is at most a few times the background level, so that lower intensity harmonics may be present but not visible.

_Optical textures and electro-optics_ –The x-ray scattering suggests that the SmZ phase should be optically biaxial, with ($x$, $y = q_M$, $z = n$) the principal axes of its optical dielectric tensor. Textures visualized using depolarized transmission optical microscopy (DTOM) in **Fig. 3** show the planar alignment achieved on cooling a $d = 3.5$ $\mu$m antiparallel-rubbed cell with a 3° angle between the buffing direction and the in-plane electrode edges. In absence of applied field, the $N - SmZ_A$ transition is difficult to observe optically, although there is a small increase in birefringence $\delta\Delta n_{yz} \approx 0.008$ across the phase transition (compare **Figs. 3A,D**). Excellent extinction is obtained for optical polarization along the buffing in both the N and $SmZ_A$ phases (compare **Figs. 3B,E**). The subsequent transition to the $N_F$, on the other hand, is quite dramatic, with no extinction observed for any sample orientation in the twisted $N_F$ state [27] (**Fig. 3G,H**). **Figs. 3C,F** show the same ~1.5 mm long section of this cell as in **Fig. 3B** but with a very slowly varying, effectively DC, field applied, at 30 V/mm. In the N phase (**Fig. 3B**), this field generates a twist Freedericksz transition, shown for $E = 30$ V/mm in the center of the electrode gap, whereas in the slowly growing-in $SmZ_A$ phase (**Fig. 3C**), this transition is suppressed.

The in-plane electric field can induce a dielectric twist Freedericksz transition in the N phase at a threshold field as small as ~ 10 V/mm. The 3° bias of the buffing direction means that the director $n$ initially makes an angle of 87° with the field $E$, so that the initial dielectric field-induced twist reorientation of $n$ is everywhere in the same azimuthal direction (the direction that reduces the angle between the director and the applied field). The initially uniform cell shows good extinction at $E = 0$ but application of the electric field causes a twist distortion of the director field that results in transmission of light that increases with field strength, as illustrated by the white-light DTOM intensity profiles plotted in **Fig. 3I**. Under the same field conditions in the SmZ$_A$ phase, however, there is no discernable optical response, as seen in **Fig. 3J**, indicating that the director reorientation is suppressed in the SmZ$_A$ phase, either by elimination of nematic dielectric torques, or by structuring that strongly limits reorientation. The former possibility can be eliminated because the nematic dielectric torques are proportional only to the dielectric anisotropy $\Delta\varepsilon$, which does not change substantially at the N–SmZ transition in DIO [4,5]. However, the latter phenomenon is quite familiar in smectic A phases. For example, the splay-bend Freedericksz reorientation observed when an in-plane electric field is applied to a homeotropically oriented cell of a nematic with $\Delta\varepsilon > 0$, is eliminated at the N–SmA phase transition by the appearance of a local potential energy that keeps $n$ along the smectic layer normal [28,29]. The restoring torque maintaining $n$ normal to the plates in the SmA phase is not that of Frank elasticity over a micron-scale cell gap $d$, but rather that of a local potential energy well (Friedel homeotropy [11]) that is typically stronger than the Frank elasticity by a large factor $\sim(d/l)^2$, where $l$ is the smectic layer thickness. This, combined with the condition that the smectic layering is essentially immovable at typical nematic Freedericksz threshold fields, suppresses molecular reorientation except under conditions of catastrophic layer reorganization at very high fields. In the SmZ phase, similar considerations explain why rotation of $n$ out of the layer plane is suppressed.

The electrooptic response to applied field can provide definitive information on how the SmZ$_A$ layers organize upon cooling from the N phase. With $n$ planar-aligned along the buffing direction, the layers will tend to order spontaneously either parallel (PA) to the plates, or in "bookshelf" geometry (BK), as depicted in **Fig. 1E** with the layers normal to the plates or nearly so. In the parallel geometry, an in-plane field readily induces in-plane reorientation of $n$, with induced dielectric polarization of the antiferroelectric structure driving twist deformation opposed by Frank elasticity. In the bookshelf geometry, on the other hand, molecular reorientation in the plane of the cell would force molecules out of the plane of the SmZ layers, a locally resisted, high energy deformation [**Error! Bookmark not defined.**,**Error! Bookmark not defined.**]. Suppression of the twist Freedericksz response in a particular cell therefore indicates that the layer normal $q_M$ is substantially parallel to the plates.

_Bookshelf and chevron layer structures_ – Further DTOM observations of the grown-by-cooling SmZ$_A$ cell textures show that the layers typically adopt bookshelf (BK) geometry at the N-SmZ$_A$ transition, with the layers normal to the plates. However, as illustrated in **Fig. 2C**, the layers contract slightly with decreasing temperature, and the layers respond by buckling (*Fig. 2C*), leading to the "chevron" (CH) variant of the lamellar bookshelf geometry (*Figs. 4D,S6-S10*), in which the layer normal exhibits a step reorientation at the chevron interface [18,19,20,21], a planar surface parallel to the plates where the layer tilt $\delta$ changes sign. This structure maintains the locations of the contact lines of the layers at the surfaces and the uniform $|\delta|$ maintains the bulk layering pitch along $y$ (in the plane of the cell), $p = 2\pi/q_M(T_{NZ})$, that the layering system had on the surfaces and in the bulk at the N-SmZ$_A$ transition.

The chevron structure and their associated zig-zag walls stand on the same footing as Friedel's focal conics as being fundamental layering defects of smectics: focal conics are enabled by step orientational discontinuities along lines, whereas chevrons are enabled by step orientational discontinuities on sheets. Zig-zags are higher energy defects which appear when the requirement in buffed cells to satisfy the constraint of uniform orientation at the surfaces does not permit focal conics. Zig-zag walls have been extensively studied in systems undergoing the N–SmA–SmC [18] and N–SmC [20] transitions, where the layer shrinkage is due to tilt of the molecular long-axis away from the layer normal in the SmC phase. Typical SmC zig-zag wall textures are shown in *Figs. S6,S7,S10*. An analogous texture of zig zag walls obtained on cooling DIO from the N to the SmZ$_A$ phase is shown in *Fig. 4*. In this case a small, in-plane square-wave electric field was applied along $q_M$ to alternate the direction of the nascent chevrons as they formed, giving a texture having an array of zig-zag walls.

The chevron is obviously vectorial along **y**, creating a class of defect lines, the "zig-zag walls" [18,19,20,21], which mediate flips in the chevron direction. The chevron tips point out from the diamond wall [⟨⟨⟨⟨⟩⟩⟩⟩], as shown in the sketch of the chevron structure on passing through a diamond wall in the upper right of *Fig. 4D*. The numberof layers per unit length on the cell surfaces is the same everywhere, so that the "empty" space in the center between the outpointing chevrons must somehow be filled with effectively thicker layers. It is filled by the diamonds, which are layer elements tilted away from $y$ by an angle $\delta_d$, larger than the chevron tilt, $\delta$, effectively making their apparent thickness larger (*Figs. 4D,S7*). Measurement of the sample orientation that gives extinction of the diamond wall line shows that $\delta_d = 23°$ (*Fig. S8*). The broad walls, which run nearly parallel to the layers (*Figs. 4C,S8B*), correspond to places where the chevron tips point toward each other [⟩⟩⟩⟩⟨⟨⟨⟨]. In these walls, the layers are tilted by less than $\delta$, making their effective thickness smaller in order to accommodate the chevron tips pointing toward the wall.

Comparing the SmZ$_A$ (*Figs. 4A,B*) and SmC zig-zag walls (*Fig. S6,S7*), one sees that the optical contrast between domains of opposite chevron direction is much more dramatic in the SmC. This is because the SmC director is on a cone about the layer normal, and the director orienta-tion in the cell adjusts to accommodate the chevron sign, boundary conditions on $n$ at the cell and chevron interfaces, and any applied field, greatly affecting the local optical appearance. In the SmZ$_A$, in contrast, $n$ is parallel to the surface and does not reorient with chevron tilt, nor in the formation of the broad walls, which, therefore have lower visibility than in the SmC, as comparison of *Figs. 4A,S8B* with *Fig S7* shows. Since the layers in the diamond walls are birefringent and they are twisted relative to the average layer orientation, they are visible when the rest of the cell is at extinction, as in *Figs. 4A,S8*.

We probed the polar characteristics of the as-grown chevron domains by applying small, in-plane probe electric fields both parallel to the layers (along $n$ ) and perpendicular to the layers (along $q_M$), the latter achieved using fringing fields near the upper edge of one of the rectangular electrodes, as shown in F*igs. 4B,C*. As can be seen, the field generated by the upper electrode edge parallel to $n$ induces a difference in hue between the chevron domains on opposite sides of zig-zag walls, indicating field-induced director rotation that is uniform over the domain, but in opposite directions in domains with opposite chevron direction. This implies a polarization of the chevron domains as indicated by the white arrows in *Figs. 4B-D*, leading to field induced director reorientation near the chevron interface. Such polarity, arising from the layer structure, is consistent with the directionality of the chevron. On the other hand, where the applied field is normal to $n$ (at the right electrode edge), there is no observable director reorientation, indicating that as the SmZ$_A$ is grown, it has no net component of polarization parallel to **n** (cyan, magenta arrows), evidence for antiferroelectricity.

In a sandwich cell, ITO electrodes cover the entire active area, enabling application of an electric field, $E_x$, normal to the cell plates. In the N phase, this produces a splay-bend Freedericksz transition by rotating $n$ out of the plane of the cell. In a bookshelf or chevron cell of the SmZ$_A$ phase, reorientation of the director takes place about the layer normal, and is therefore resisted only by splay-bend Frank elasticity, as in the N phase. The splay-bend Freedericksz transition is readily observed in the SmZ$_A$ phase as a field-induced reduction in the effective in-plane birefringence $\Delta n_{BK} = n_n - n_q$ , as the molecules stand up normal to the plates. Once this reorientation is saturated, the induced polarization in the tilted layers is still not fully parallel to the applied field, so that the field produces torque on the layers that tends to stand them up normal to the plates. The resulting dilative strain on the layering system generates, irreversibly, massive arrays of broad-wall zig-zag defects that fill nearly the entire cell, as shown in *Fig. S9B,C*, These walls develop internal periodic birefringent structures (*Fig. S10D-F*) suggesting that the broad wall can break up into small zig-zag loops, or that the dilation causes a local strain-induced periodic rotation of $n$ out of the plane of the layers, as would occur for the smectic dilative layer undulation instability [30]. These defect arrays are comparable to those obtained in ferroelectric SmC* cells at high applied field, as seen in *Fig. S10*.

_Observation of antiferroelectricity of the SmZ$_A$ phase_ – Increasing the field to ~10X the Freedericksz threshold in **Fig. S9F** reorients the director and toward being normal to the plates. If a field in this range is applied to the sandwich cell filled with zig-zags of **Fig. S9B,C** or if the bookshelf alignment is obtained by cooling from the N phase with a comparable field applied, the chevron layering defects, which rarely disappear on their own, can be forced out, and monodomain bookshelf alignment with the layers uniformly oriented normal to the plates achieved (**Fig. S9E**). These monodomains also exhibit the splay-bend Freedericksz transition at low field (**Fig. S9F**), but at higher field, they provide key evidence for the antiferroelectric nature of the SmZ$_A$ phase. The I-V characteristics of a $d$ = 100 $\mu$m, bookshelf ITO-sandwich cell with a 5 Hz, 50V/100$\mu$m amplitude triangle wave electric field applied along $n$ during an N–SmZ$_A$-N$_F$ cooling scan are shown in **Fig. 5A**. At high temperatures, in the N phase ($T$ > 84 ºC), the current shows the usual cell capacitance step and an ion bump. In the SmZ$_A$ phase (84ºC > $T$ > 68ºC), new polarization peaks appear at the highest voltages, growing in area and with their peak center voltages $V_{FA}$ moving to lower magnitudes as $T$ decreases. This is typical antiferroelectric behavior, the peaks marking the transition at finite voltage between the equilibrium antiferroelectric (A) state and the field-induced ferroelectric (F) state. The time integral of this current, $Q$ = $\int i(t) \mathrm{d}t$ gives the polarization reversal in the antiferroelectic–ferroelectric transition, $P_{FA}(T)$ = $Q(T)/S$, where $S$ is the electrode area. Both the polarization and the transition voltages, $V_{FA}(T)$ are plotted in **Fig. 5B**. The open squares give the first order SmZ$_A$–N$_F$ phase boundary in the ($E_z$–$T$) plane, and the open circles the polarization change $P_{FA}$ at this transition (**Fig. S12**). The transition entropy, $\Delta S$, decreases along this line to zero at the maximum $E_z$. Also shown is the product $P_{FA}(T) \times V_{FA}(T)$, proportional to the stabilization energy of the antiferroelectric state, which is maximum in the middle of the SmZ$_A$ $T$ range. These observations clearly establish that the SmZ$_A$ phase is antiferroelectric. The SmZ$_A$ to N$_F$ transition occurs between 69ºC and 68ºC, with the current transitioning to a single peak near $V$ = 0, as seen in **Fig. 5A**, corresponding to the field-induced, Goldstone-mode reorientation of a macroscopic polarization, $P_F$. The measured polarization in the N$_F$ phase is comparable to values previously obtained for DIO [7]. Recently, Brown et al. have reported evidence for "local antiferroelectric" behavior in the M2 phase of DIO [31], and Nacke et al. have observed antiferroelectric polarization behavior in an intermediate phase between N and N$_F$ in a different fluorinated material [32].

_Bookshelf to-parallel layering transformation_ – If a low frequency (20 mHz) in-plane triangle electric field is applied to the bookshelf SmZ$_A$ cell and its amplitude increased above about 80 V/mm, ferroelectrohydrodynamic flow is generated [7] which disrupts the SmZ$_A$ layering. The field-induced polarization tends to orient along the field direction, and in doing so restructures the layers in some areas to be parallel to the plates. When the field is reduced, these areas anneal into highly ordered monodomains filling the thickness of the cell, with the SmZ$_A$ layers parallel to the plates as sketched for the parallel geometry in **Fig. 1D**. The director $n$ remains along the buffing direction, so that the layers have effectively been rotated by 90º about *n* from the original bookshelf geometry.  Near the N–SmZ$_A$ transition, a typical cell treated in this way can simultaneously exhibit N, SmZ$_A$ bookshelf, and SmZ$_A$ parallel domains, as seen in *Fig. 6A*.

As in the N phase, the SmZ$_A$ parallel domains respond to in-plane applied fields by exhibiting a field-induced azimuthal reorientation of *n*(x) about *x*, the normal to the cell plates.  EO observations with in-plane fields normal to *n*, are summarized in **Fig. 7** showing this orientational response to be a strictly dielectric-driven, twist Freedericksz transition as in the N phase.  This confirms that the bulk polarization density along *n* is zero, i.e., that, at fields well below those inducing the AF to F transition, the SmZ$_A$ phase has no net ferroelectric polarization and its response to field is dielectric.

Applying a field to the cell of *Fig. 6A* generates a Freedericksz transition in both the parallel and N regions, but it is suppressed in the bookshelf region, as shown earlier in *Fig. 3F*.  Below the Freedericksz thresholds, the bookshelf, parallel and N regions all relax to highly extinguishing, uniform monodomains, as shown in *Figs. 6B,C*.  The Freedericksz thresholds in the parallel and N regions are similar but it is notable that once the field is removed, the relaxation of the field-induced twisted state in the SmZ$_A$ parallel regions back to the uniform state is ~ 10× faster than in the N phase.

*Biaxiality and the modulation structure of the SmZ$_A$ phase* – The simultaneous presence of bookshelf and parallel domains enables a measurement of features of the optical biaxiality of the SmZ$_A$ phase.  Referring to *Fig. 4* the principal axes of the optical dielectric tensor of the SmZ$_A$ are (*l*,*q*$_M$,*n*) such that in the bookshelf  (BK) geometry, the effective in-plane birefringence  is $\Delta n_{BK} = n_n - n_q = [(1/(2\langle n \rangle))]\Delta \varepsilon_{BK}$, where the optical dielectric anisotropy  is $\Delta \varepsilon_{BK} = \varepsilon_{nn} - \varepsilon_{qq}$  and $\langle n \rangle$ is the average refractive index.  In the parallel (PA) domains, we have $\Delta n_{PA} = n_n - n_l$.  The difference between these measurements yields the weak biaxial birefringence in the plane normal to *n*, $\Delta n_{biax} \equiv n_q - n_l = \Delta n_{PA} - \Delta n_{BK}$ that distinguishes the modulated SmZ$_A$ from the uniaxial nematic, which effectively has $n_l = n_q$.  With a cell thickness $d = 3.5$ μm and $\Delta n_{BK} \approx \Delta n_{PA} \approx \Delta n_{Nem} \approx 0.18$, the path difference for each of these is $\Delta nd$ ~ 630 nm, near the magenta-to-purple-to-blue-to-green band for increasing $\Delta nd$ on the Michel-Levy chart [33].  The DTOM image in *Fig. 6D* shows $\Delta n_{BK}$ giving a purple color and $\Delta n_{PA}$ blue, implying that $\Delta n_{BK} < \Delta n_{PA}$ and therefore that $\Delta n_{biax} > 0$.  Use of a Berek compensator enables simultaneous measurement of $\Delta n_{BK}$ and $\Delta n_{PA}$, with the results for these and the resulting $\Delta n_{biax}$ presented in *Fig 6E*.  The values of $\Delta n_{biax}$ can be interpreted in the context of spatial modulation of the nematic.  If this modulation is primarily through molecular long-axis reorientation, $\Delta n_{biax}(y)$, then, given that *q*$_M$ is normal to *n*, the choices for such reorientation are only twist $\delta n_x(y)$, or splay $\delta n_y(y)$, as illustrated for periodic waves with wavevector parallel to *q*$_M$ in *Fig. 4D*.  These sketches show immediately that splay increases $n_q$, making $\Delta n_{biax} > 0$ in accord with experiment, whereas the twist wave makes $\Delta n_{biax} < 0$. These changes can be related to $\langle\delta\psi(y)^2\rangle$, where $\delta\psi(y)$ is the rotation of the local director about $l$ in the $\delta n_y(y)$ splay deformation, which changes $\Delta n_{biax}(\delta\psi) = [(1/(2\langle n\rangle)]\Delta\varepsilon_{biax}(\delta\psi)$ according to $\Delta\varepsilon_{biax}(\delta\psi) = \Delta\varepsilon_{PA} - \Delta\varepsilon_{BK} = c\varepsilon_{qq}(\delta\psi) - \varepsilon_{ll}$. Taking $\delta\psi$ to be a function of $y$, averaging over this dependence, and applying a 2D rotational similarity transformation in the $(q_M, l)$ plane to get $\varepsilon_{qq}(\delta\psi)$, gives $\langle\varepsilon_{qq}(\delta\psi)\rangle = \langle\varepsilon_{nn}\sin^2\delta\psi + \varepsilon_{ll}\cos^2\delta\psi\rangle$ and therefore, since $\Delta n_{biax} << \Delta n_{Nem}$, $\Delta\varepsilon_{biax}(\delta\psi) = (\varepsilon_{nn} - \varepsilon_{ll})\langle\sin^2\delta\psi(y)\rangle$, or $\Delta n_{biax} \approx \Delta n_{Nem}\langle\sin^2\delta\psi(y)\rangle$. Upon inversion we find $\delta\psi_{RMS} \approx \sqrt{\langle\delta\psi^2(y)\rangle} = \sqrt{(\Delta n_{biax}/\Delta n_{Nem})} \approx \sqrt{(\Delta n_{biax}/0.185)}$, with the result shown in **Fig. 6E**. At a particular $T$, the range of values of $\delta\psi(y)$ as a periodic function will depend on the waveform of $\delta\psi_S(y)$. For example, if $\delta\psi(y)$ is a square wave, then its peak amplitudes $\delta\psi_p = \delta\psi_{RMS}$ and $\delta\psi_p \sim 4^o$. If $\delta\psi(y)$ is sinusoidal $\delta\psi_p = (\sqrt{2})\delta\psi_{RMS}$, and the peak amplitudes will be $\delta\psi_p \sim +/-6^o$.

This birefringence measurement does not distinguish ferroelectric [34] from antiferroelectric [4,26, 35] splay modulation. In the present case, the EO observations in the bookshelf and parallel SmZ$_A$ domains indicate that the response to low fields is dielectric rather than ferroelectric, and therefore that the low-field state has zero net polarization, *i.e.*, is antiferroelectric.

### DISCUSSION: MODELING THE N − SmZ$_A$ − N$_F$ PHASE DIAGRAM

Our observations identify the SmZ$_A$ to be a kind of spontaneously modulated phase, of which there are many examples in soft materials [36,37], liquid crystals [26,34,35,38,39], and ferroelectric or ferromagnetic) solids [40,41,42,43,44]. Such modulated phases have been explored by theory and simulation of equilibrium states in systems of molecular-scale (*e.g.*, spin) variables, as well as with coarse-grained and Landau modeling of their mesoscopic averages as fields.

The observation of both the N − N$_F$ and N − SmZ$_A$ − N$_F$ phase sequences in the RM734/DIO binary phase diagram of **Fig. 1** presents a new paraelectric − antiferroelectric − ferroelectric experimental landscape, one that must be viewed theoretically as a feature of some common ordering process. Among modulated states found in ferroelectric materials are the antipolar-ordered stripe phases, which appear in a wide variety of theoretical and experimental systems exhibiting frustration between competing short- and long-range (power law) interactions [36,37,41,42,43,44,45,46,47]. A particularly rich class of experimental systems are the antiferromagnetic and antiferroelectric phases found in crystals with sufficiently strong dipolar interactions. Their ground state structures are ferroelectric, or exhibit various periodic patterns of alternating polarity [45], and at finite temperature they are observed to order in patterns that are both multi-unit-cellular and incommensurate, having periods that are not rational fractions of the unit cell period of the underlying crystal lattice [43,48]. In this literature, the principal theoretical approaches are either Landau models in which the mesoscopic ordering field is $P(r)$ in an appropriate, symmetry-allowed free energy functional [40], or models starting with an Ising Hamiltonian comprising molecular dipoles, typically described by a sum of binary variables interacting with ferro-like local interactions and long-range dipolar interactions [41,42,45,46,49,50], inherently incorporating the frustration leading to stripes.

A promising Landau-based direction for modeling the $N_F$ system is the simple free energy shown in *Eq. 1* proposed by Shiba and Ishibashi [51,52] and first applied by these authors to describe incommensurate antiferroelectricity in thiourea [53] and in sodium nitrite ($NaNO_2$) [54,55]:

$$F_{IS} = \frac{A(T)}{2}P_z^2 + \frac{B}{4}P_z^4 + \frac{C}{6}P_z^6 + \frac{\kappa}{2}\left(\frac{\partial P_z}{\partial y}\right)^2 + \frac{\lambda}{2}\left(\frac{\partial^2 P_z}{\partial y^2}\right)^2 + \frac{\eta}{2}P_z^2\left(\frac{\partial P_z}{\partial y}\right)^2 - P_z E_z \qquad Eq. 1$$

Here ($\kappa < 0$; $\lambda$, $\eta > 0$), and (B < 0, C > 0) or (B > 0, C = 0), and the polarization inverse susceptibility $\chi(T)^{-1} = A(T) = a(T - T_{ZF})$. This free energy functional has been used in the analysis of soft matter modulation instabilities [37] and a variety of structural and dynamic experiments on incommensurate phases [56,57,58,59,60,61,62,63,64,65], and can be obtained as a mean-field approximation form of the dipolar Ising Hamiltonian [50]. Of relevance here is the application of *Eq. 1* to the paraelectric – incommensurate modulated antiferroelectric – ferroelectric (P – A – F) phase sequence in thiourea [66,67,68] and $NaNO_2$ [61,62], as this sequence exhibits key features consistent with our observations in the N – $SmZ_A$ – $N_F$ system, in particular the emergence of the modulated antiferroelectric $SmZ_A$ phase as an ordered state of a paraelectric free energy well, as follows:

(*i*) *The $N_F$ phase is homogeneous and unmodulated* – In the RM734/DIO mixtures [9], the $N_F$ is continuous across the phase diagram, a homogeneous (unmodulated) fluid ferroelectric nematic having: (*a*) a nematic director, *n(r)*, which is locally free to reorient in any direction, subject only to the mutual and field-induced interactions (no underlying lattice as in solid-state ferroelectrics); (*b*) polarization *P(r)*, locally parallel to *n(r)*, forming a coupled *n(r)*-*P(r)* orientation variable. *n(r)* does not differ observably in direction from *P(r)* in the $N_F$ phase; (*c*) large saturated polar order parameter $p > 0.9$; (*d*) strong polar director reorientational responses to applied electric field and polar surfaces; (*e*) Frank elastic behavior with the addition of polarization-generated space-charge interactions, which increase the energy cost of director splay. $F_{IS}$ describes an unmodulated, ferroelectric, low-temperature state of its finite-$P$ free energy well, which is the basic ingredient of all of these features.

(*ii*) *The transition to the $N_F$ phase is an expulsion of splay* – It was argued previously [7] that extreme shape anisotropy of the polarization correlation fluctuations is an important feature of the paranematic pretransitional behavior of RM734 [4,5], providing evidence for a competition be-tween short-range ferroelectric interaction, and long-range (dipole-dipole) electrostatic interaction, behavior demonstrated theoretically for the magnetic case by Aharony [69], and observed in dipolar ferromagnets [70]. The effects of the long-range dipolar interactions are evident in the polarization fluctuations of paranematic RM734, as shown in **Fig. S11**, with the longitudinal fluctuations of $P(r)$ (those in which $\partial P_z / \partial z$ is nonzero), energetically suppressed by polarization space charge, whereas transverse (curl) fluctuations ($\partial P_z / \partial y$) are not charged or suppressed, selectively extending correlations along the **z, n** direction [7]. An extreme limit of this behavior in the electric case is the SmZ$_A$, the ordering of the $P(r)$ into stripe or layer arrays having both $\langle P(r) \rangle = 0$ and $\partial P_z / \partial z = 0$ that induces the complete expulsion of longitudinal space charge from the paraelectric phase. On the DIO side of the phase diagram the polarization modulation of stripes of alternating sign of $P(r)$ in the SmZ$_A$ phase is necessarily accompanied by splay modulation of $P(r)$, as discussed in the previous section and illustrated in **Figs. 2E,4D**. Considering the SmZ$_A$ to be a kind of spatially ordered paraelectric phase, we have, on the RM734 side, a spatially disordered paraelectric in which DTOM observations indicate similar splay variations in the form of irregular, fluctuating, stripe-like domains with alternating sign of $P(r)$ extended along $n(r)$, as in **Fig. S11** [5,7,8]. Across the entire phase diagram, therefore, the transition to the N$_F$ phase is accompanied by an expulsion of splay, making the N$_F$ ground state uniform, with the average magnitude of the director splay modulation deformation reduced from the large value of $|\nabla \cdot n(r)| \sim 0.01 / \text{nm}$ in the SmZ$_A$ phase of DIO to zero in the N$_F$ ground state.

**(iii)** <u>*The SmZ$_A$ is an ordered paraelectric state*</u> - This structural analogy, coupled with the linearity of the phase boundary and the weak variation of the N–N$_F$ (SmZ$_A$–N$_F$) transition entropy across the phase diagram, suggests that the N and SmZ$_A$ phases should be considered to be paraelectric and the N – SmZ$_A$ phase change a transition within the paraelectric state. This is the $F_{IS}$ prediction: a paraelectric phase with a susceptibility for polarization diverging at the transition to the N$_F$ in which the typical pretransition scenario approaching the transition to a ferroelectric phase is interrupted by an instability to an ordered paraelectric (modulated antiferroelectric) state, in our case the SmZ$_A$. This instability occurs entirely within the $P = 0$ paraelectric free energy well of $F_{IS}$ [61], with the approach to the N$_F$ transition, and the instability both controlled by $A(T)$.

The free energy of **Eq. 1** thus gives either an N –($T_{NF}$)– N$_F$ or an N –($T_{NZ}$)– SmZ$_A$ –($T_{ZF}$)– N$_F$ phase sequence, depending on parameters, where the transition to the N$_F$ phase is first-order, and the N – SmZ$_A$ second-order or weakly first-order, as confirmed by the optical observations in **Fig. 3F** and by DSC [6] in DIO. Importantly, the antiferroelectric incommensurate (SmZ$_A$) phase is produced by frustration between the terms forcing slope ($\kappa$, $\eta$) and that resisting curvature ($\lambda$) of $P_z(y)$, producing a nearly temperature-independent modulation wavevector $q_M$. Upon cooling, near T $\sim T_{NZ}$ the initially weak modulation of $P_z(y)$ is sinusoidal with wavevector $q_M = \sqrt{2}$, de-termined only by $\kappa$ and $\lambda$, independent of the magnitude of $P_z$. **Eq. 1** correctly predicts the temperature dependence of $q_M$, including the weak decrease of $q_M$ with $T$ through the SmZ$_A$ range (**Fig. 2F**), and, notably, the lack of a $q_M \propto \sqrt{(T_{NZ} - T)}$ anomaly at high $T$ in the SmZ$_A$, a feature of Landau models in which $q_M$ is controlled directly by the coefficient of the $P^2$ term [4,5]. The higher order $\eta$ term contributes to the 5% decrease in $q_M$ over the SmZ$_A$ range (**Fig. 2F**).

The weak variation of $q_M(T)$ over the entire SmZ$_A$ range in **Fig. 2H** is rather remarkable, given the decrease with $T$ of the antiferroelectric – ferroelectric transition threshold field $E_{FA}(T)$ (**Fig. 5B**) and the increase of the induced polarization $P_{FA}$ (**Fig. 5B**), which indicate a transformative change in layer structure over the same range. The SAXS scans exhibit only the fundamental peak of the scattering electron density modulation, which would generally be interpreted in smectics to indicate a sinusoidal electron density modulation. However, although the peak intensity $I_p(T)$ decreases for both large and small $T$ (**Fig. 2G**), the layer structures must be very different near the two limits, in a way not describable by simply varying a sine-wave amplitude. If higher sinusoidal harmonics of $P_z(y)$ are included in the $F_{IS}$ model, then $P_z(y)$ evolves on cooling from sinusoidal to approaching a solitonic or square waveform for $T \sim T_{ZF}$, in which alternating +/- plateaus of maximum $|P_z(y)|$ and high electron-density are separated by narrow domain walls of width $d(T)$ where the sign of $P_z(y)$ changes and electron density is lower [57,58,59,60]. The decrease in the non-resonant SAXS peak intensity as $T$ approaches $T_{ZF}$ in **Fig. 2G** will depend on $d(T)$ as $I_p(T) \propto I_{pmax}\sin^2(\pi\, d(T)/w_M)$. The resonant peak intensity, the fundamental harmonic of $P(r)$, is nearly constant through the SmZ$_A$ range (**Fig. S5G**). The question of the constancy of $q_M(T)$ has also been considered by Kats, who speculated that the M2 phase could be a modulated antiferroelectric [71] and suggested that the N to M2 transition would then be a weak crystallization phenomenon, with $q_M(T)$ constrained as in "weak crystallization theory" [72].

(*iv*) *The temperature – electric field (T, E_z) phase diagram* - **Fig. 5A** shows that the transition to the N$_F$ phase can be induced by an electric field applied along the director, with **Fig. 5B** giving $E_z(T)_{ZF}$, the field required to induce the transition, vs. $T$. A similar field-induced shift in the transition to the ferroelectric phase has been observed in thiourea and sodium nitrite, and is a prominent feature of the P – A – F experimental phase diagrams obtained by Shiba and Ishibashi and the theoretical fits calculated from **Eq. 1** [61,62] and shown in **Fig. S12**. The (T, E) thiourea phase diagram [62], obtained using **Eq. 1** with B > 0 and C = 0, matches the optical [67] and neutron scattering [68] data for thiourea, and describes well the DIO (T, $E_z$) phase behavior with appropriate temperature and field scaling. An interesting feature of both the incommensutate (IC) data in NaNO$_2$ and the predictions of **Eq. 1** is that in the entire IC area of the (T,E) phase diagram, the modulation wavevector is nearly constant at $q_M = \sqrt{\kappa/2\lambda}$, having only a very weak dependence on applied field in addition to being insensitive to temperature [61,62]. Application of the $F_{IS}$ model to the $SmZ_A$ case, would therefore suggest that in the $N_F - SmZ_A$ phase coexistence obtained upon reducing the field in **Fig. 5**, the $SmZ_A$ domains have a nearly field-independent periodicity $q_M \sim 2\pi/(9 \text{ nm})$. This has not yet been tested experimentally.

(**v**) _Concentration dependence of $F_{IS}$_ - We can expect the parameters of the free energy of **Eq. 1** to depend on concentration, $c$, in the DIO/RM734 binary mixtures of **Fig. 1**. Their ideal mixing behavior in the transition to the $N_F$ phase indicates that the transition temperature extrapolates linearly between the DIO and RM734 values, and that $a$, $B$, $C$, are only weakly dependent on $c$, with the transition entropy $\Delta S = aB/C$ constant, and polarization in the $N_F$ phase $P = \sqrt{-B/C}$ changing little across the phase diagram [9]. $\kappa(c)$ and $\lambda(c)$ also determine the temperature range of the $SmZ_A$ according to $T_{NZ} - T_{ZF} = \kappa^2/2a\lambda$, suggesting that $\kappa(c)$ changes sign at the $SmZ_A$ endpoint at $c \sim 40\%$, and that measurement of $q_M(c) = \sqrt{\kappa(c)/2\lambda(c)}$ will be a key experiment for determining the dependence of these parameters on $c$. In any case, given that non-zero $\partial P_z/\partial y$ is favored in DIO, it may be that $\partial P_z/\partial y$ will at least be not strongly suppressed generally in the paranematic phase of ferroelectric nematic materials. Such a tendency is already evident in RM734, in the dominance of $\partial P_z/\partial y$ fluctuations in its paranematic phase (**Fig. S11**), behavior observable as the extension along $z$ of its director and polarization fluctuation correlations [7], also due to the fact that in RM734 the $\partial P_z/\partial z$ fluctuations are strongly suppressed by polarization space charge [7,70]. The IC phase is the corresponding extreme limit: $|\partial P_z/\partial z|$ is simply set to zero.

(**vi**) _Modification of $F_{IS}$_ – The principal limitation of $F_{IS}$ in its application to the ferroelectric nematic realm is that it models only the $y$ dependence of $P_z$. Therefore, desirable enhancements would include: (**a**) a generalization to enable the 3D vectorial spatial variation of $\boldsymbol{P}(\boldsymbol{r})$, including incorporation of the 3D polarization self-energy term, $U_P = \frac{1}{2}\int dv \ \nabla\cdot\boldsymbol{P}(\boldsymbol{r})\nabla\cdot\boldsymbol{P}(\boldsymbol{r'})[1 \ / \ |(\boldsymbol{r}-\boldsymbol{r'})|]$. In the $F_{IS}$ model of **Eq. 1** [61,62,51], the sinusoidal stripe phase is assumed as the solution and its free energy calculated. This assumption effectively sets $P_x(z)$ constant, introducing a large (infinite) energy cost for $\partial P_z/\partial z$ and illustrating the need for a $(\partial P_z/\partial z)^2$ term in the free energy. The previous paragraph and the Aharony calculation [69] document the need to include $U_P$ in the treatment of the N phase. (**b**) the introduction of $\boldsymbol{n}(\boldsymbol{r})$ and its flexoelectric coupling to $\boldsymbol{P}(\boldsymbol{r})$. At high RM734 concentrations, the N phase exhibits pretransitional softening of the Frank splay elastic constant, $K_s$, as $T$ approaches $T_{NF}$, indicating the importance of the flexoelectric coupling, $\boldsymbol{n}\cdot\boldsymbol{P}(\nabla\cdot\boldsymbol{n}) = P_z(\partial n_y/\partial y)$ between polarization and director splay [4,5]. This coupling also produces a splay-modulated $N_F$ phase, which is not observed, additional evidence for the need to include $U_P$ in the description of the $N_F$, since a simple estimate [7] shows that $U_P$ will strongly suppress polarization splay modulation. Additionally, the spatial derivative term $\boldsymbol{P}^2(\nabla\cdot\boldsymbol{P})$ is allowed by symmetry in a 3D Landau free energy describing the ordering of electric polarization [38,39,40,48, 73].

<u>*MATERIALS AND METHODS*</u>

DIO was synthesized using published literature schemes as described in Ref. [9], and has the phase sequence on cooling: Iso – 174ºC – N – 84ºC – SmZ$_A$ – 69ºC – N$_F$ – 34ºC – X, where N is the nematic, SmZ$_A$ the antiferroelectric phase, originally named M2 in Ref. [6], and N$_F$ the ferroelectric nematic, originally named MP in Refs. [6,8]. DIO was studied using standard liquid crystal phase analysis techniques, including Depolarized Transmission Optical Microscopy (DTOM), differential scanning calorimetry (DSC), and x-ray scattering (SAXS and WAXS), as well as ferroelectric nematic-specific polarization measurement/electro-optic techniques, previously described [7], for establishing the appearance of spontaneous ferroelectric polarization, determining its magnitude, and measuring electro-optic response.

<u>*Non-resonant SAXS & WAXS*</u> measurements were carried out in-house and on the microfocus Soft Matter Interfaces beamline 12-ID at NSLS II at Brookhaven. DIO was filled into 1 mm-diameter, thin-wall quartz capillaries and the director magnetically aligned normal to the beam using a field of ~2 kGauss from rare-earth magnets. <u>*Resonant SAXS*</u> measurements were carried out at the Carbon K$_\alpha$-edge on beamline 11.0.1.2 at the Advanced Light Source, Lawrence Berkeley National Laboratory, on powder samples in ~5 $\mu$m-thick cells with 100 nm-thick silicon nitride windows.

<u>*Electro-optics*</u> – For making electro-optical measurements, DIO was filled into planar-aligned, in-plane switching test cells with unidirectionally buffed alignment layers arranged antiparallel on the two plates, which were uniformly separated by $d$ in the range 3.5 $\mu$m $< d < 8$ $\mu$m. In-plane ITO electrodes were spaced by a 1 mm wide gap and the buffing was almost parallel to the gap. Such surfaces give a quadrupolar alignment of the N and SmZ$_A$ directors along the buffing axis and polar alignment of the N$_F$ at each plate. The antiparallel buffing makes *ANTI-POLAR* cells in the N$_F$, generating a director/polarization field parallel to the plates, and with a π-twist between the plates [9]. ITO Sandwich cells with $d$ =4.6 $\mu$m for studying the splay-bend Freedericksz transition in the N phase were also used.

<u>*Polarization measurement*</u> – The ferroelectric polarization density, $P$ , was measured by applying an in-plane square-wave electric field to generate polarization reversal while measuring the induced current between the electrodes, the REV mode in Refs. [7,9]. A 50 Hz, 104 V peak-to-peak square-wave voltage was applied across a 1 mm wide electrode gap in the sample cell. The polarization current, $P(t)$, was obtained by monitoring the voltage across a 55 kΩ resistor connected in series with the cell. Polarization current vs. time data for different temperatures are shown in *Fig. 5*. The polarization reversal is symmetric, with the +/– and –/+ reversals giving essentially identical data sets. In the isotropic, nematic and smectic Z$_A$ phases, the current consists only of a small initial peak, appearing following the square-wave voltage sign-reversal and then decaying exponentially. This peak is the RC circuit response of the cell and resistor. Upon entering the $N_F$ phase, an additional, much larger, current signal appears at longer times which comes from the field-induced polarization reversal. The switched polarization charge $Q$ is obtained as the time integral of the polarization current, and the $N_F$ polarization density is given by $P = Q/2A$, where $A$ is the cross-sectional area of the liquid crystal sample in the plane normal to the field and passing through the mid-line of the electrode gap.

### *ACKNOWLEDGEMENTS*

This work was supported by NSF Condensed Matter Physics Grants DMR 1710711 and DMR 2005170, and by Materials Research Science and Engineering Center (MRSEC) Grant DMR 1420736. This research used the microfocus Soft Matter Interfaces beamline 12-ID of the National Synchrotron Light Source II, a U.S. Department of Energy (DOE) Office of Science User Facility operated for the DOE Office of Science by Brookhaven National Laboratory under Contract No. DE-SC0012704. We also acknowledge the use of beam line 11.0.1.2 of the Advanced Light Source at the Lawrence Berkeley National Laboratory supported by the Director of the Office of Science, Office of Basic Energy Sciences, of the U.S. Department of Energy under Contract No. DE-AC02-05CH11231.

*FIGURES*

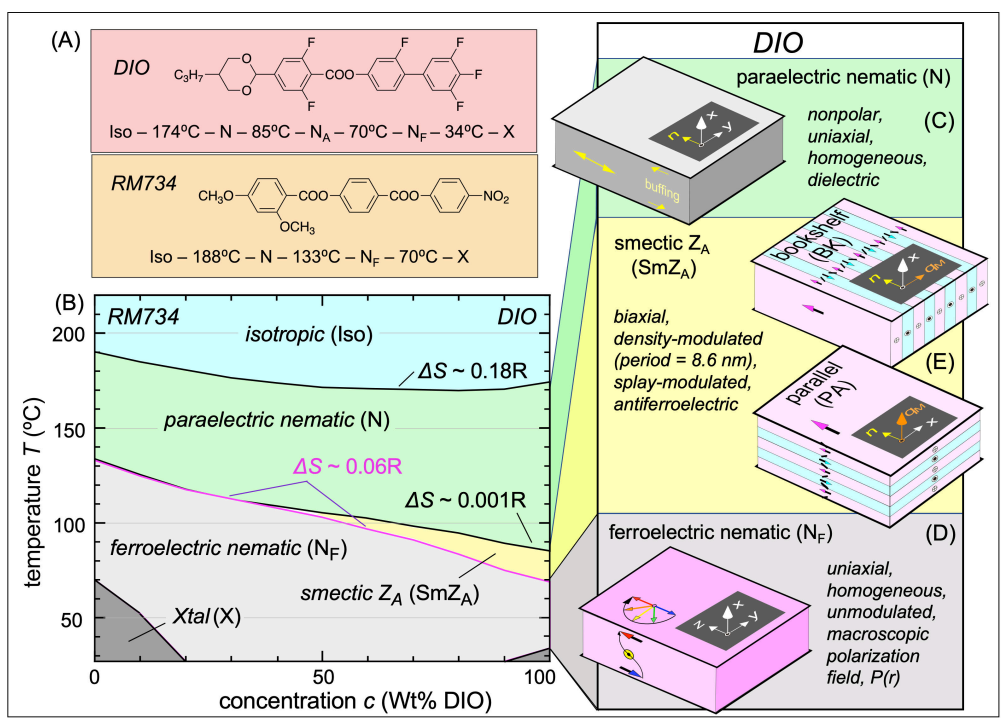

***Figure 1***: (***A***) Representative members RM734 and DIO of nitro- and fluoro-based molecular families of molecules that exhibit the ferroelectric nematic phase. This behavior is attributable in part to their rod-like molecular shape, large longitudinal dipole moment, and a common tendency for head-to-tail molecular association. (***B***) Phase diagram of DIO and RM734 and their binary mixtures exhibiting: (***C***) Paraelectric nematic (N) phase, in which the molecular long axes are aligned in a common direction but with no net polarity, making a nonpolar phase with a large dielectric response. (***D***) Ferroelectric nematic ($N_F$) phase, in which the polar orientation of dipoles is unidirectional and long-ranged, with a large polar order parameter (~0.9). The $N_F$ phase is uniaxial and spatially homogeneous, with a macroscopic polarization/director field ***P(r)***. The phase diagram shows that the transition temperature into the $N_F$ depends nearly linearly on concentration *c*, indicating that these molecules exhibit ideal mixing behavior of the transition into the ferroelectric nematic ($N_F$) phase. This transition is weakly first-order, with entropy $\Delta S \approx 0.06R$. (***E***) Intermediate phase observed in neat DIO and in DIO-rich mixtures. Originally called the M2 [6], we find this phase to be electron density-modulated into ~9 nm thick, polar layers, in which the director and polarization are parallel to the layer plane and alternate in direction from layer to layer, as sketched. We have termed this lamellar, antiferroelectric phase the smectic $Z_A$ ($SmZ_A$). The $SmZ_A$ can be prepared either with the layers normal to the plates (bookshelf geometry), parallel to the plates, or with the chevron texture shown in ***Figs. 2,4***.

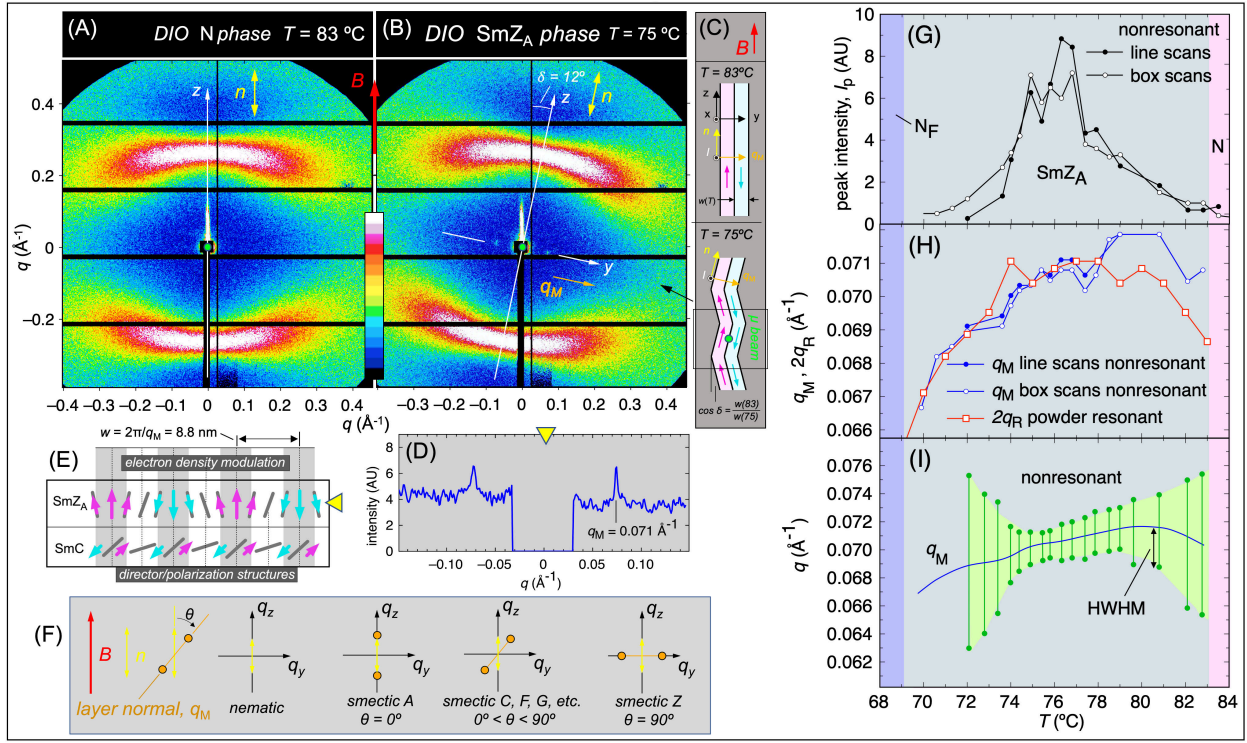

***Figure 2***: SAXS scattering from the N and SmZ$_A$ phases of DIO. These diffractograms are from a sample in a 1 mm-diameter capillaries in which the director, ***n***, is aligned along the magnetic field ***B*** (red), studied on the SMI microbeam line at NSLS II. (***A,B***) Non-resonant diffractograms obtained on cooling show a diffuse scattering arc from head-to-tail stacking of the molecules in the N and SmZ$_A$ phases. In the SmZ$_A$ phase, equatorial Bragg spots appear, indicating the presence of an electron density wave of $w_M$ = 8.8 nm periodicity, with wavevector $q_M$ normal to ***n***. These peaks disappear upon transitioning to the N$_F$ phase (***G,S3***). (***C***) The scattering pattern rotates from (***A***) to (***B***) because of layer reorientations caused by a small reduction in layer thickness upon cooling in the SmZ$_A$ phase (see text). (***D***) Line scan through the non-resonant scattering peaks. The central part of the beam is blocked by the beam stop. (***E***) Since the sample is a powder in orientation about the B field, the observed scattering pattern could be indicative of either lamellar or hexagonal columnar positional ordering. Since non-resonant scattering is sensitive only to electron density, the layers (thickness $w_M$) of opposite polarization scatter identically. Layer boundaries (white stripes) have different electron density than the layer centers (gray stripes), making $w_M$ the period observed in non-resonant scattering. However, Carbon-edge resonant scattering exhibits the half-order peak at $q_R = q_M/2$, showing conclusively that the SmZ$_A$ is lamellar and bilayer. The DTOM experiments show that it is lamellar and antiferroelectric, with layers of alternating polarization, as sketched. Other smectic structures (*e.g.*, the SmC, also sketched) have spontaneous antiferroelectric polarization in the tilt plane [74,75] as well as splay modulation [76,77], but in different geometries. (***F***) Com-

parative geometry of various smectics. (**G-I**) Dependence on $T$ of the Bragg peak parameters from data as in **Fig. S5** [non-resonant peak intensity ($I_p$), non-resonant peak position ($q_M = 2\pi/w_M$), resonant peak position ($q_R = 2\pi/2w_M$, and non-resonant half width at half-maximum (HWHM$_M$)]. (**H**) Layers of opposite **P** have different Carbon K-edge resonant scattering cross sections, and therefore Bragg scatter at the full antiferroelectric period $2w_M$. (**I**) The HWHM$_M$ values of the mid-range peaks ($T \sim 75ºC$) are SAXS-resolution limited.

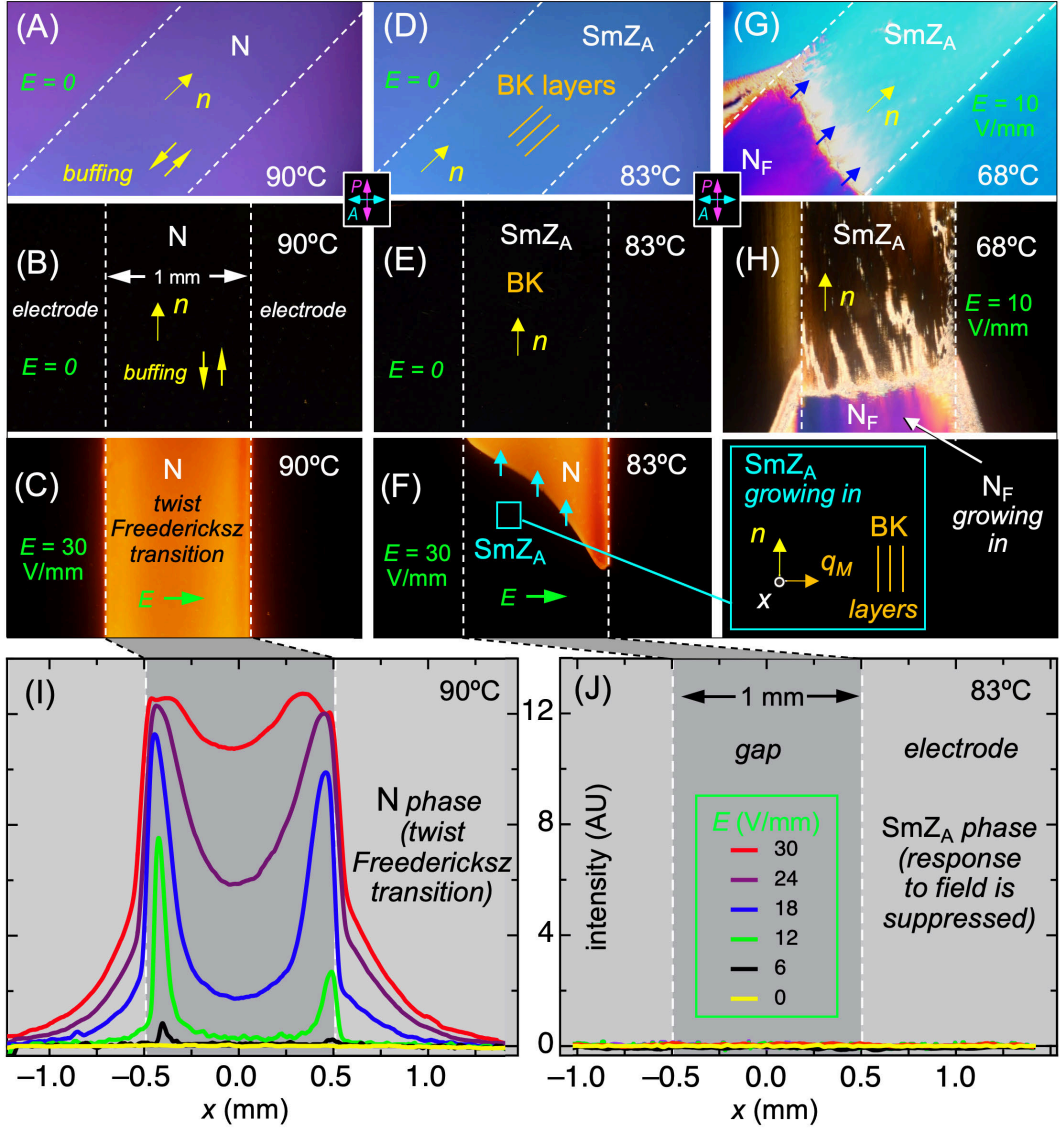

***Figure 3***: Textures of DIO in a $d = 3.5$ μm cell with in-plane electrodes spaced by 1 mm and anti-parallel buffing parallel to the electrode edges, cooled at -1ºC/min and viewed in DTOM with polarizer and analyzer as indicated. (***A-C***) Planar-aligned nematic monodomain with ***n*** parallel to the plates and to the electrode edges, showing (*A*) a birefringence color corresponding to $d = 3.5$ and $\Delta n = 0.18$, (*B*) excellent extinction with ***n*** parallel to the polarizer, and (*C*) the transmission due to twist of ***n*** during a field-induced twist Freedericksz transition. (***D,E***) Bookshelf-aligned smectic $Z_A$ monodomain formed on cooling. The smectic layers are parallel to ***n*** and normal to the cell plates. The birefringence is slightly larger than in the nematic phase and excellent extinction is again obtained with ***n*** parallel to the polarizer. (***F***) Optical response of coexisting nematic and SmZ$_A$ phases to an applied 200 Hz square wave field of peak amplitude, *E*. The nematic region undergoes a twist Freedericksz transition but in the smectic Z$_A$ phase, rotation of the director is suppressed by the layering. (***G,H***) N$_F$ phase growing into the SmZ$_A$ upon

further cooling. The antiparallel buffing stabilizes a π-twist state in the polarization/director field of the $N_F$, which does not extinguish between crossed polarizer and analyzer at any cell orientation. (**I,J**) Optical intensity scans along the magenta dashed lines across the bottom of the images in (C,F), probing the electric field-induced in-plane twist Freedericksz reorientation in the $N_F$ phase and showing that there is no observable electro-optic effect in the $SmZ_A$ phase. In this geometry, where the layers have bookshelf (BK) alignment, an in-plane field would tend to rotate the director out of the smectic layer planes, a deformation that is locally resisted by the $SmZ_A$ structure.

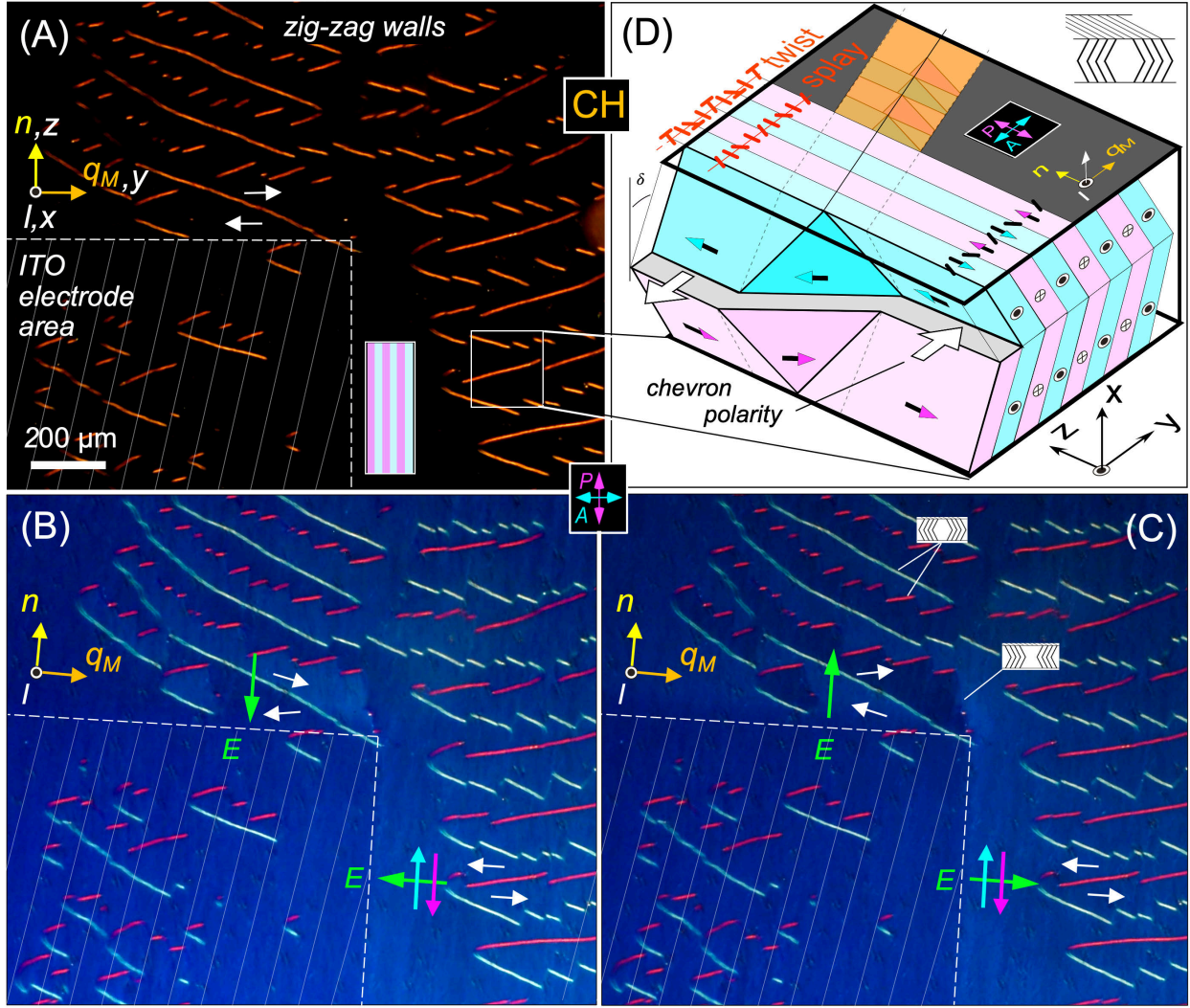

*Figure 4*: Zig-zag defects in a SmZ$_A$ cell with chevron layering confirm the lamellar structure of the SmZ$_A$. Zig-zag walls appear in smectic cells originally with the layers substantially normal to the cell plates (bookshelf geometry) when the layers are caused to shrink in thickness. A common response to accommodate layer shrinkage in cells is for the layers to tilt, forming the chevron variation (CH) of bookshelf alignment, and zig-zag walls are the layering defects which mediate binary change in the pointing direction of the chevron tip. (**A-C**) Zig-zags generated in the SmZ$_A$ phase as it is grown by cooling in a weak electric field (1 Hz square-wave, ~10 V/mm). Zig-zag walls are of two types, sketched in (*C*): "diamond" walls [⟨⟨⟨⟨⟩⟩⟩⟩], detailed in (*D*), where the chevrons point out from the wall center, and which run nearly along $q_M$, i.e., nearly normal to the layers; and "broad" walls [⟩⟩⟩⟩⟩⟨⟨⟨⟨⟨], where the chevrons point in toward the defect wall, which run parallel to the layers and to the director, **n**. These zig-zag walls are analogous to those found in planar-aligned SmC and SmC* cells (see ***Fig. S7***). (**D**) Detailed structure of a diamond wall in which the normal to the diamond-shaped layer elements is ro-

-23-

tated to make an angle with the cell normal that is larger than the layer tilt angle, $\delta$, of the chevrons, enabling the wall to space out the chevron tips in opposite directions [21]. The orientation and birefringence of the diamonds walls causes them to transmit light when the overall chevron structure is at or near extinction, as in ($A$-$C$). The orientation of the diamond elements can be determined from their extinction angle (see **Fig. S8C**). The broad wall structure is less visible in the SmZ$_A$ than in the SmC cells shown in **Fig. S6** because in the SmZ$_A$ the rotations of layer elements in the wall do not result in reorientation of $n$. The presence of fringing fields near the electrodes [marked with white dashed lines in ($B,C$)] enables $E$ fields to be applied either parallel or and normal to $n$ in separate regions of the  cell. The difference in optical saturation observed in these regions for different signs of field applied parallel to $n$ ($B,C$) reveals that the polarity of the chevron structure generates an interfacial polarization that is indicated by the white arrows: the chevron areas are polar along $y$, with the white arrows rotating in opposite directions on opposite sides of a zig-zag wall, indicating a polar rotation of the director near the chevron interface when a field $E_z$ is applied. In contrast, with the field along $y$ there is no polar reorientational response of the director to change of field sign ($B,C$), indicating that the antiferroelectricity cancels the net polarization in the bulk (magenta, cyan arrows).

.

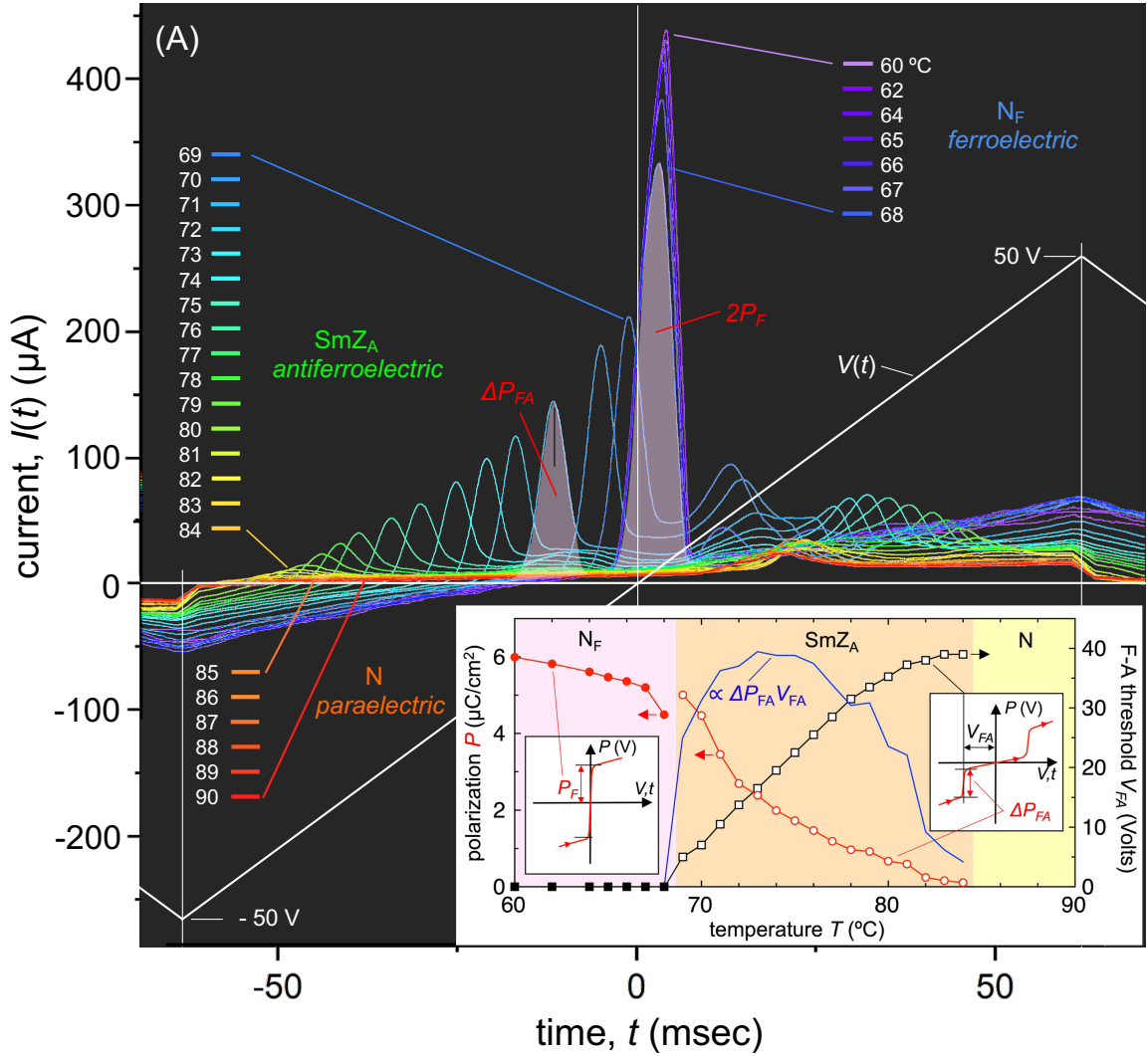

***Figure 5***: (***A***) Measured $I(t)$-$V(t)$ characteristics of DIO as a function of temperature, revealing a large field-induced shift in the SmZ$_A$–N$_F$ phase transition temperature for electric field applied along ***n***. The plot shows the current response $i(t)$ of DIO in a $d = 100\ \mu$m ITO-sandwich cell with bookshelf layering, to a 5 Hz, 50 V amplitude triangle wave ($v(t)$, white curve) applied during an N–SmZ$_A$–N$_F$ cooling scan. In the N phase ($T > 84$ºC), the current shows the expected cell capacitance step and an ion bump. In the SmZ$_A$ phase ($84$ºC $> T > 68$ºC), new polarization peaks appear at the highest voltages, growing in area, with their peak center voltages $V_{FA}$ becoming smaller on cooling. This is typical antiferroelectric behavior, the peaks marking the transition at finite voltage between the field-induced ferroelectric (F) state and the equilibrium antiferroelectric (A) state. (***B***) Polarization values $P_{FA}(T)$ [open circles] and ferroelectric-to-antiferroelectric depolarization voltage $V_{FA}(T)$ [open squares] obtained from the depolarization current response (in the time range $t < 0$), where there is the least interference from ion flow). The corresponding field $E_{FA}(T) = V_{FA}(T)/100\mu$m. The open squares give the first order SmZ$_A$–N$_F$

phase boundary in the ($E_z$–$T$) plane, and the open circles the polarization change $P_{FA}$ at this transition. The transition entropy, $\Delta S$, also decreases along this line to zero at the maximum $E_z$ with increasing $E_z$. The blue curve is proportional to the product $P_{FA}(T)V_{FA}(T)$, and therefore proportional to the stabilization energy of the antiferroelectric state, which is maximum at $T$ = 73ºC. The SmZ$_A$ to N$_F$ transition occurs between T = 69ºC and 68ºC, with the current transforming to a single peak centered around $V$ = 0, typical of field-induced, Goldstone-mode reorientation of macroscopic polarization, $P_F$ [red solid circles]. This peak area corresponds to a net polarization comparable to that measured in the N$_F$ phase of RM734 [7].

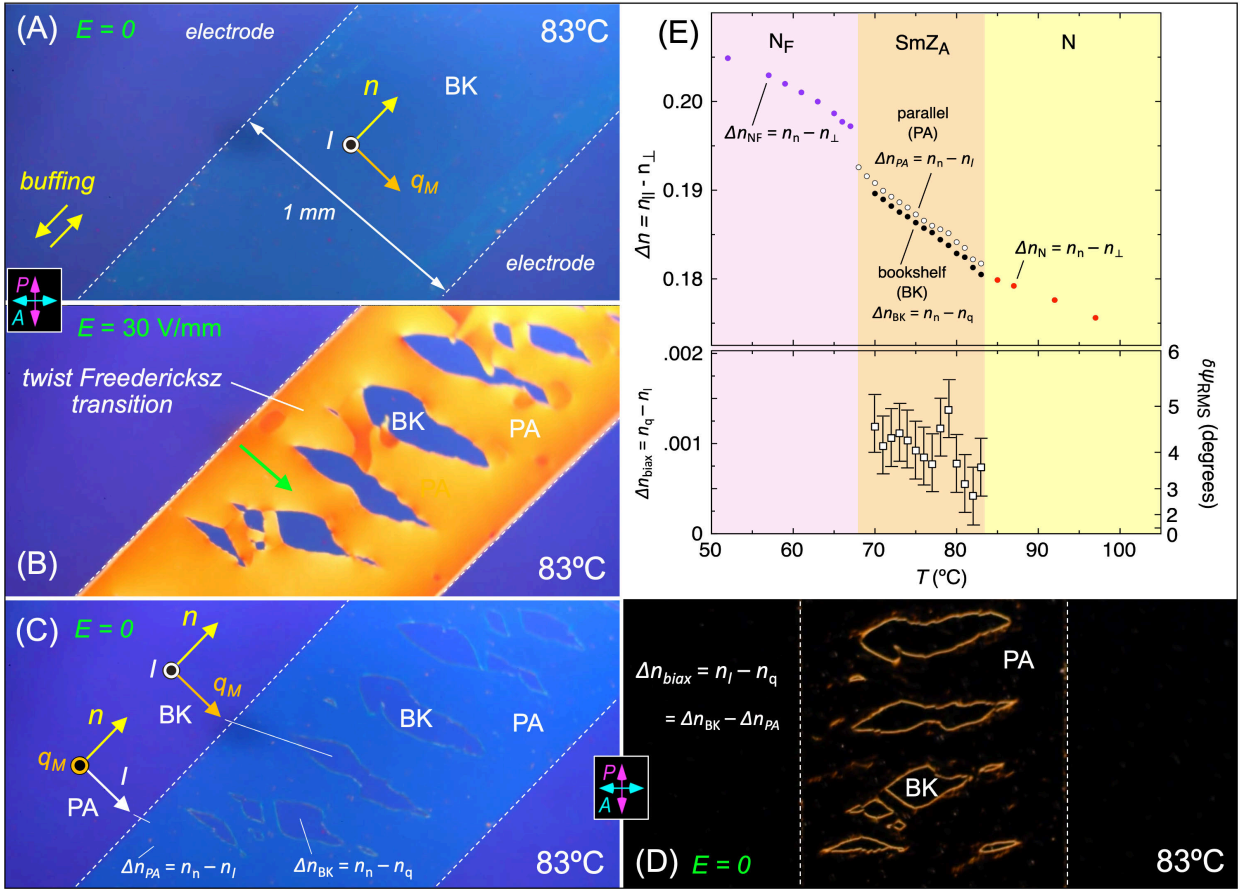

**Figure 6**: Coexisting SmZ$_A$ domains in the cell of **Fig. 3** with bookshelf (BK) and parallel (PA) layer orientation, enabling measurement of the optical biaxiality of the SmZ$_A$ phase. (**A**) Orientation of ($l$,$q_M$,$n$), the principal axes of the optical dielectric tensor $\varepsilon$ of the SmZ$_A$ in a BK domain. (**B**) Field-induced growth of a PA domain of SmZ$_A$. (**C**) Comparative geometry of the BK and PA domains. The difference in the birefringence of the BK and PA two regions yields the biaxiality, $\Delta n_{biax} \equiv n_l - n_q$, that distinguishes the modulated SmZ$_A$ from the uniaxial nematic. (**D**) When the field is removed, both BK and PA domains adopt uniform director alignment. (**A-D**) With $d = 3.5$ μm and since $\Delta n \sim 0.18$ in both domains, the path difference is $\Delta nd \sim 600$ nm, corresponding to birefringence colors in a magenta-to-purple-to-blue-to-green band for increasing $\Delta nd$ on the Michel-Levy chart [33]. With the director oriented at 45° to the polarizers, the BK regions are purple while the planar-aligned regions are blue, indicating that $\Delta n_{BK} < \Delta n_{PA}$ and therefore that $n_q > n_l$, indicating that the director modulation of the layering is splay, rather than twist (**Fig. 4D**). (**E**) Use of a Berek compensator enables measurement of $\Delta n_{BK}$ and $\Delta n_{PA}$, from which $\Delta n_{biax}(T)$ can be calculated under the assumption that the biaxiality is due to splay modulation of the director field. The resulting RMS director tilt of the periodic layering is in the range 3° < $\sqrt{\langle \delta \psi^2(y) \rangle}$ < 5°. The peak amplitude of $\delta \psi$ will depend on the profile $\delta \psi(y)$.

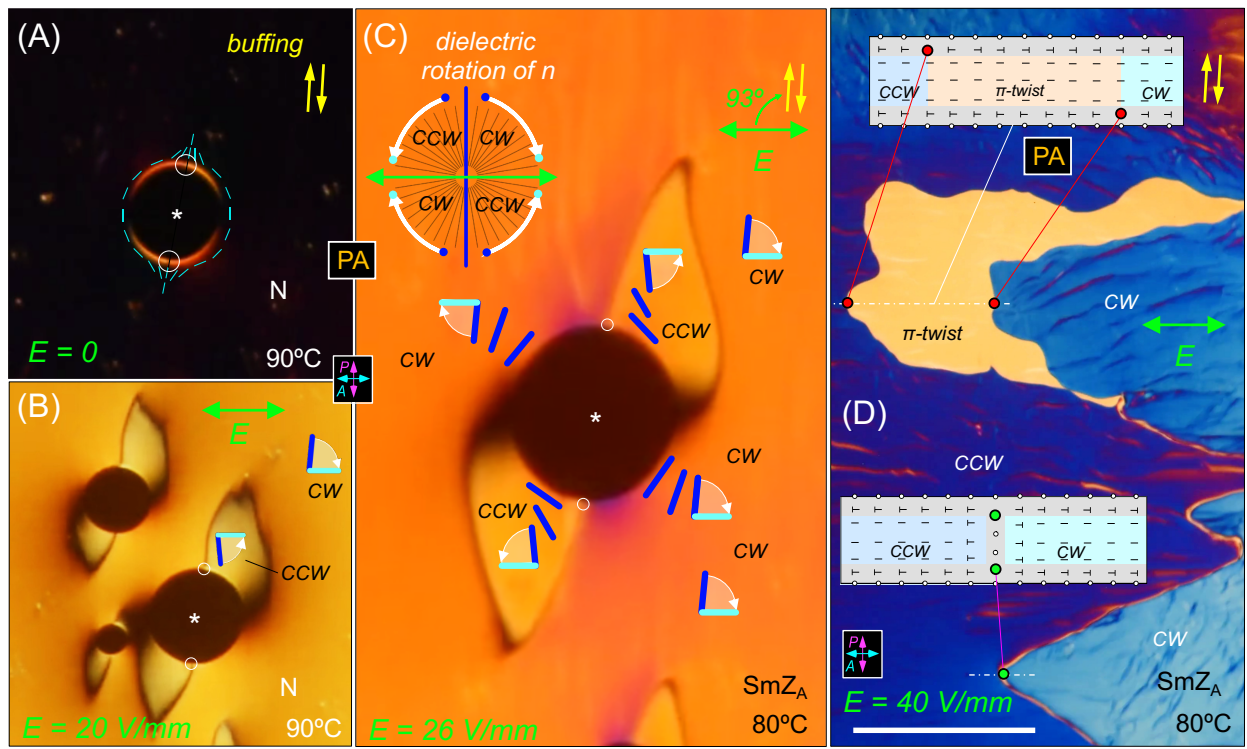

**Figure 7:** Twist Freedericksz response to applied in-plane field in the N and parallel-aligned (PA) SmZ$_A$ phases. The buffing orientation is 93º from the applied field direction. (***A***) Extinction of the starting planar-aligned nematic state. (***B,C***) In the N phase, the dielectric torque-induced orientational Freedericksz response gives a clockwise (CW) rotation of the director everywhere except near bubble inclusions (black circles), where the anchoring conditions induce CCW orientation in the wing-shaped domains at the first and third quadrants of the circle. The starred bubble has a diameter of 105 $\mu$m. (***C***) This dielectric behavior is duplicated exactly in the SmZ$_A$ phase, showing that the response of the parallel-aligned SmZ$_A$ to in-layer fields parallel to the layers is strictly dielectric, and that the phase is therefore antiferroelectric. (***D***) At higher field the bulk orientation becomes more uniform and therefore optically more extinguishing. This decouples the $\pi$-surface domain walls (magenta dots), enabling them to move apart (red dots), leaving a $\pi$-twist domain in between. Scale bar = 100 $\mu$m.

## REFERENCES

1  P. Debye, Einige Resultate einer kinetischen Theorie der Isolatoren. *Physikalische Zeitschrift* **13**, 97-100 (1912).

2  M. Born, Über anisotrope Flüssigkeiten. Versuch einer Theorie der flüssigen Kristalle und des elektrischen Kerr-Effekts in Flüssigkeiten. *Sitzungsber. Preuss. Akad Wiss.* **30**, 614-650 (1916).

3  R.J. Mandle, S.J. Cowling, J.W. Goodby, A nematic to nematic transformation exhibited by a rod-like liquid crystal. *Phys. Chem. Chem. Phys.* **19**, 11429–11435 (2017). DOI: 10.1039/C7CP00456G

4  A. Mertelj, L. Cmok, N. Sebastián, R.J. Mandle, R.R. Parker, A.C. Whitwood, J.W. Goodby, M. Čopič, Splay nematic phase. *Phys. Rev. X* **8**, 041025 (2018). DOI: 10.1103/PhysRevX.8.041025

5  N. Sebastian, L. Cmok, R.J. Mandle, M. Rosario de la Fuente, I. Drevenšek Olenik, M. Čopič, A. Mertelj, Ferroelectric-ferroelastic phase transition in a nematic liquid crystal. *Phys. Rev. Lett.* **124**, 037801 (2020). DOI: 10.1103/PhysRevLett.124.037801

6  H. Nishikawa, K. Shiroshita, H. Higuchi, Y. Okumura, Y. Haseba, S. Yamamoto, K. Sago, H. Kikuchi, A fluid liquid-crystal material with highly polar order. *Adv. Mater.* **29**, 1702354 (2017). DOI: 10.1002/adma.201702354

7  X. Chen, E. Korblova, D. Dong, X. Wei, R. Shao, L. Radzihovsky, M.A. Glaser, J.E. Maclennan, D. Bedrov, D.M. Walba, N.A. Clark, First-principles experimental demonstration of ferroelectricity in a thermotropic nematic liquid crystal: spontaneous polar domains and striking electro-optics, *Proceedings of the National Academy of Sciences of the United States of America* **117**, 14021–14031 (2020). DOI: 10.1073/pnas.2002290117

8  J. Li, H. Nishikawa, J. Kougo, J. Zhou, S. Dai, W. Tang, X. Zhao, Y. Hisai, M. Huang, S. Aya, Development of ferroelectric nematic fluids with giant-ε dielectricity and nonlinear optical properties, *Science Advances* **7** (17), eabf5047 (2021). DOI: 10.1126/sciadv.abf5047

9  X. Chen, Z. Zhu, M.J. Magrini, E. Korblova, C.S. Park, M.A. Glaser, J.E. Maclennan, D.M. Walba, N.A. Clark, Ideal mixing of paraelectric and ferroelectric nematic phases in liquid crystals of distinct molecular species. arXiv: 2110.1082 (2021).

10 R.J. Mandle, S.J. Cowling, J.W. Goodby, Structural variants of RM734 in the design of splay nematic materials. DOI: 10.26434/chemrxiv.14269916.v1

11 G. Friedel, Les états mésomorphes de la matière. *Ann. Phys. (Paris)* **18**, 273–474 (1922). DOI: 10.1051/anphys/192209180273

12 G. Friedel, F. Grandjean, Cristaux liquides. Réponse à MM. G. Friedel, et F. Grandjean a M. O. Lehmann *Bulletin de Minéralogie* **33**, 300–317 (1910). DOI : 10.3406/bulmi.1910.3440

13 H. Sackmann, D. Demus, The polymorphism of liquid crystals, *Molecular Crystals and Liquid Crystals* **2**, 81–102, (1966). DOI: 10.1080/15421406608083062

14 N.A. Clark, Antiferroelectric Smectic Ordering as a Prelude to the Ferroelectric Nematic: Introducing the Smectic Z Phase. Plenary Lecture 2, 18th International Conference on Ferroelectric Liquid Crystals (FLC2021), Ljubljana, Slovenia (2021).

15 C.R. Safinya, M. Kaplan, J. Als-Nielsen, R.J. Birgeneau, D. Davidov, J.D. Litster, D.L. Johnson M.E. Neubert, High-resolution x-ray study of a smectic-A – smectic-C phase transition. *Physical Review B* **21**, 4149–4153 (1980). DOI: 10.1103/PhysRevB.21.4149o

16 T.P. Rieker, N.A. Clark, G.S. Smith, D.S. Parmar, E.B. Sirota, C.R. Safinya, 'Chevron' local layer structure in surface-stabilized ferroelectric smectic-C cells, *Physical Review Letters* **59**, 2658–2661 (1987). DOI:10.1103/PhysRevLett.59.2658

17 Smectic C 'chevron', a planar liquid-crystal defect: implications for the surface-stabilized ferroelectric liquid-crystal geometry, N.A. Clark, T.P. Rieker, *Physical Review A: Rapid Communications* **37**, 1053–1056 (1988). DOI: 10.1103/physreva.37.1053

18 N.A. Clark, T. P. Rieker, J.E. Maclennan, Director and layer structure of SSFLC cells, *Ferroelectrics* **85**, 79–97 (1988). DOI: 10.1080/00150198808007647

19 T.P. Rieker, N.A. Clark, G.S. Smith, C.R. Safinya, Layer and director structure in surface stabilized ferroelectric liquid crystal cells with non-planar boundary conditions, *Liquid Crystals* **6**, 565–576 (1989). DOI: 10.1080/02678298908034176

20 T.P. Rieker, N.A. Clark, C.R. Safinya, Chevron layer structures in surface stabilized ferroelectric liquid crystal (SSFLC) cells filled with a material which exhibits the chiral nematic to smectic C* phase transition. *Ferroelectrics* **113**, 245–256 (1991). DOI: 10.1080/00150199108014067

21 S.T. Lagerwall, *Ferroelectric and antiferroelectric liquid crystals*, (Wiley VCH, Weinheim, 1999) ISBN 3-527-2983 1-2.

22 Y. Takanishi, Y. Ouchi, H. Takezoe. Chevron layer structure in the smectic-A phase of 8CB. *Japanese Journal of Applied Physics Part 2-Letters* **28**, L487–L489  (1989). DOI: 10.1143/JJAP.28.L487

23 X. Cheng, M. Prehm, M. K. Das, J. Kain, U. Baumeister, S. Diele, D. Leine, A. Blume, C. Tschierske, Calamitic bolaamphiphiles with (semi)perfluorinated lateral chains: polyphilic block molecules with new liquid crystalline phase structures. *J. Am. Chem. Soc.* **125** (36), 10977–10996 (2003). DOI: 10.1021/ja036213g

24 J.O. Rädler, I. Koltover, T. Salditt, C.R. Safinya, Structure of DNA-cationic liposome complexes: dna intercalation in multilamellar membranes in distinct interhelical packing regimes. *Science* **275** (5301), 810–814 (1997). DOI: 10.1126/science.275.5301.810.

25 R.A. Reddy, C. Tschierske, Bent-core liquid crystals: polar order, superstructural chirality and spontaneous desymmetrisation in soft matter systems. *J. Mater. Chem.* **16**, 907 (2006). DOI: 10.1039/B504400F

26 J.V. Selinger, Director deformations, geometric frustration, and modulated phases in liquid crystals. arXiv:2103.03803 (2021).

27 X. Chen, E. Korblova, M.A. Glaser, J. E. Maclennan, D.M. Walba, N.A. Clark, Polar in-plane surface orientation of a ferroelectric nematic liquid crystal: polar monodomains and twisted state electro-optics, *Proceedings of the National Academy of* Sciences **118**, e2104092118 (2021). DOI: 10.1073/pnas.2104092118

28 P.G. de Gennes, An analogy between superconductors and smectics A. *Solid State Communications* **10**, 753–756 (1972). DOI: 10.1016/0038-1098(93)90291-T

29  H. Birecki, R. Schaetzing, F. Rondelez, J.D. Litster Light-scattering study of a smectic-A phase near the smectic-A – nematic transition. *Phys. Rev. Lett.* **36**, 1376–1379 (1976). DOI: 10.1103/PhysRevLett.36.1376

30  N.A. Clark, R.B. Meyer, Strain-induced instability of monodomain smectic A and cholesteric liquid-crystals. *Applied Physics Letters* **22**, 494–493 (1973). DOI: 10.1063/1.1654481

31  S. Brown, E. Cruickshank, J.M.D. Storey, C.T. Imrie, D. Pociecha, M. Majewska, A. Maka, E. Gorecka, Multiple polar and non-polar nematic phases. *Chemphyschem* **22**, (2021). DOI: 10.1002/cphc.202100644

32  P. Nacke, A. Manabe, M. Klasen-Memmer, M. Bremer, F. Giesselmann, New example of a ferroelectric nematic phase material. Poster P2, 18th International Conference on Ferroelectric Liquid Crystals (FLC2021), Ljubljana, Slovenia (2021).

33  M. Magnus, *Michel-Levy Color Chart* (Carl Zeiss Microscopy GmbH, 07745 Jena, Germany) *URL: www.zeiss.com/microscopy*

34  D.A. Coleman, J. Fernsler, N. Chattham, M. Nakata,Y. Takanishi, E. Körblova, D.R. Link, R.-F. Shao, W.G. Jang, J.E. Maclennan, O. Mondainn-Monval, C. Boyer, W. Weissflog, G. Pelzl, L.-C. Chien, D.M. Walba, J. Zasadzinski, J. Watanabe, H. Takezoe, N.A. Clark, Polarization modulated smectic liquid crystal phases. *Science* **301**, 1204–1211 (2003). DOI: 10.1126/science.1084956

35  R.D. Kamien, J.V. Selinger, Order and frustration in chiral liquid crystals. *J. Phys.: Condens. Matter* **13**, R1–R22 (2001). DOI: 10.1088/0953-8984/13/3/201

36  M. Seul and D. Andelman, Domain shapes and patterns: the phenomenology of modulated phases. *Science,* **267** 476–483 (1995). DOI: 10.1126/science.267.5197.476

37  M. Marder, H.L. Frisch, J.S. Langer, H.M. McConnell, Theory of the intermediate rippled phase of phospholipid bilayers. *Proceedings of the National Academy of* Sciences **81**, 6559–6561 (1984). DOI: 10.1073/pnas.81.20.6559

38  A.E. Jacobs, G. Goldner, D. Mukamel, Modulated structures in tilted chiral smectic films. *Physical Review A* **45**, 5783-5788 (1992). DOI: 10.1103/PhysRevA.45.5783

39   G.A. Hinshaw, Jr., R.G. Petschek, Transitions and modulated phases in centrosymmetric ferroelectrics: Mean-field and renormalization-group predictions. *Physical Review B* **37,** 2133-2155 (1988). DOI: 10.1103/PhysRevB.37.2133

40  J.W. Felix, D. Mukamel, R.M. Hornreich, Novel class of continuous phase transitions to incommensurate structures, *Physical Review Letters* **57**, 2180-2183 (1986). DOI: 10.1103/PhysRevLett.57.2180

41  K. De'Bell, A.B. MacIsaac, J.P. Whitehead, Dipolar effects in magnetic thin films and quasi-two-dimensional systems.  *Reviews of Modern Physics* **72**,  225-257 (2000). DOI: 10.1103/RevModPhys.72.225

42  G. Szabo, G. Kadar Magnetic hysteresis in an Ising-like dipole-dipole model. *Physical Review B* **58**, 5584-5587 (1998). DOI: 10.1103/PhysRevB.58.5584

43  H.Z. Cummins, Experimental studies of structurally incommensurate crystal phases. *Physics Reports* **185**, 211–409 (1980). DOI: 10.1016/0370-1573(90)90058-A

44  R. Blinc, A.P. Levanyuk, <u>Incommensurate phases in dielectrics.</u> Volume 1 Fundamentals; Volume 2 Materials. Elsevier Science Publishers, Amsterdam (1986).

45  A.B. MacIsaac, J.P. Whitehead, M.C. Robinson, K. De'Bell, Striped phases in two-dimensional dipolar ferromagnets. *Physical Review B* **51**, 16033-16045 (1995). DOI: 10.1103/PhysRevB.51.16033

46  M. Grousson, G. Tarjus, P. Viot, Phase diagram of an Ising model with long-range frustrating interactions: A theoretical analysis. *Physical Review E* **62**,7781-7792(2000). DOI: 10.1103/physreve.62.7781

47  M. Grousson, G. Tarjus, P. Viot, Monte Carlo study of the three-dimensional Coulomb frustrated Ising ferromagnet. *Physical Review E* **64**,036109 (2000). DOI: 10.1103/PhysRevE.64.036109

48  R.M. Hornreich, M. Luban, S. Shtrikman, Critical Behavior at the Onset of k-Space instability on the $\lambda$ line. *Physical Review Letters* **35**, 1678-1681 (1975). DOI: 10.1103/PhysRevLett.35.1678

49  S.A. Pighín, S.A. Cannas, Phase diagram of an Ising model for ultrathin magnetic films: Comparing mean field and Monte Carlo predictions. P*hysical Review E* **75,** 224433 (2007). DOI: 10.1103/PhysRevB.75.224433

50  M.Y. Choi, Domain-wall pinning in the incommensurate phase of sodium nitrite. *Physical Review B* **37**, 5874-5876 (1988).  DOI: 10.1103/PhysRevB.37.5874

51  Y. Ishibashi, H. Shiba, Successive phase-transitions in ferroelectric $NaNO_2$ and $SC(NH_2)_2$. *Journal of the Physical Society of Japan* **45**, 409–413 (1978).  DOI: 0.1143/JPSJ.45.409

52  Y. Ishibashi, H. Shiba, Incommensurate-commensurate phase-transitions in ferroelectric substances *Journal of the Physical Society of Japan* **45**, 1592–1599 (1978). DOI: 10.1143/JPSJ.44.159

53  Y. Shiozaki, Satellite x-ray scattering and structural modulation of thiourea. *Ferroelectrics* **2**, 245–260 (1978).  DOI: 10.1080/00150197108234099

54  S. Tanisaki, I. Shibuya, Microdomain structure in paraelectric phase of $NaNO_2$. *Journal of the Physical Society of Japan* **16**, 579–579 (1961).  DOI: 10.1143/JPSJ.16.579

55  Y. Yamada, S. Hoshino, I. Shibuya, Phase transition in $NaNO_2$. *Journal of the Physical Society of Japan* **18**, 1594–1603 (1963).  DOI:10.1143/JPSJ.18.1594

56  M. Iwata, H. Orihara, Y. Ishibashi, Phenomenological theory of the linear and nonlinear dielectric susceptibilities in the type-II incommensurate phase. *Journal of the Physical Society of Japan* **67**, 3130-3136 (1998). DOI: 10.1143/jpsj.67.3130

57  S.V. Berezovsky Soliton regime in the model with no Lifshitz invariant. arXiv:cond-mat/9909079 (1999).

58  M.Y. Choi, Domain-wall pinning in the incommensurate phase of sodium nitrite. *Physical Review B* **37**, 5874-5876 (1988).  DOI: 10.1103/PhysRevB.37.5874

59  I. Aramburu, G. Madariaga, and J.M. Perez-Mato, Phenomenological model for type-II incommensurate phases having a soliton regime: Thiourea case. *Physical Review B* **49**, 802-814 (1994).  DOI: 10.1103/PhysRevB.49.802

60  I. Aramburu, G. Madariaga, J.M Perez-Mato, A structural viewpoint on the sine-Gordon equation in incommensurate phases. *J. Phys.: Condensed Matter* **7**, 6187-6196 (1995). DOI: 10.1088/0953-8984/7/31/003

61  D. Durand, F. Denoyer, D. Lefur, Neutron diffraction study of sodium nitrite in an applied electric field. *Journal de Physique* **44**  L207–L216 (1983). DOI : 10.1051/jphyslet:0183004405020700

62  D. Durand, F. Denoyer, R. Currat, M. Lambert Chapter 13 - Incommensurate phase in NaNO₂. <u>Incommensurate phases in dielectrics: 2 Materials</u>, R. Blinc and A.P. Levanyuk, Ed. (Elsevier  Science Publishers, Amsterdam 1986).  DOI: 10.1016/B978-0-444-86970-8.50010-X

63  G.H.F. van Raaij, K.J.H. van Bemmel, T. Janssen, Lattice models and Landau theory for type-II incommensurate crystals. *Physical Review B* **62**, 3751-3765 (2000). DOI: 10.1103/PhysRevB.62.3751

64  A.E. Jacobs, C. Grein, F. Marsiglio, Rippled commensurate state: A possible new type of incommensurate state. *Physical Review B* **29**, 4179-4181 (1984). DOI: 10.1103/PhysRevB.29.4179

65  A.E. Jacobs, Intrinsic domain-wall pinning and spatial chaos in continuum models of one-dimensionally incommensurate systems. *Physical Review B* **33**, 6340-6345 (1986). DOI: 10.1103/PhysRevB.33.6340

66  P. Lederer, C.M. Chaves, Phase diagram of thiourea at atmospheric pressure under electric field: a theoretical analysis. *Journal de Physique Lettres* **42**, L127 - L130  (1981). DOI: 10.1051/jphyslet:01981004206012700

67  J.P. Jamet, Electric field phase diagram of thiourea determined by optical birefringence. J. *J. Physique - Letteres* (1981) **42**, L123 - L125  (1981).   DOI: 10.1051/jphyslet:01981004206012300

68  F. Denoyer, R. Currat, Chapter 14 - Currat, Modulated Phases in Thiourea - <u>Incommensurate phases in dielectrics: 2 Materials</u>, R. Blinc and A.P. Levanyuk, Ed. (Elsevier  Science Publishers, Amsterdam 1986).  DOI: 10.1016/B978-0-444-86970-8.50010-X

69  A. Aharony, Critical behavior of magnets with dipolar interactions. V. Uniaxial magnets in *d*-dimensions. *Physical Review B* **8**, 3363-3370 (1973). DOI: 10.1103/PhysRevB.8.3363

70  J. Als-Nielsen, Experimental test of renormalization group theory on the uniaxial, dipolar coupled ferromagnet LiTbF₄. *Physical Review Letters* **37**, 1161-1164 (1976). DOI: 10.1103/PhysRevLett.37.1161

71  E.I. Kats, Stability of the uniform ferroelectric nematic phase. *Physical Review E* **103**, 012704 (2021).  DOI: 10.1103/PhysRevE.103.012704

72  E.I. Kats, V.V. Lebedev, and A.R. Muratov, Weak crystallization theory. *Phys. Rep.* **228**, 1-91 (1993).  DOI: 10.1016/0370-1573(93)90119-X

73  A.E. Jacobs, D. Mukamel, Universal incommensurate structures. *Journal of Statistical Physics* **58**, 503-509 (1990). DOI: 10.1007/BF01112759

74  D.R. Link, J.E. Maclennan, and N.A. Clark, Simultaneous observation of longitudinal and transverse ferroelectricity in freely suspended films of an antiferroelectric liquid crystal. *Physical Review Letters* **77**, 2237–2240 (1996).  DOI: 10.1103/PhysRevE.52.2120

75  K.H. Kim, K. Ishikawa, H. Takezoe, A. Fukuda, Orientation of alkyl chains and hindered rotation of carbonyl groups in the smectic C*phase of antiferroelectric liquid-crystals studied by polarized fourier-transform infrared-spectroscopy. *Physical Review E* **51**, 2166–2175 (1995).  DOI: 10.1103/PhysRevE.51.2166

76  R. Bartolino, J. Doucet and G. Durand, Molecular tilt in the smectic C phase: a zigzag model. *Annale de Physique* **3**, 389–395 (1978).  DOI: 0.1051/anphys/197803030389

77  D.M. Walba and N.A. Clark, Model for the molecular origins of the polarization in ferroelectric liquid crystals, *Proceedings of the Society of Photo-Optical Instrumentation Engineers* **825**, 81– 87 (1987).  DOI: 10.1117/12.941988

Antiferroelectric Smectic Ordering as a Prelude to the Ferroelectric Nematic:
Introducing the Smectic $Z_A$ Phase

Xi Chen[1], Vikina Martinez[1], Eva Korblova[2], Guillaume Freychet[3], Mikhail Zhernenkov[3],
Matthew A. Glaser[1], Cheng Wang[4], Chenhui Zhu[4], Leo Radzihovsky[1],
Joseph E. Maclennan[1], David M. Walba[2], Noel A. Clark[1]*

[1]Department of Physics and Soft Materials Research Center,
University of Colorado, Boulder, CO 80309, USA

[2]Department of Chemistry and Soft Materials Research Center,
University of Colorado, Boulder, CO 80309, USA

[3]Brookhaven National Laboratory, National Synchrotron Light Source-II
Upton, NY 11973, USA

[4]Advanced Light Source, Lawrence Berkeley National Laboratory, Berkeley, CA 94720, USA

*Abstract*

We have structurally characterized the liquid crystal phase that appears as an intermediate state when a dielectric nematic, having polar disorder of its molecular dipoles, transitions to the almost perfectly polar-ordered ferroelectric nematic. This intermediate phase, which fills a 100-year-old void in the taxonomy of smectics and which we term the "smectic $Z_A$", is antiferroelectric, with the nematic director and polarization oriented parallel to smectic layer planes, and the polarization alternating in sign from layer to layer. The period of this polarization wave (~180 Å) is mesoscopic, corresponding to ~40 molecules side-by-side, indicating that this lamellar structure is collectively stabilized. A Landau free energy, originally formulated to model incommensurate antiferroelectricity in crystals, describes the key features of the nematic–$SmZ_A$–ferroelectric nematic phase sequence.

<u>*TABLE OF CONTENTS*</u>

## SECTION S1 – Materials and Methods

*Synthesis of DIO* – First reported by Nishikawa et al. [1], *DIO* (2,3',4',5'-tetrafluoro-[1,1'-biphenyl]-4-yl 2,6-difluoro-4-(5-propyl-1,3-dioxane-2-yl)benzoate, **Fig. S1,** compound **3**) is a rod-shaped molecule about 20 Å long and 5 Å in diameter, with a longitudinal electric dipole moment of about 11 Debye. The synthesized compound was found to melt at $T = 173.6^\circ$C and have an isotropic (I) phase and two additional phases with nematic-like character. The transition temperatures on cooling were I – 173.6°C – N –84.5°C – M2 – 68.8°C –$N_F$ – 34°C – X, very similar to the temperatures reported by Nishikawa.

Our synthetic scheme, shown in **Fig. S1,** is based on a general synthetic reaction. The key intermediate **1** was purchased from Manchester Organics Ltd., UK and intermediate **2** from Sigma-Aldrich Inc., USA. Reactions were performed in oven-dried glassware under an atmosphere of dry argon. Purification by flash chromatography was performed with silica gel (40–63 microns) purchased from Zeochem AG. Analytical thin-layer chromatography (TLC) was performed on silica gel 60 $F_{254}$ TLC plates from Millipore Sigma (Darmstadt, Germany). Compounds were visualized using short-wavelength ultra-violet (UV). Nuclear magnetic resonance (NMR) spectra were obtained using a Bruker Avance-III 300 spectrometer. NMR chemical shifts were referenced to deuterochloroform (7.24 ppm for $^1$H, 77.16 ppm for $^{13}$C).

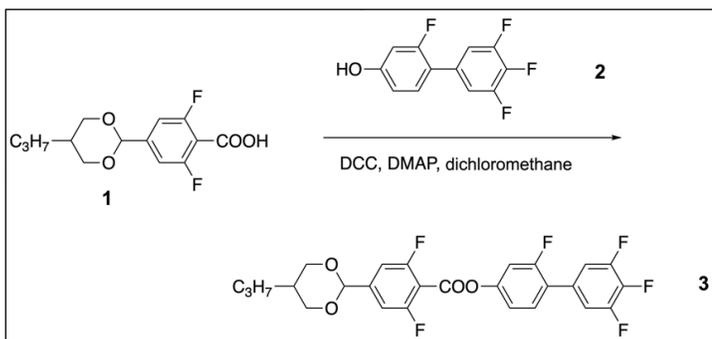

***Figure S1***: Synthesis scheme for DIO.

To a suspension of compound **1** (3.44 g, 12 mmol) and the intermediate **2** (2.91 g, 12 mmol) in $CH_2Cl_2$ (125 mL) was added DCC (4.95 g, 24 mmol) and a trace of DMAP.

The reaction mixture was stirred at room temperature for 4 days, then filtered, washed with water, and with brine, dried over $MgSO_4$, filtered, and concentrated at reduced pressure.

The resulting product was purified by flash chromatography (silica gel, petroleum ether/10% ethyl acetate). The crude product was crystallized by dissolving in boiling 75 mL petroleum ether/20% ethyl acetate solvent mixture, followed by cooling down to -20°C for 1 hour, yielding 2.98 g (49%) white needles of compound **3**.

$^1$H NMR (300 MHz, Chloroform-*d*) δ 7.64 – 7.35 (m, 1H), 7.24 – 6.87 (m, 6H), 5.40 (s, 1H), 4.42 – 4.14 (m, 2H), 3.54 (ddd, J = 11.6, 10.3, 1.5 Hz, 2H), 2.14 (tddd, J = 11.4, 9.2, 6.9, 4.6 Hz, 1H), 1.48 – 1.23 (m, 2H), 1.23 – 1.01 (m, 2H), 0.94 (t, J = 7.3 Hz, 3H).

$^{13}$C NMR (75 MHz, Chloroform-*d*) δ 162.50, 162.43, 160.85, 159.08, 159.00, 157.52, 150.87, 150.72, 145.60, 145.47, 130.51, 130.46, 117.99, 117.94, 113.22, 113.17, 113.02, 112.93, 112.88, 110.65, 110.30, 110.26, 110.00, 109.95, 98.65, 98.62, 98.59, 72.41, 33.72, 30.05, 19.35, 14.01.

## SECTION S2 – X-ray Diffraction from RM734 and DIO

*Figure S2*: Comparison of WAXS from RM734 and DIO. The samples have the nematic director magnetically-aligned along **z** by a ~1 Tesla magnetic field. The color gamuts are linear in intensity, with the (black) minima corresponding to zero intensity. Scattered intensity line scans $I(q_z)$, from such WAXS images are shown in *Figure S7*. The WAXS scattering patterns that appear in RM734 and DIO upon cooling into the N phase are strikingly similar, and do not change very much on cooling into the lower temperature phases, apart from the appearance of the SmZ$_A$ layering peaks (also visible in SAXS images [2]). (*A-D*) The WAXS patterns exhibit familiar nematic diffuse scattering features at $q_z \sim 0.25$ Å$^{-1}$ and $q_y \sim 1.4$ Å$^{-1}$, arising respectively from the end-to-end and side-by-side pair-correlations, that are typically generated by the steric rod-shape of the molecules and are located respectively at ($2\pi$/molecular length $\sim 0.25$ Å$^{-1}$) and ($2\pi$/molecular width $\sim 1.4$ Å$^{-1}$) [1,2]. In contrast to typical nematics, RM734 also exhibits a series of scattering bands for $q_y < 0.4$ Å$^{-1}$ and $q_z > 0.25$ Å$^{-1}$, initially reported in RM734 and its homologs [2,3,4,5]. Interestingly, DIO presents a qualitatively very similar scattering pattern (*A,B,D*), but with an even more well-defined peak structure, likely a result of the higher variation of excess electron density along the molecule associated with the fluorines. Also notable is that the $q_z \sim 0.25$ Å$^{-1}$ feature in RM734 is weak compared to that found in typical nematics such as 5CB and all-aromatic

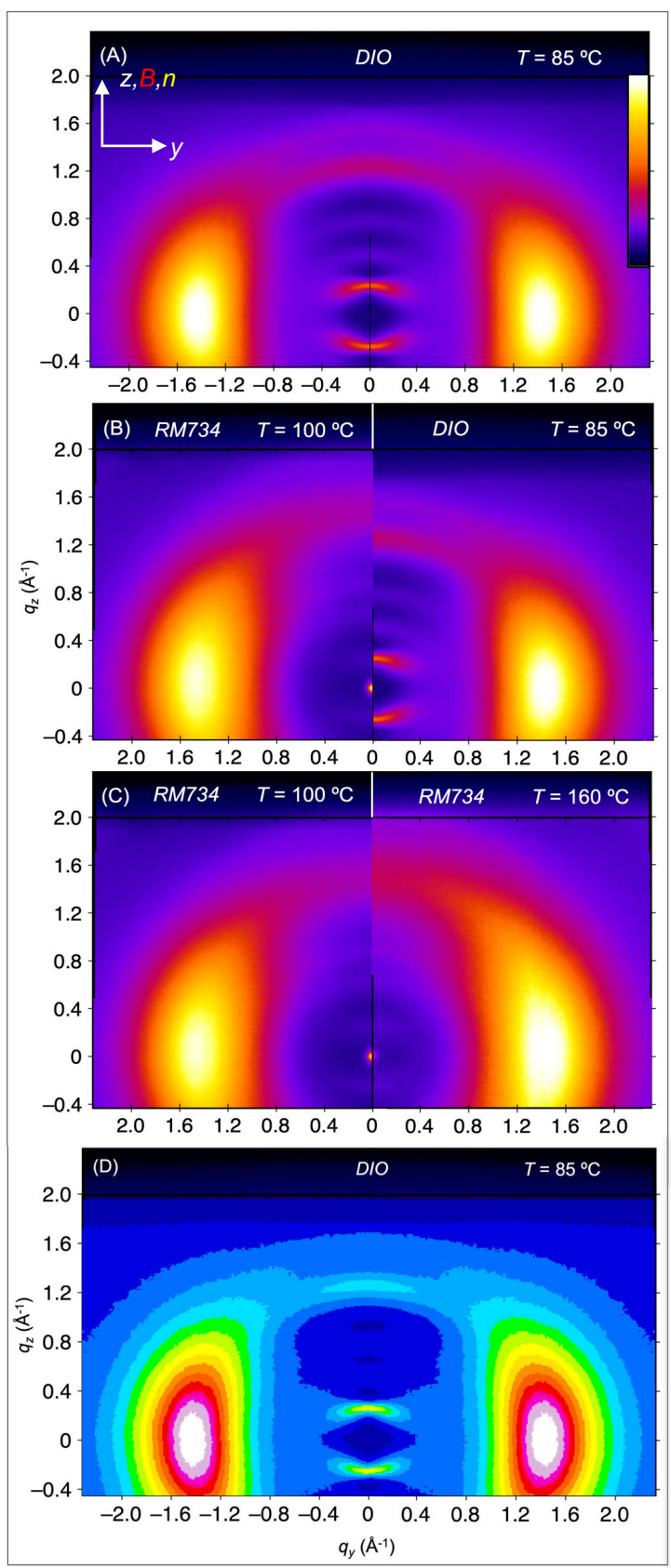

LCs [6,7].  We attribute this weak scattering to the head-to-tail electrostatic adhesion in RM734, which makes the molecular correlations along $z$ more polar and chain-like, reducing the tail-to-head gaps between molecules.  The resulting end-to-end correlations are then like those in main-chain LC polymers, where there are no gaps, and, as a result, the scattering along $q_z$ is weaker than in monomer nematics [8,9,10]. In DIO, on the other hand, the trifluoro group at the end of the molecule generates large electron density peaks that periodically mark chain-like correlations along $z$, even if there are no gaps, resulting in strong scattering at $q_z \sim 0.25$ Å$^{-1}$.  The signal around $q = 0$ (for $q < 0.1$ Å$^{-1}$) in $B,C$ is from stray light.

***Figure S3***: Line scans, $I(q_z)$, of WAXS images of scattering from DIO and RM734 similar to those in ***Figure S2*** along $q_z$ at $q_y$ = 0.004 Å⁻¹. (*A*) Comparison of RM734 at $T$ = 160°C and DIO at $T$ = 85°C, the temperatures where the peak structures of $I(q_z)$ are the strongest, reveals common features. These include the previously reported intense diffuse scattering features at $q_z \sim 0.25$ Å⁻¹ and $q_y \sim 1.4$ Å⁻¹ [1,2], and the multiplicity of diffuse peaks along $q_z$ previously observed in the RM734 family [2,3,4,5]. The pairs of similarly colored dots show analogous peaks for the two compounds. The diffuse peaks located at $q_{zp} \approx 0.25$ Å⁻¹ (white dots) correspond to short-ranged order with a quasi-periodic spacing of $2\pi / q_{zp} = p \approx 24$ Å in both materials, comparable to the molecular lengths of DIO and RM734, and, in RM734, to the periodicity of the molecular spacing along the director in head-to-tail assemblies seen in simulations [11].

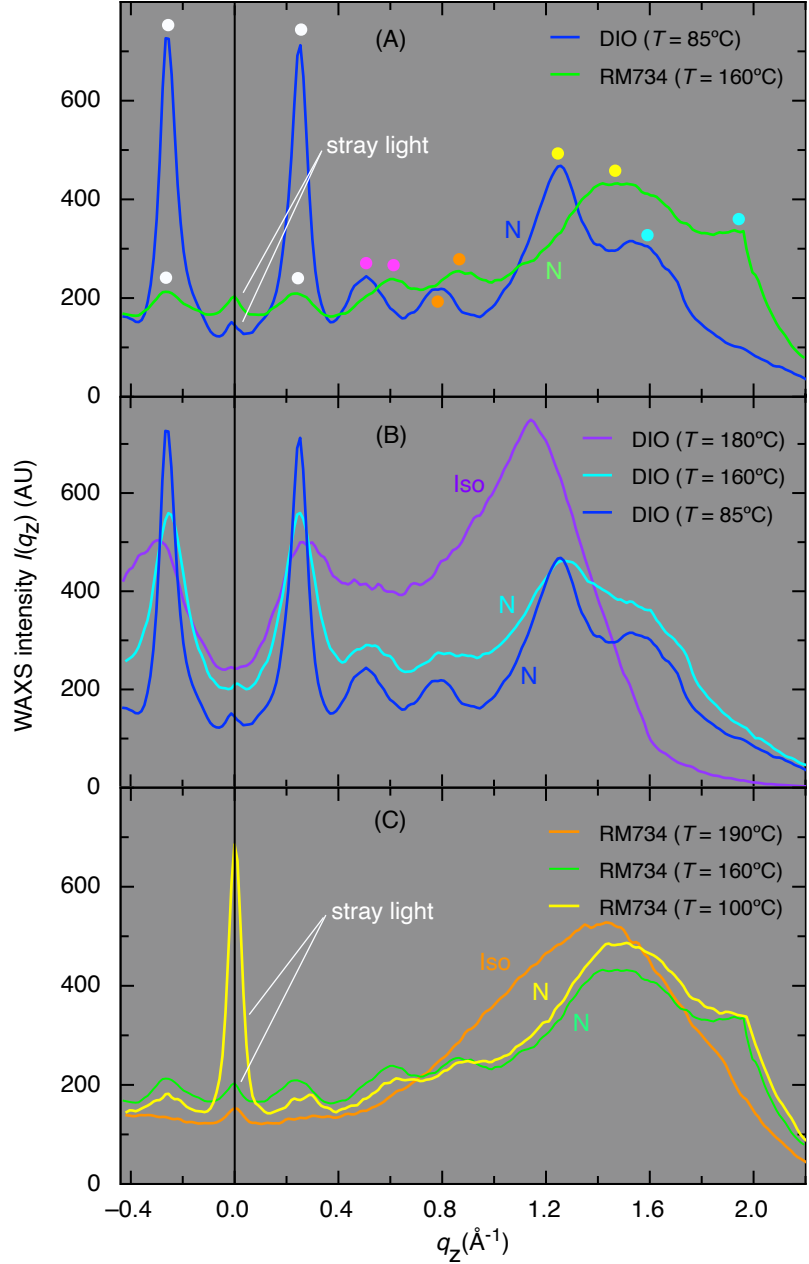

Viewing such head-to-tail assemblies as one-dimensional chains with displacement fluctuations along the chains, the root mean square relative displacement of neighboring molecules along the chain, $\sqrt{\langle \delta u^2 \rangle}$, can be estimated from the ratio of the half-width at half maximum of the scattering peak at $q_{zp}$ (0.04 Å⁻¹) to $q_{zp}$ [see Ref. 12, Supplementary Figure S13]. This ratio is 0.2, which gives $\sqrt{\langle \delta u^2 \rangle}/p \sim 0.25$ and $\sqrt{\langle \delta u^2 \rangle} \sim 5$ Å. This is somewhat larger than the rms displacement found in atomistic computer simulations of ~400 RM734 molecules [11], implying that longer length-scale fluctuations may also be contributing to the peak width. (*B,C*) Temperature dependence of WAXS in DIO and RM734. In the N phase of RM734, the peaks in $I(q_z)$ become better defined with *increasing* temperature, unusual behavior in agreement with the results of [7]. In DIO, in contrast, $I(q_z)$ looks the same through most of the N range and below, with the well-defined peaks appearing at $T = 85^\circ$C (*B*), and broadening with increasing temperature only near the N–Iso transition, before disappearing in the Iso phase. In the N phase, the sequence of peak positions in **DIO** at $q_z$ = [0.25, 0.50, 0.78, 1.25, 1.58 Å$^{-1}$] are in the ratios $q_z/q_{zp}$ = [1, 2.0, 3.1, 5.0, 7.8] which can be indexed approximately as one-dimensional periodicity, giving a harmonic series of multiples of $q_{zp} \approx 0.25$ Å$^{-1}$. In the case of RM734, at $T = 100^\circ$C similar indexing of the peak sequence $q_z$ = [0.28, 0.60, 0.85, 1.46, 1.96 Å$^{-1}$] is possible, with $q_z/q_{zp}$ = [1, 2.2, 3.1, 5.3, 7.0], but there are significant deviations from harmonic behavior at T = 160$^\circ$C, as observed in other members of the RM734 family [13]. In the Iso phase this structure is lost altogether. The nature of the correlations that produce this multiband structure, and their relation to the ferroelectric ordering are currently not understood.

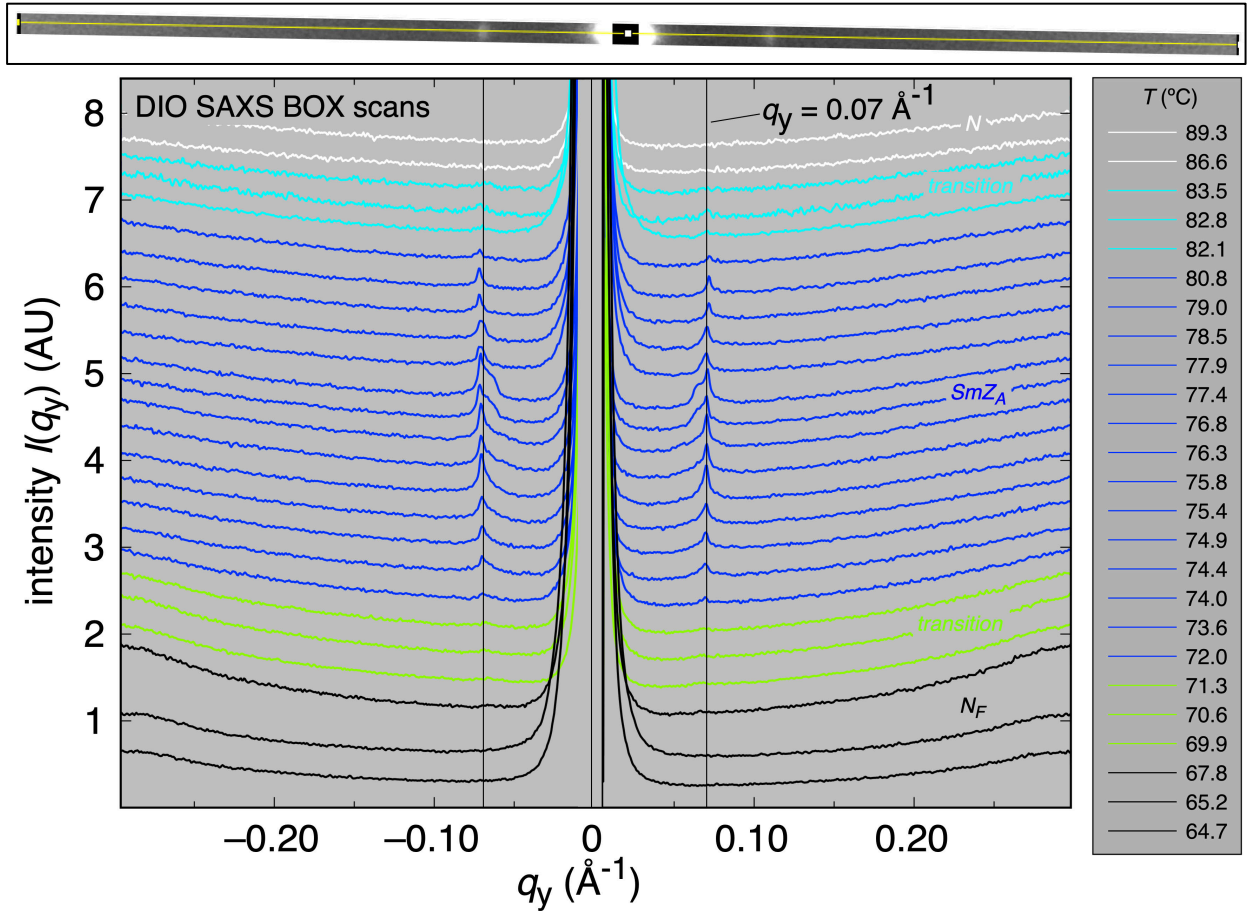

*Figure S4*: BOX scans of the SAXS images of DIO showing the temperature dependence of the lamellar reflection from the SmZ$_A$ layering. $I(q_y)$ is obtained by measuring $I(q_y, q_z)$ over a rectangular area [(-0.3 Å$^{-1}$ < $q_y$ < +0.3 Å$^{-1}$), (-0.009 Å$^{-1}$ < $q_z$ < 0.009 Å$^{-1}$)] centered on the peaks and then averaging over the $q_z$ range. The scans are displaced vertically for clarity, by equal intensity increments. The lowest temperature scan ($T$ = 64.7°C) is not displaced, showing that the amplitude of the layering peak is comparable to the background at this temperature. The shoulders appearing at $T \sim 76$°C are due to spontaneous rearrangement of the layers upon cooling. Peak amplitude and position are plotted in *Figs. 2F,G*. The inset above shows typical BOX and LINE (yellow) scan areas of a SAXS image.

*Figure S5 (below)*: LINE scans of the SAXS images of DIO showing the temperature dependence of the lamellar reflection from the SmZ$_A$ layering. $I(q_y)$ is obtained by measuring $I(q_y, q_z)$ along a line passing through the peak centers. The profiles obtained from BOX scans are shown at the highest and lowest temperatures. The transmitted beam is blocked around $I(q_y) = 0$ by the beamstop. The shoulders appearing for $T \sim 76$°C are due to spontaneous rearrangement of layers upon cooling. The principal peak amplitude and position are plotted vs. $T$ in *Figs. 2F,G*, and the peak half-width at half maximum (HWHM) in *Fig. 2H*.

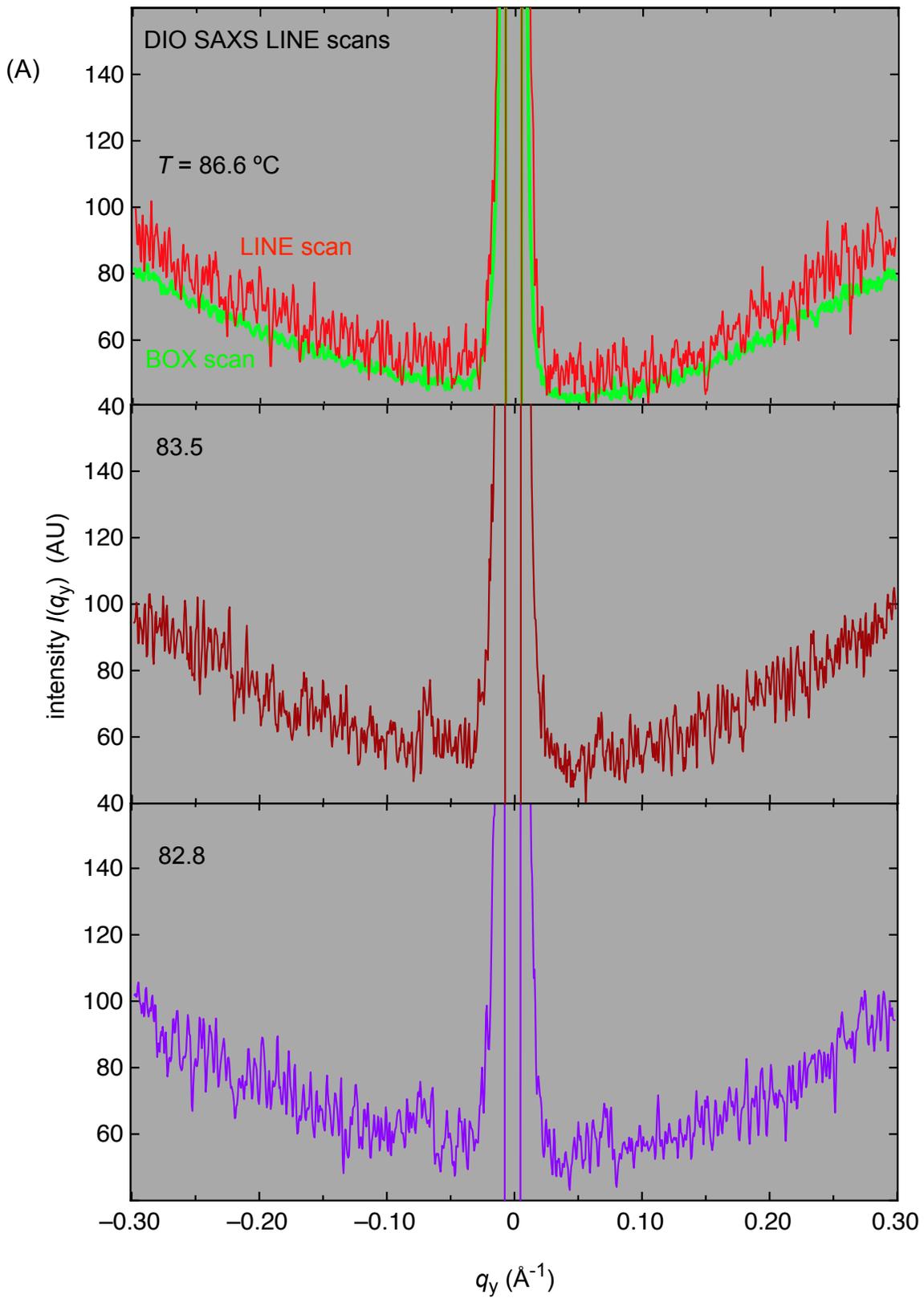

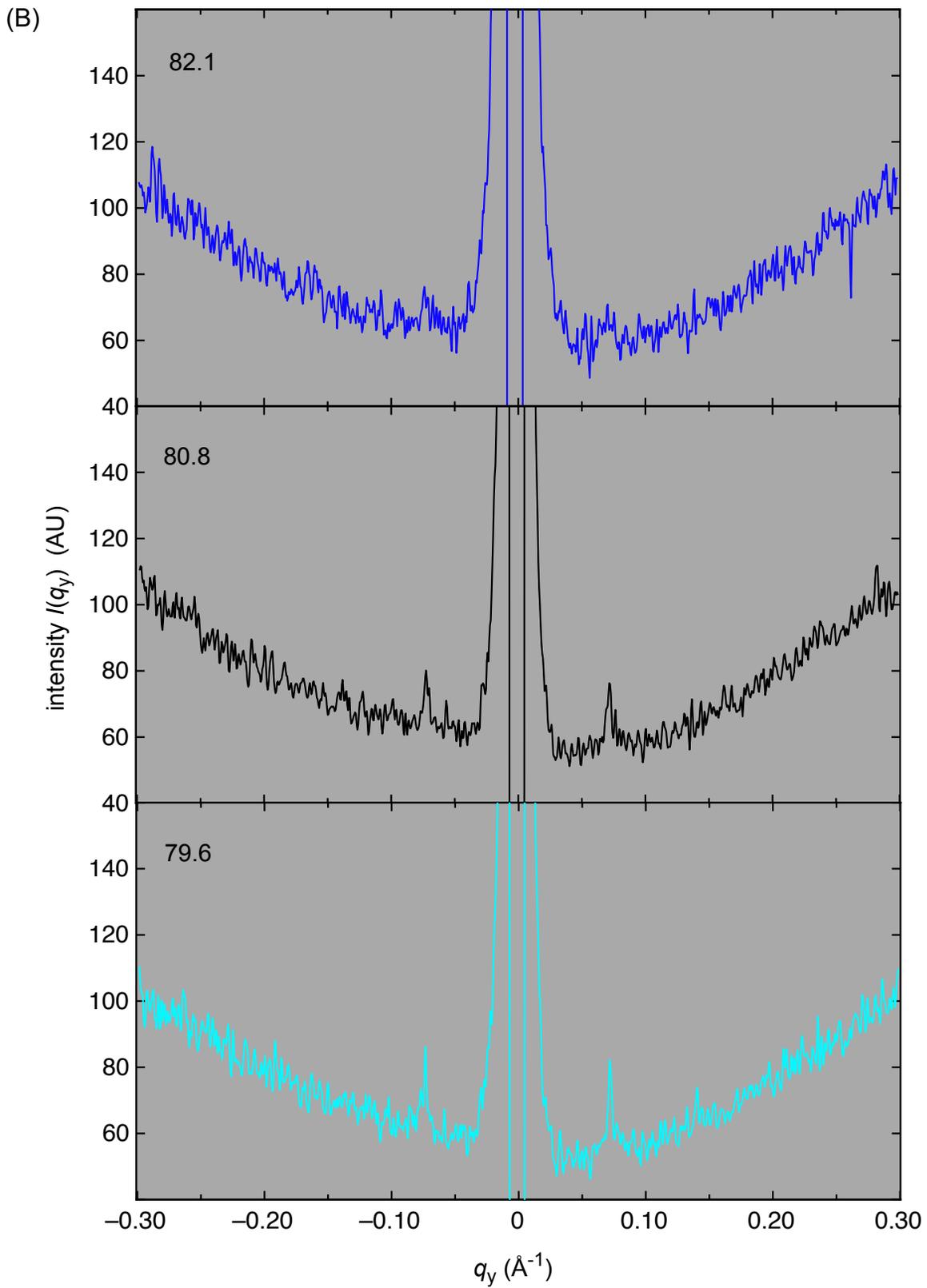

(C)

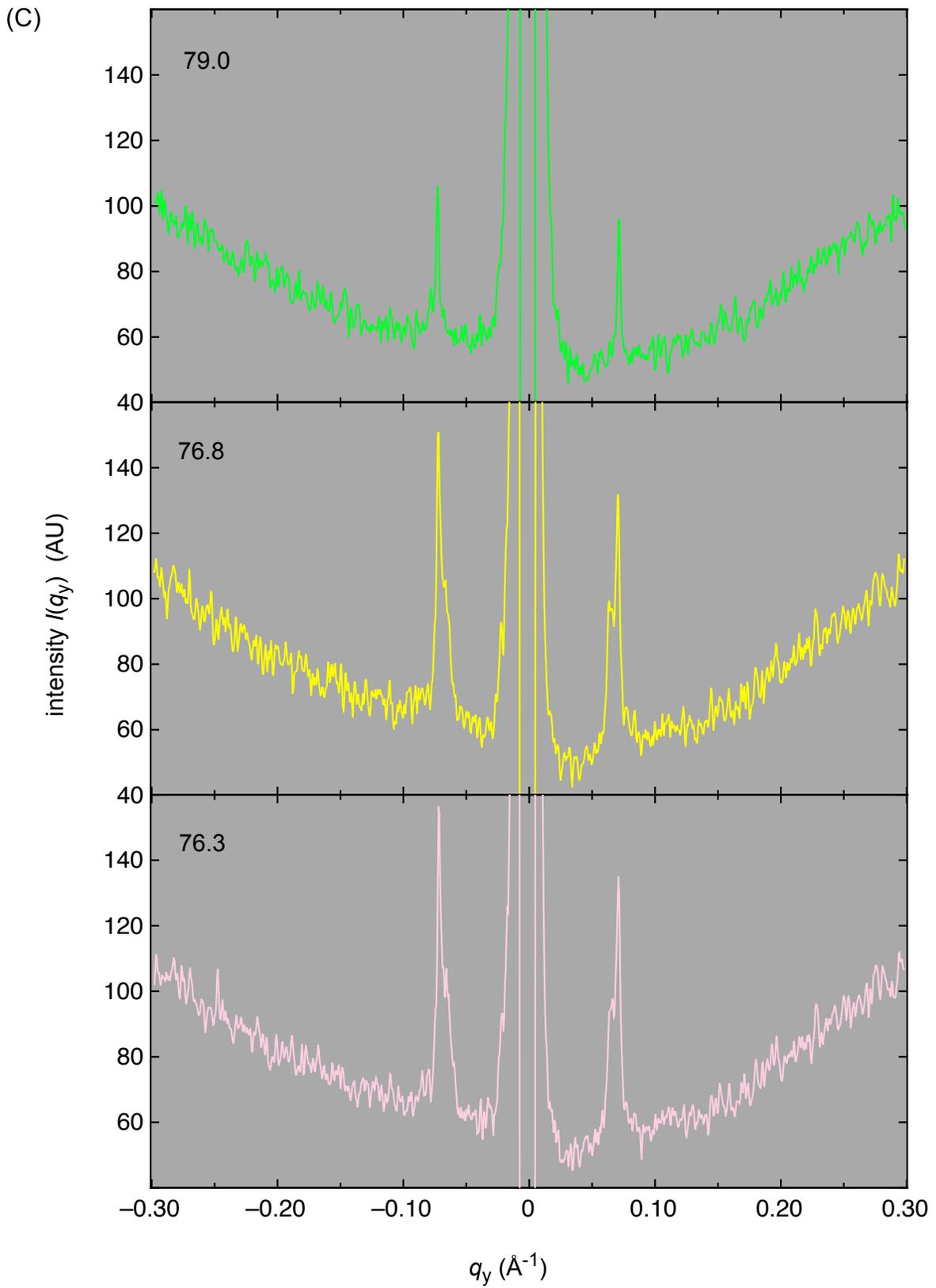

(D)

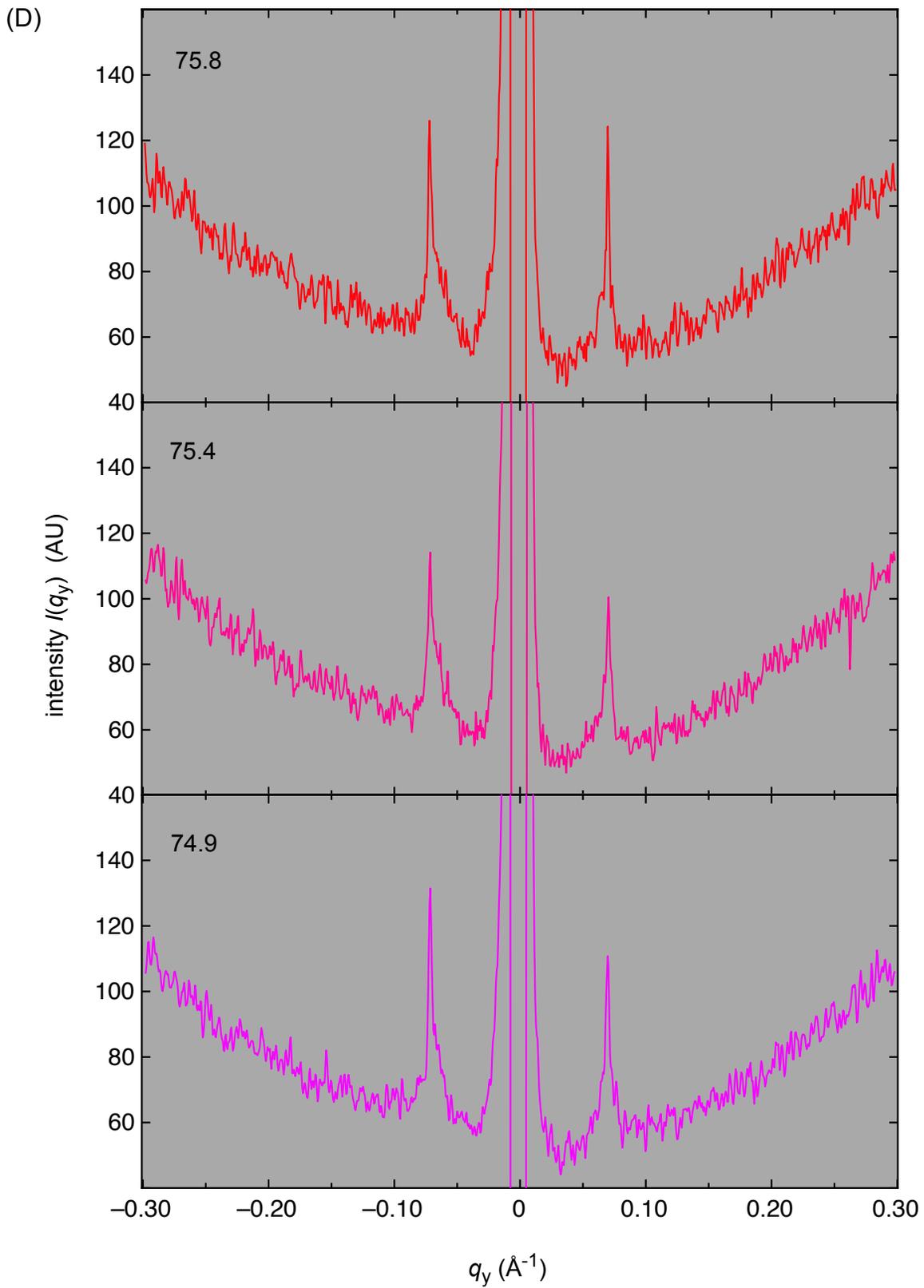

(E)

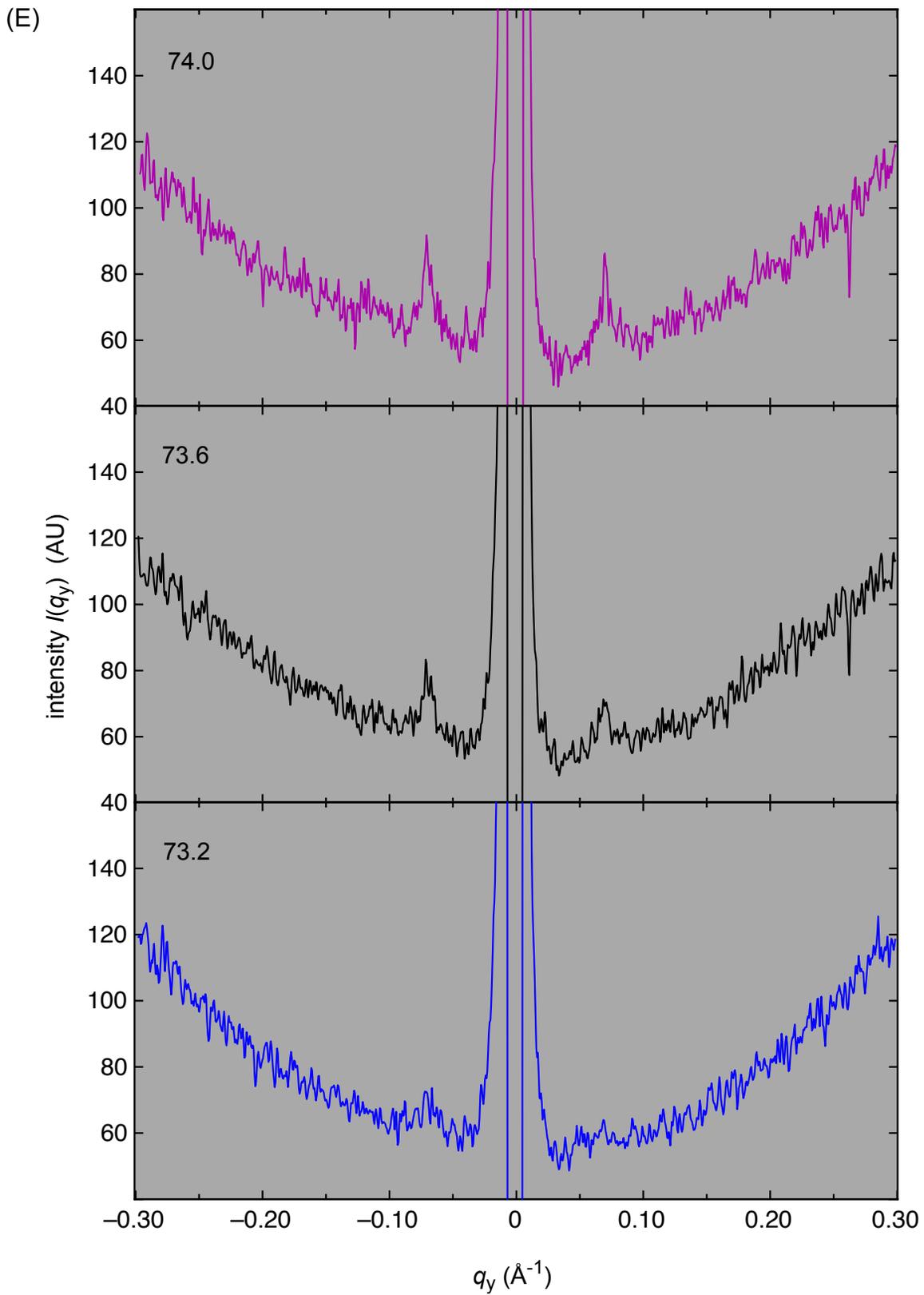

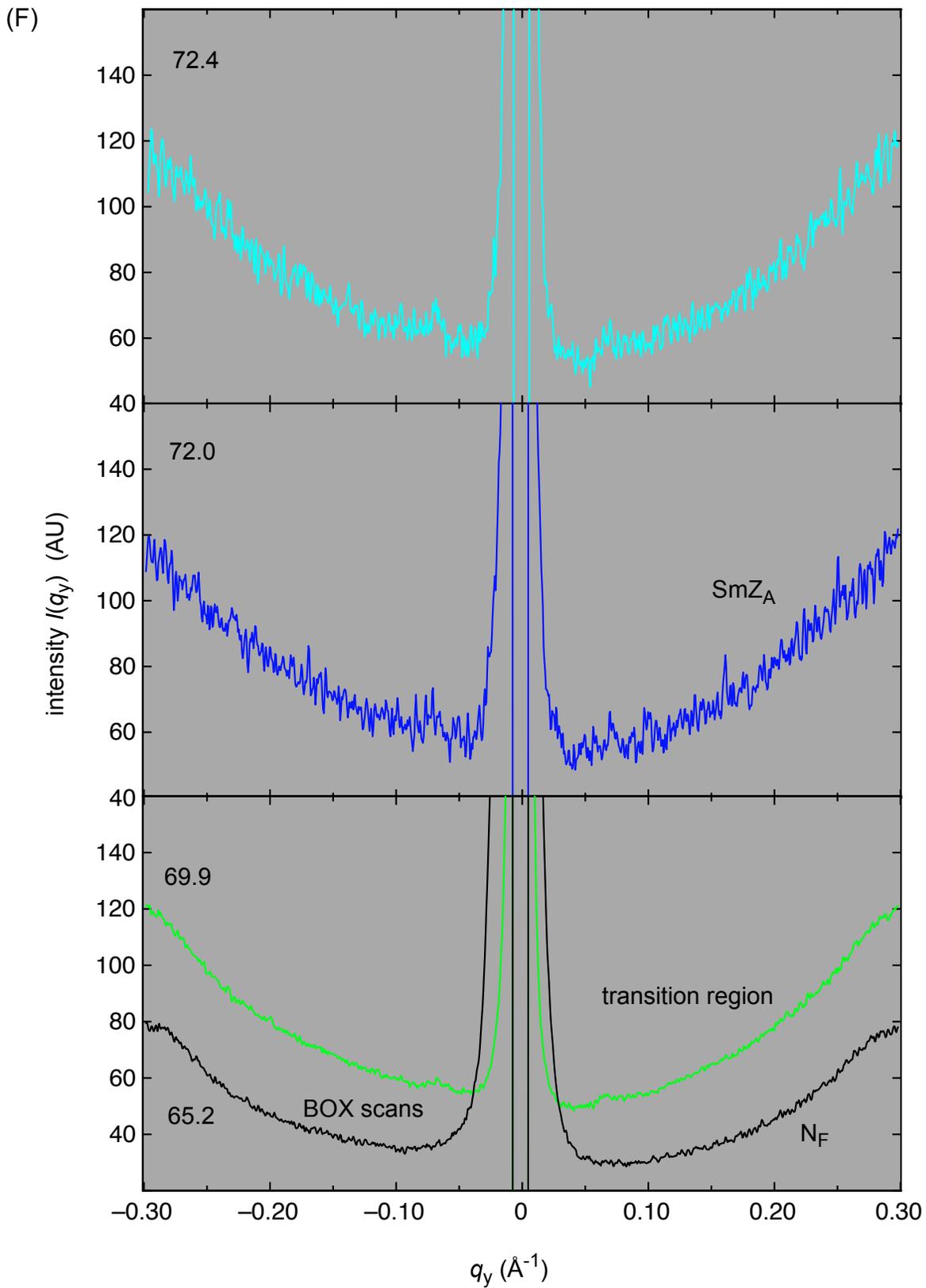

(F)

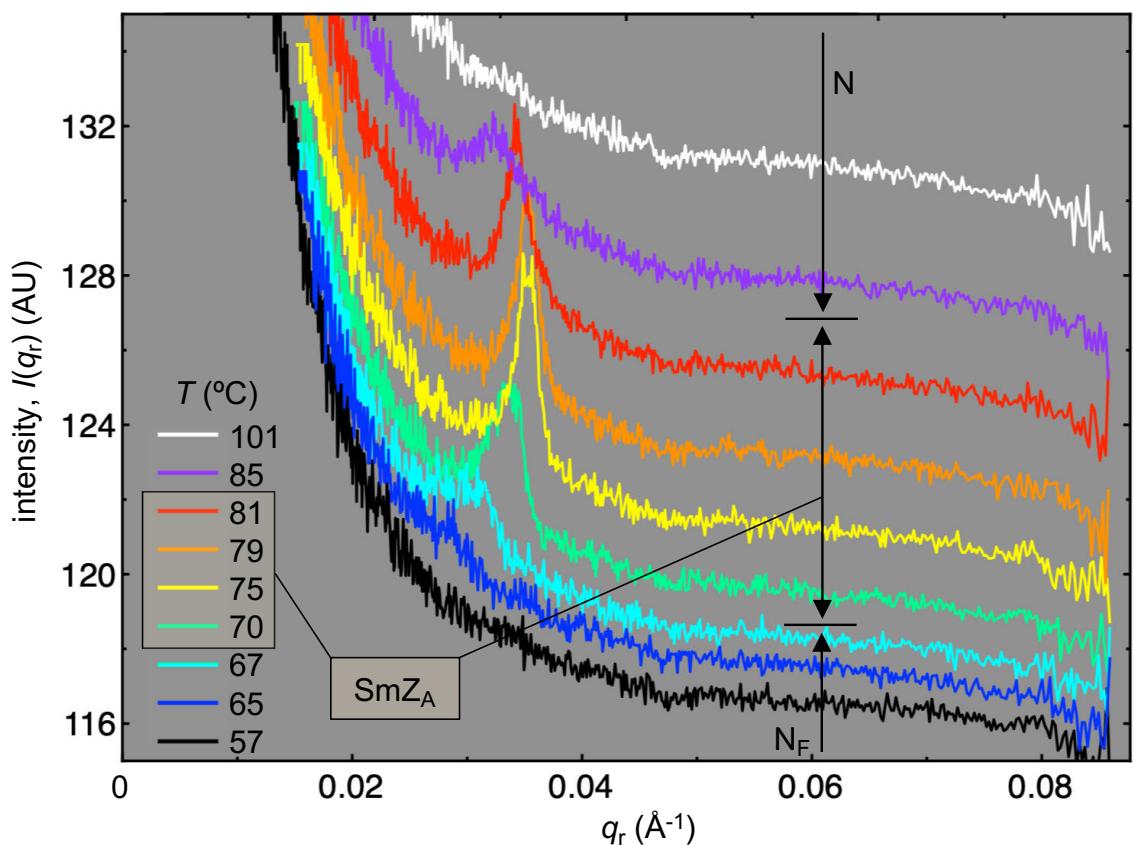

***Figure S5:*** (**G**) Resonant carbon K$_\alpha$-edge x-ray scattering from DIO (incident x-ray wavelength = 44 Å). The DIO sample was filled by capillarity into a $d \sim 5$ μm thick gap between untreated 100 nm thick SiN windows. Depolarized transmission optical microscopy (DTOM) showed random planar alignment of DIO between the plates in thc N, SmZ$_A$, and N$_F$ phases. The x-ray scattering patterns were two-dimensional power rings, which were circularly averaged to give the scattered intensity vs. radial wavevector $q_r$ and temperature curves plotted here. Scans are shifted vertically for clarity. In the SmZ$_A$ temperature range, the scattered intensity $I(q,T)$ shows a single scattered peak in the wavevector range ($0 < q_r < 0.08$ Å$^{-1}$). This peak is located at $q_R \sim 0.035$ Å$^{-1}$, comparable to that expected for Bragg scattering associated with a cell doubling of the periodicity of the lamellar electron density modulation at $q_M \sim 0.070$ Å$^{-1}$. Plots of $2q_R$ and $q_M$ show very similar dependence on temperature over the SmZ$_A$ range, as seen in ***Fig. 2H***. This indicates that the SmZ$_A$ lamellar structure is bilayer, comprising layers which have identical electron-density modulation, but different resonant scattering cross-sections. We attribute this difference to an alternation of polarization direction in adjacent layers. The resonant peak intensity, the fundamental harmonic of $P(r)$, is nearly constant through the SmZ$_A$ range. The SmZ$_A$ - N$_F$ transition exhibits phase coexistence. The N - SmZ$_A$ transition exhibits phase coexistence with a pretranslational diffuse peak in the N phase.

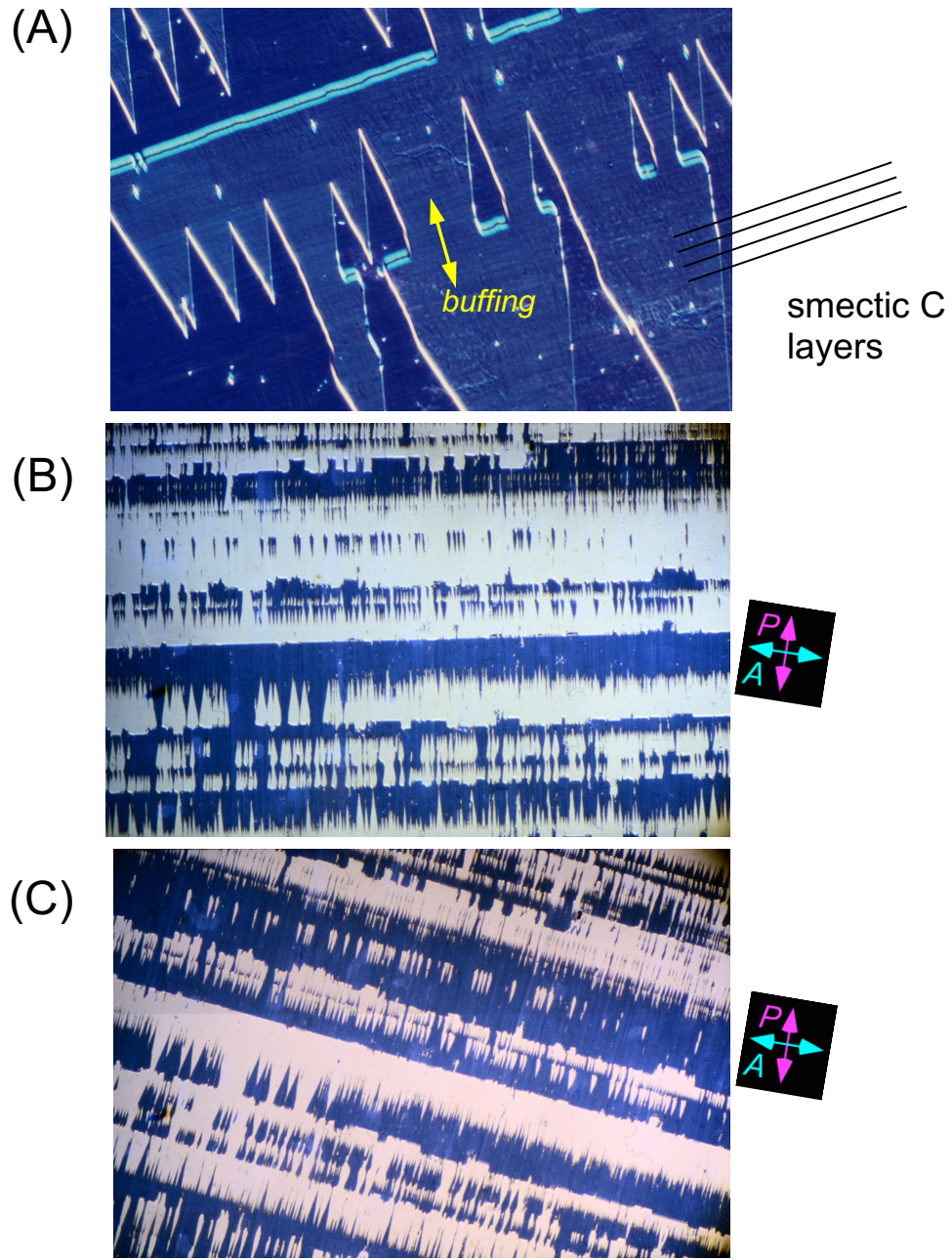

**Figure S6**: Zig-zag defect lines in SmC cells mediating changes in the direction of the chevron bend of the layers. Zig-zags are a typical smectic layering defect in thin smectic C LC samples between glass plates treated to impose a preferred uniform molecular orientation (yellow arrow). Regions with uniform chevron orientation, located within an area where the layer chevrons point the other way are bounded by closed loops [14], the detailed structure of which is shown in **Fig. S7.** Chevron bends point toward the broad walls that run parallel to the layers, and away from the diamond walls running obliquely to the layers. Image width: *A*- 300 *μm*; *B,C* - 2 mm.

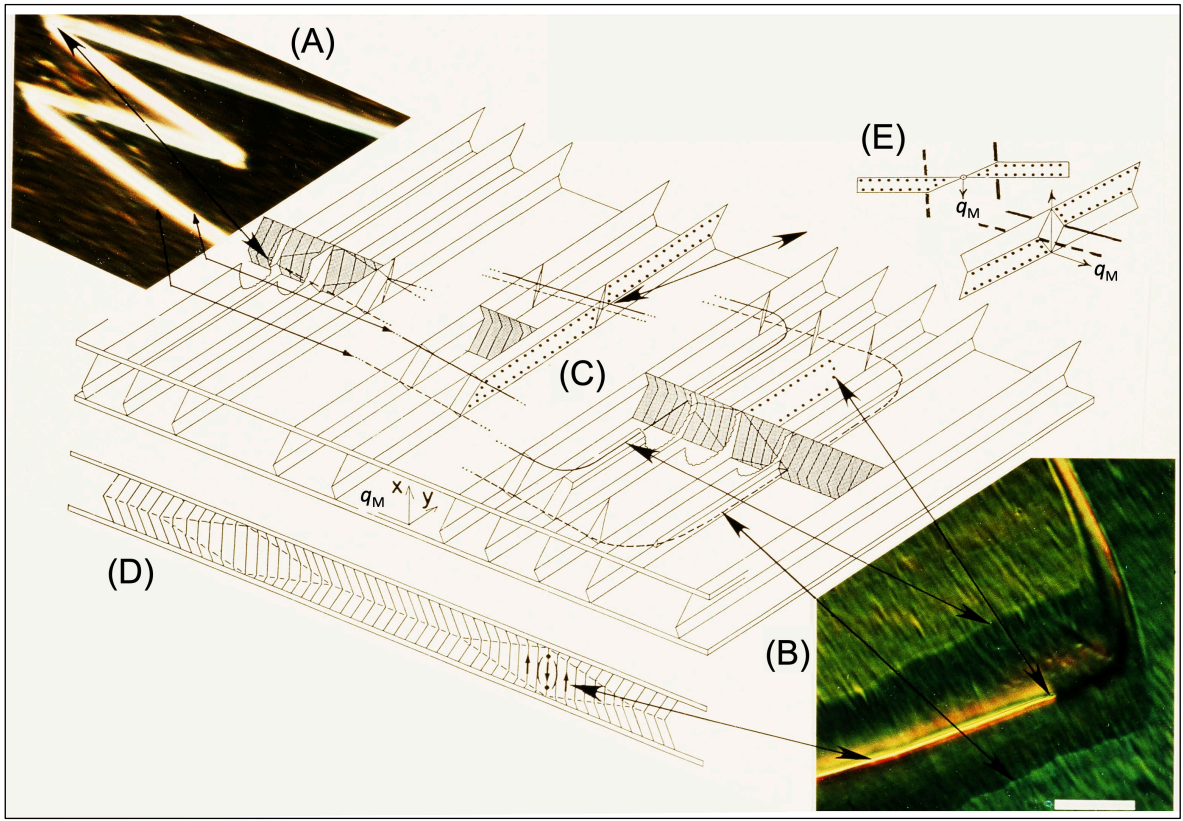

***Figure S7***: Geometrical features of the chevron layer structure and zig-zag defect loops in smectic C cells. (***A***,***B***) Photomicrographs of the narrow and broad walls of zig-zag defects in cells 3 to 4 μm thick. (***C***) The three-dimensional structure of a closed zig-zag defect loop, the spontaneously formed defect separating one direction of chevron layering from the other in a SmC, here in a sample between solid plates that was cooled from the smectic A phase. The loop consists of narrow (diamond) walls running nearly parallel to ***z*** and a broad wall running parallel to ***y***. These walls are delimited by the lines which focus sharply in (***A***) and (***B***), also indicated by the heavy solid and dashed lines in (***C***). In this example, the chevron layer-bend plane is displaced from the cell mid-plane. The defect structure is an assembly of planar layer elements continuously connected to one another by sharp layer-bend discontinuities. The mean layer pitch $p = d$, established by layer anchoring at the surfaces, is the same everywhere. (***D***) Section showing layer structure change upon passing through the diamond (<<<>>>) and broad (>>><<<) walls for the symmetric case where the chevron interface is at the cell center. (***E***) Schematic of a single layer in the diamond wall. The bar in (***B***) is 15 μm long. From [15].

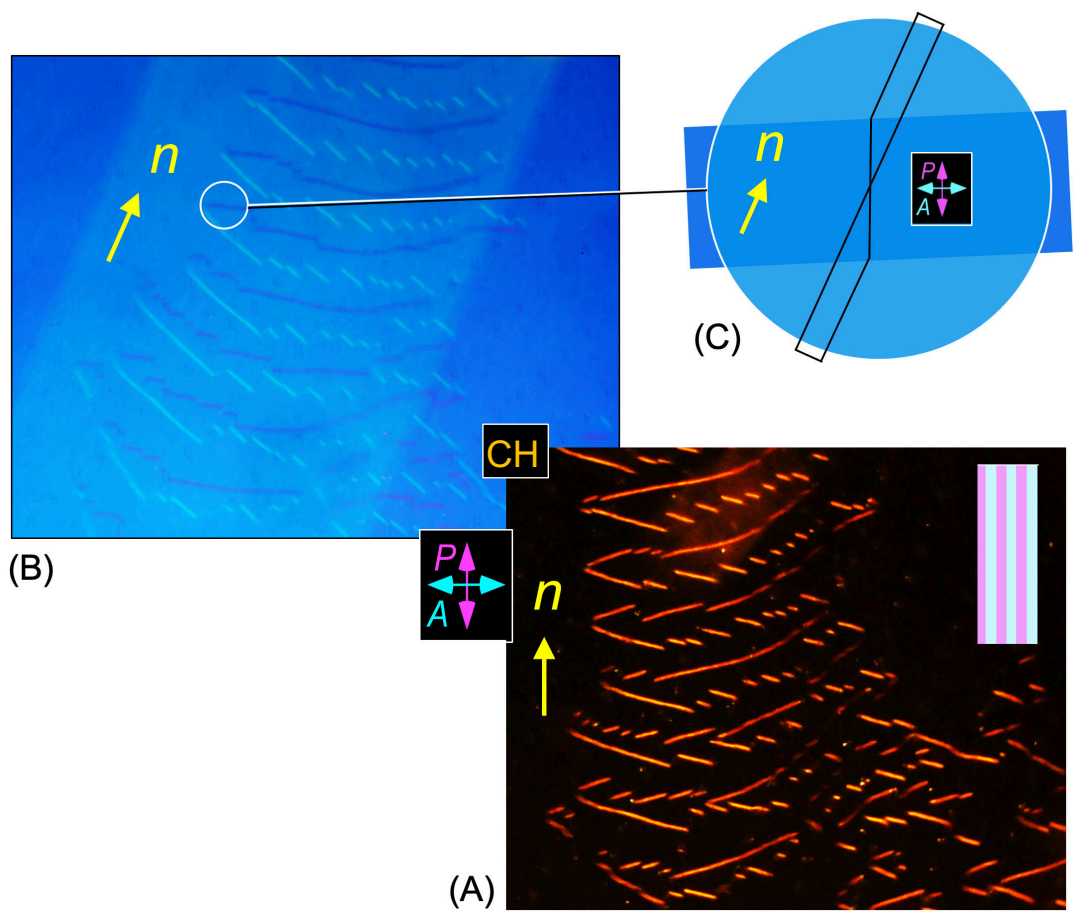

*Figure S8*: Diamond walls in the SmZ$_\text{A}$ phase of DIO in the cell of *Figs. 3,4* viewed using DTLM. (*A*) Extinction obtained when *n* is along the polarizer direction. The diamond zig-zag walls are bright because the diamond layer elements at the core of the walls have a different azimuthal orientation than the layers (and hence the bulk director, yellow) in the rest of the sample. (*B*) The cell reoriented to the extinction orientation for one set of diamond walls. (*C*) Geometry of the diamond wall, showing the top view of a single layer passing through the wall. The tip of the chevron bend discontinuity (cyan), the diamond plane, normal to the surface (red), and the intersection of the part of the layer above the chevron with the top surface (black) are highlighted. In this cell orientation, the diamond plane is aligned with the optical polarizer and gives extinction. The white arrows indicate the chevron polarity, as in *Fig. 4D*.

*Figure S9 below*: (**A-E**) DTLM images of cell texture changes induced when a 200 Hz square wave electric field of peak field $E_p = V_p/d$ is applied normal to the cell plates to a tilted-layer/chevron layer/bookshelf cell filled with DIO. This cell has $d = 4.6$ $\mu$m and an antiparallel buffed, planar-aligned, sandwich electrode geometry giving peak field $[E_p(V/\mu m)] = 0.22[V_p(V)]$. As cooled from the isotropic, the N and SmZ phases grew in as well-aligned planar monodomains. In the N phase they exhibited the splay-bend Freedericksz transitions, with threshold $V_p$ values given in (**F**). Upon cooling into the SmZ$_A$ phase the layers grow in as a BK geometry, with some evidence for zig-zag walls indicating the presence of some chevron BK layer structure with layer normal $\mathbf{q}_M$, normal to the buffing direction, $\boldsymbol{n}$. The field, normal to the cell plates, induces the splay-bend Freedericksz transition also the SmZ$_A$, with a threshold field is somewhat higher than in the N phase ($F$). (**A-C**) The field-induced start of the splay-bend Freedericksz director reorientation generates the formation of additional zig zag defects, which develop into massive arrays filling the cell. Mechanisms for generating zig-zags are (***i***) *field-induced uprighting of the layers* as the polarization $\boldsymbol{P}$ is rotated about the layer normal such that $\boldsymbol{P}$ in adjacent layers is rotated away from their antiferroelectric opposition. The non-zero $\boldsymbol{E}_x$ couples to the resulting induced $\delta\boldsymbol{p}$ to apply a torque $\boldsymbol{\tau}_z = \delta\boldsymbol{p} \times \boldsymbol{E}_x$ to the LC that tends to stand the layers up normal to the plates, a process that dilates the layer system and creates zig-zag walls; (***ii***) *expulsion of layers* if the field-induced ferroelectric state is layerless. The intensive creation of broad walls, evident in ($B$,C) may be evidence for the antiferroelectric-to-ferroelectric transition, the layering may simply disappear in the induced ferroelectric state, especially at lower temperatures in the SmZ$_A$. The diamond walls are darker in the images and the broad walls, running parallel to the layers, are brighter. (**D**) As the field is increased to ~ $V_p = 10$V, well above the Freedericksz threshold, the zig-zag defects anneal and coarsen to leave a bookshelf monodomain in the cell, with layers and the director nearly normal to the plates. (**E**) When the field is returned to zero, the director reorients to be everywhere parallel to the plates, and the layering must return, a compressive transition on the layering system so the zig-zags disappear. Image width: 2.2 mm. (**F**) Freedericksz threshold values of $V_p$ vs. $T$.

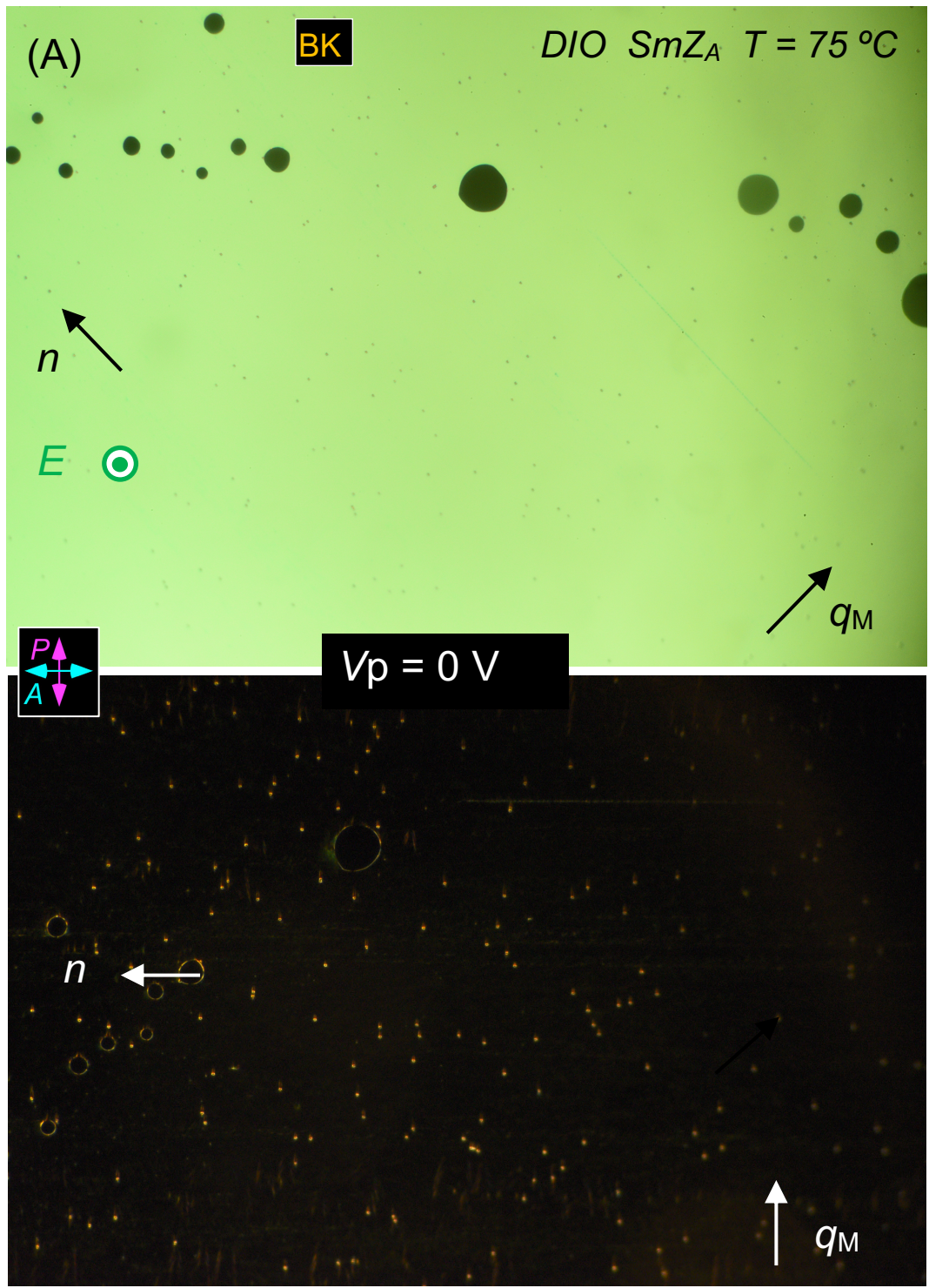

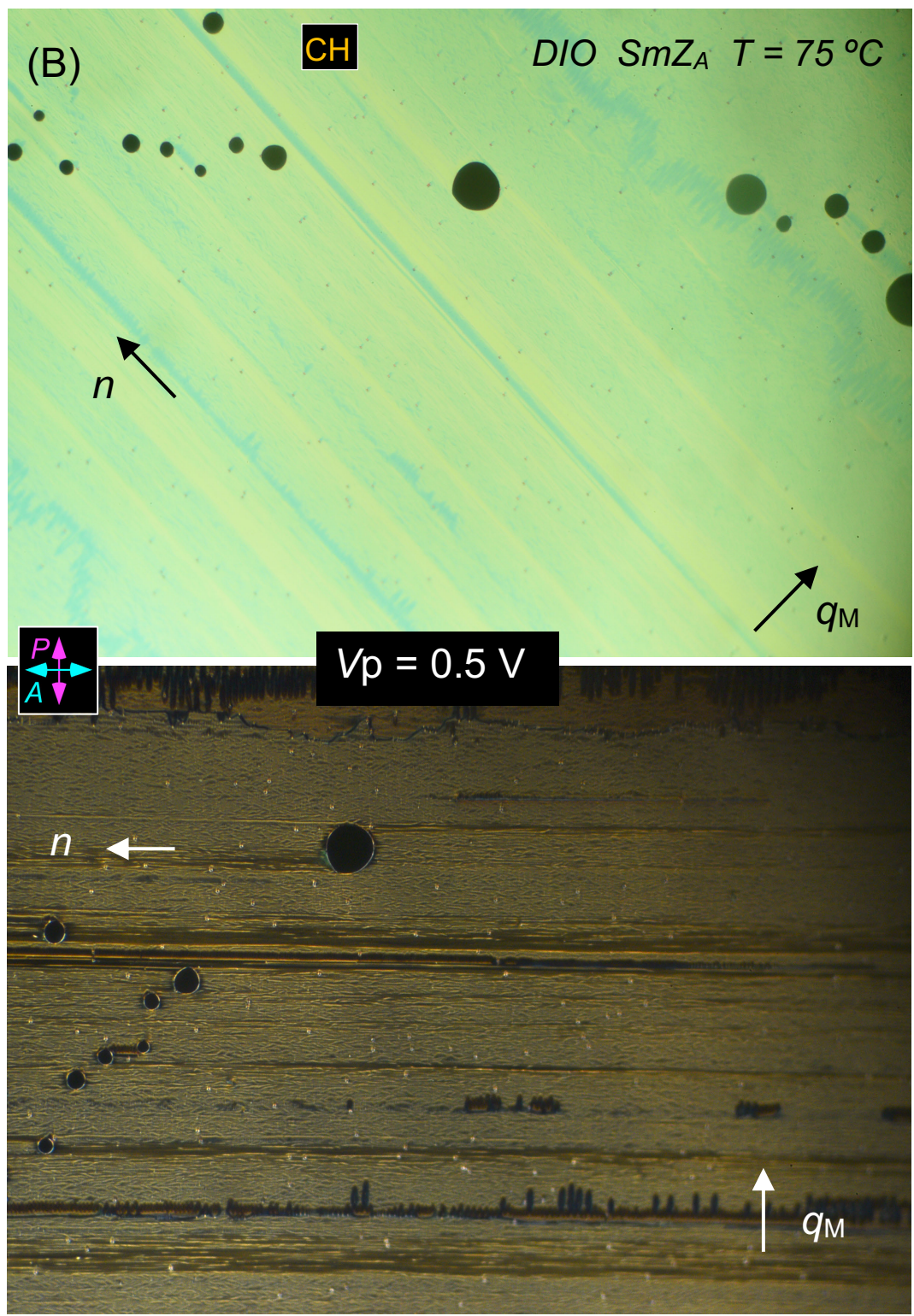

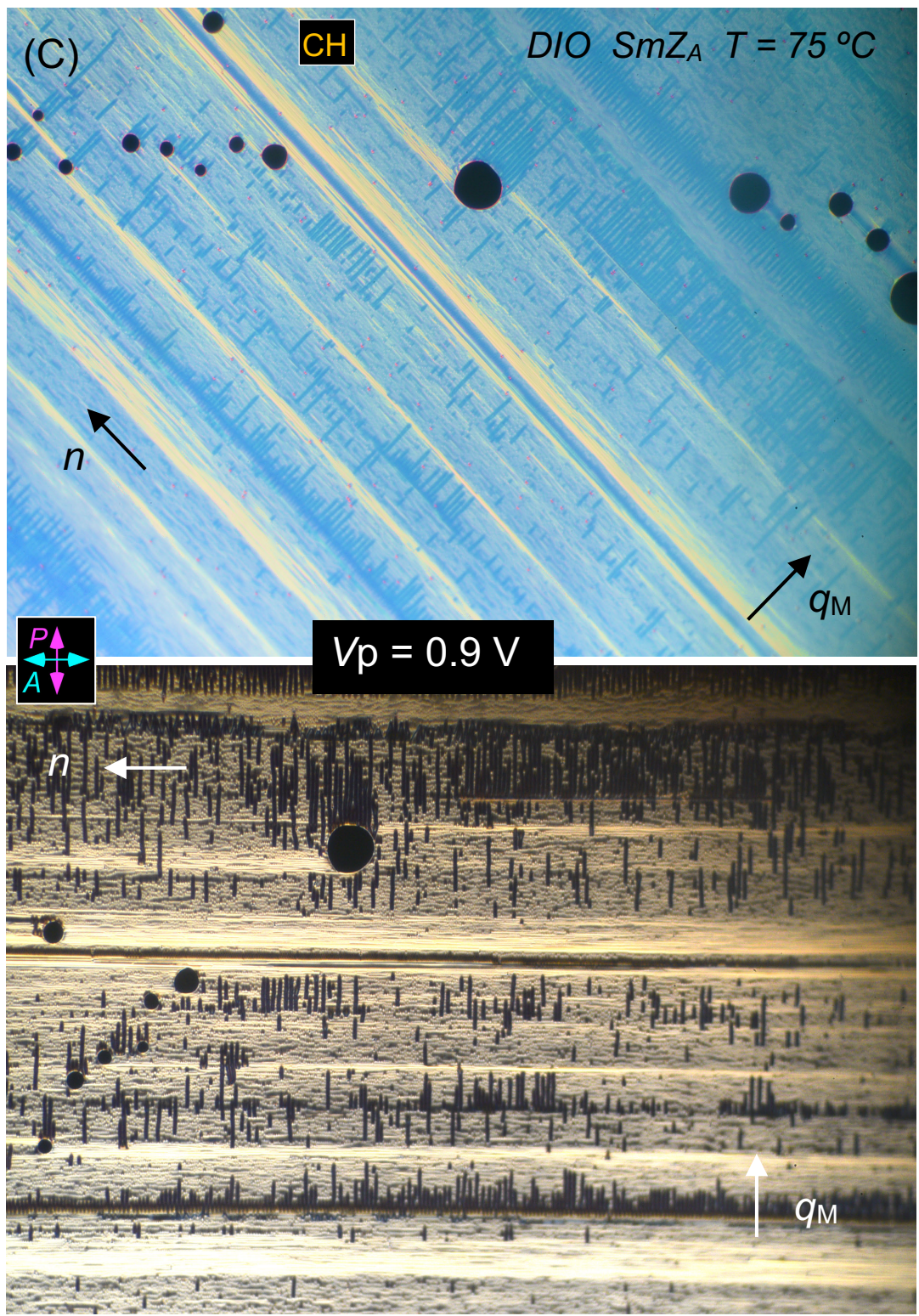

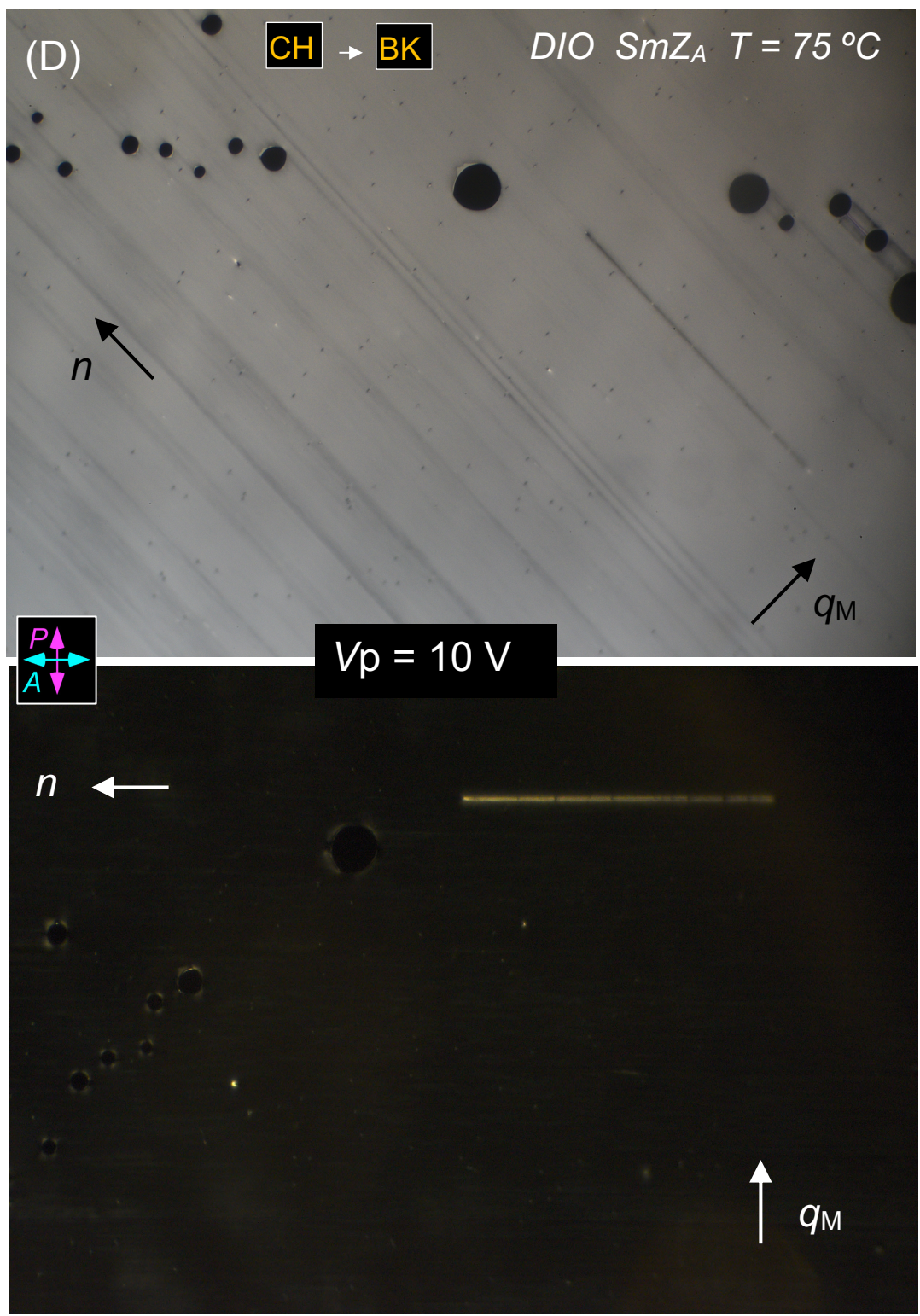

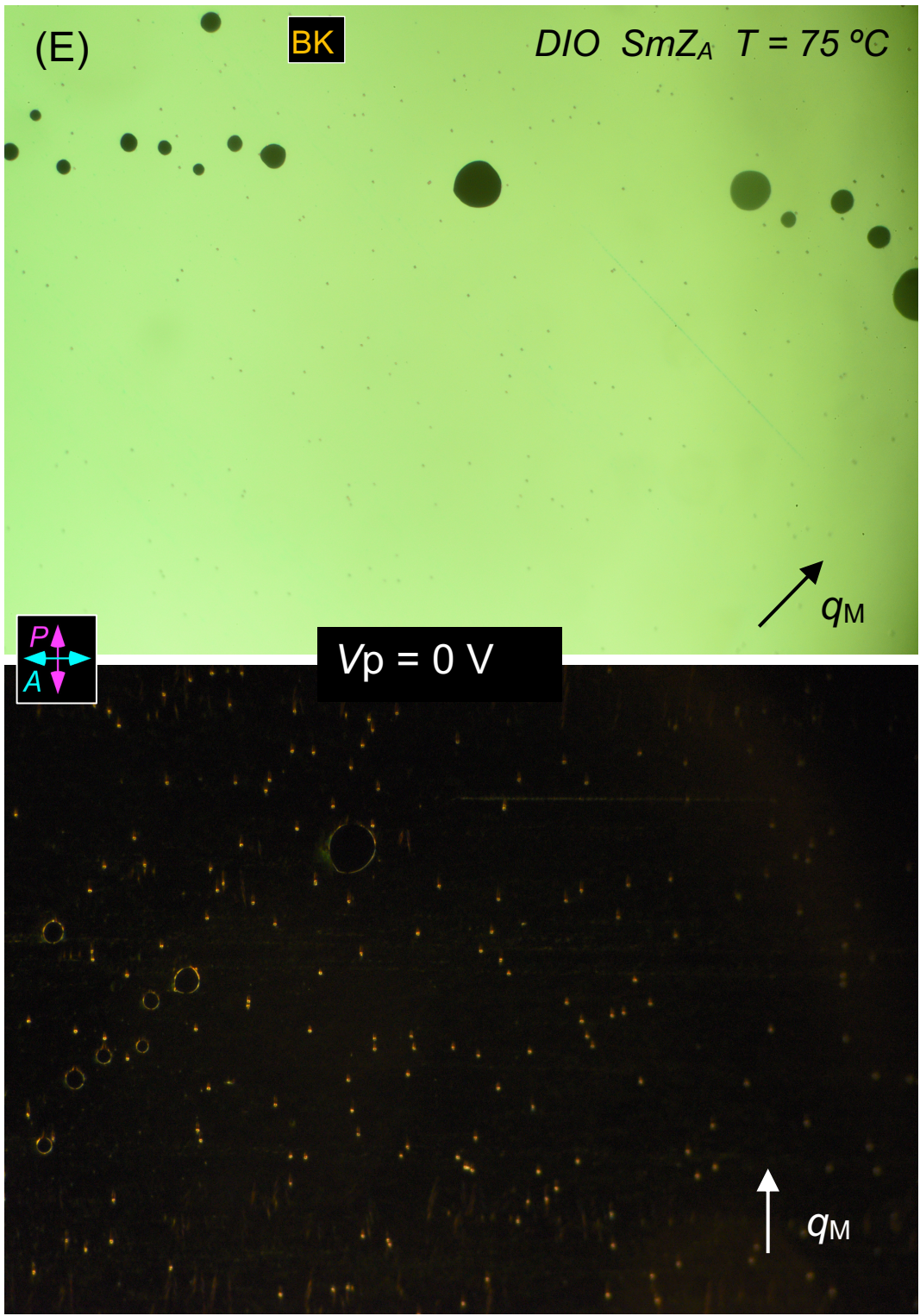

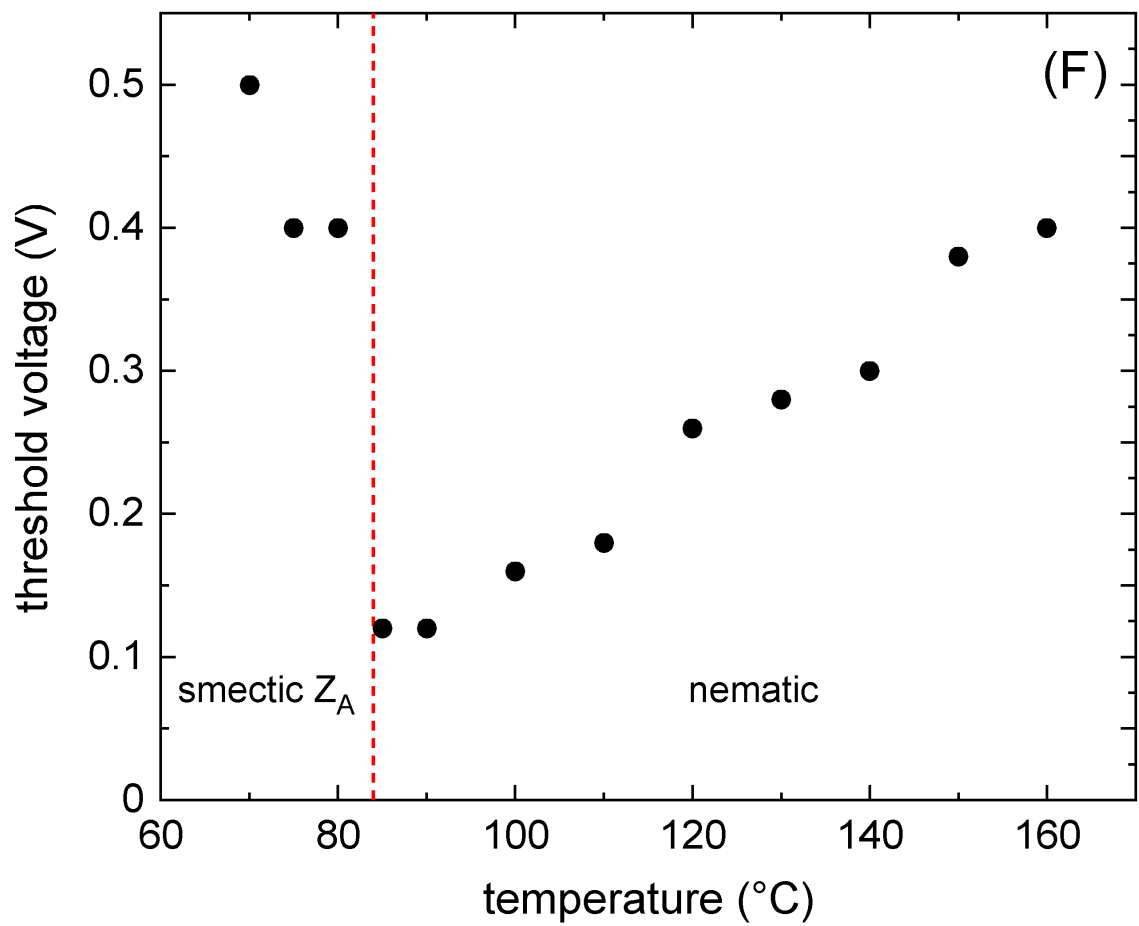

|   | SmC | | SmZ$_A$ DIO |   |
|---|---|---|---|---|
| (A) | | | | (D) |
| (B) | | | | (E) |
| (C) | | | | (F) |

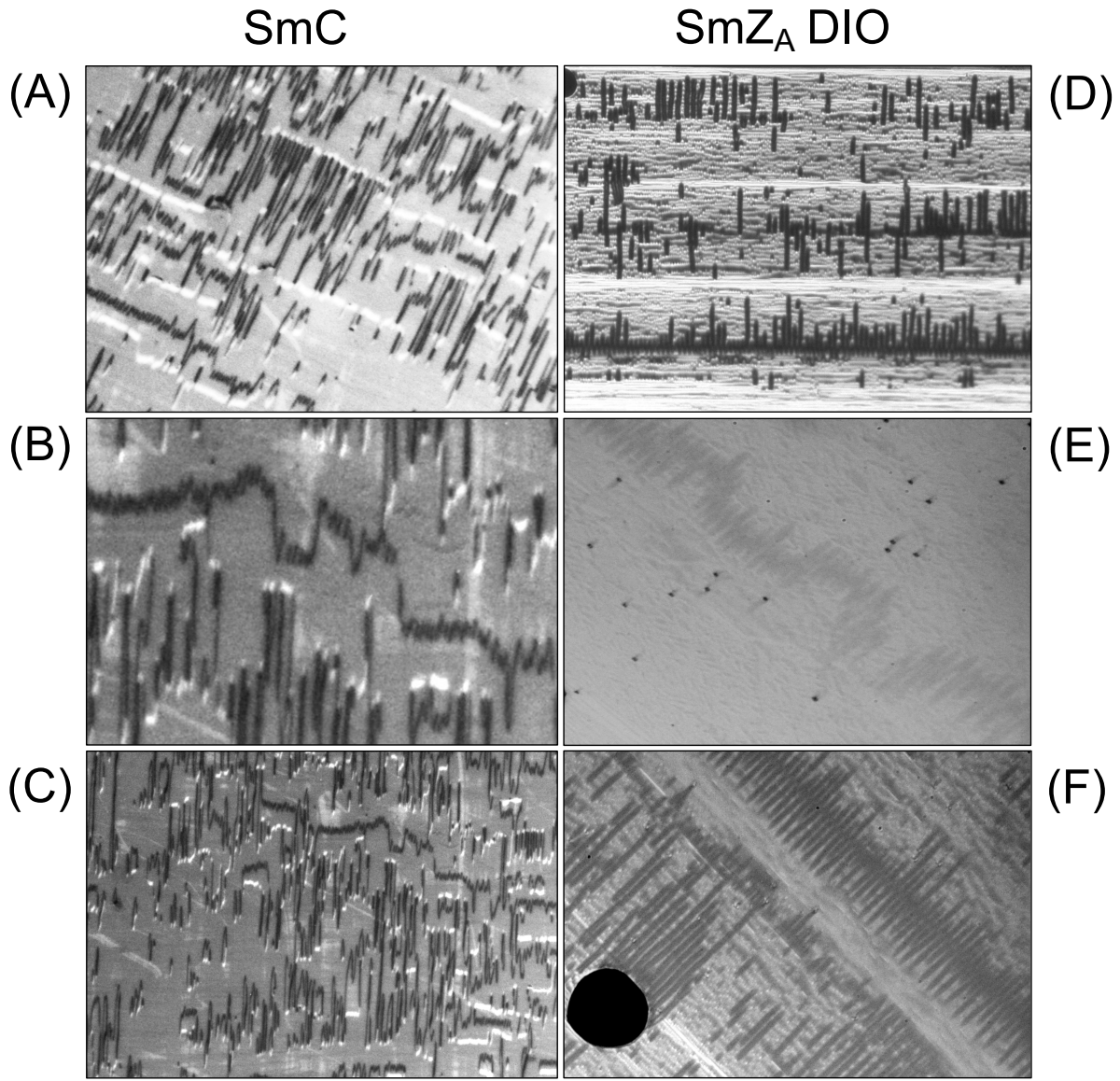

*Figure S10*: Photomicrographs showing structural similarities of arrays of zig-zag defects induced by electric field in chevron/tilted layer cells, SmC on the left, and SmZ$_A$ on the right. The black lines are the diamond walls and the brighter lines running across the images are the broad walls, parallel to the layers. The cell thickness is comparable to the broad wall width. Image width: *A-C,E,F*- 200 μm; *D* – 600 μm. . In (*D-F*) some of the broad walls have broken up into strings of small defects. These may be zig-zag loops, or evidence of a smectic periodic layer dilative undulation instability [16].

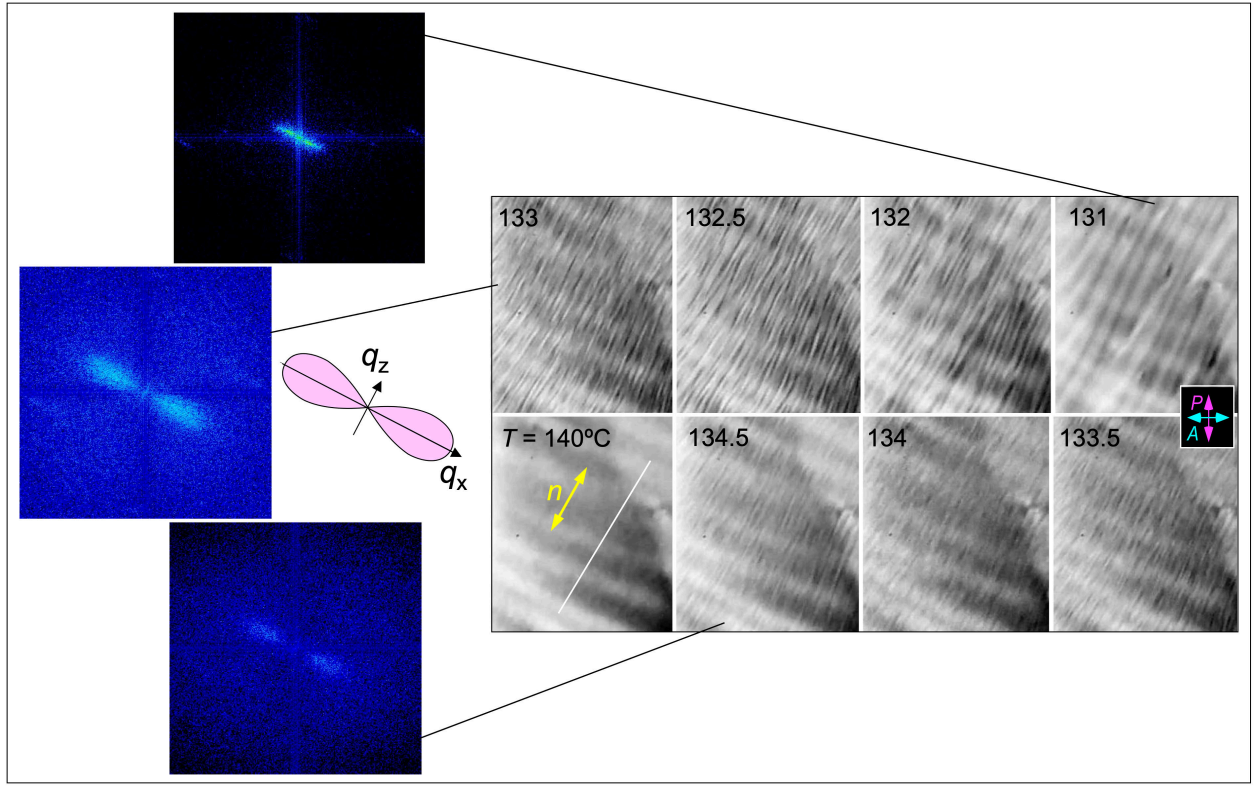

*Figure S11*: Grayscale DTLM images of a sample area of the RM734 texture and selected optical Fourier transforms near the N–N$_F$ phase transition. In the N phase (T < 133ºC), the images represent an optical-resolution integration of the birefringent phase accumulated by the light upon traversing the sample thickness, *t*. The resulting averaging captures the extension of the domain shape along *n*, which gives a wing-shaped optical Fourier transform. The pink inset shows a contour of $\chi_P(\boldsymbol{q})$, the susceptibility for scattering by polarization fluctuations given by the Aharony model [17,18]:

$$\chi_P(\boldsymbol{q}) \propto \langle P_z(\boldsymbol{q})P_z(\boldsymbol{q})^* \rangle \propto [\tau(T)(1 + \xi(T)^2 q^2) + (2\pi/\varepsilon)(q_z/q)^2]^{-1},$$

where $\tau(T) \propto (T\text{-}T_{NF})/T_{NF}$. $\chi_P(\boldsymbol{q})$ describes the anisotropy in the polarization fluctuations in the N phase generated by short range ferroelectric interactions, giving the Ornstein-Zernicke term about $q = 0$, and the last term from long-range polarization space charge interactions. The favorable qualitative comparison with the optical Fourier transform at $T = 133$ºC indicates that charge stabilization is as important a factor governing fluctuations of the N phase of RM734 as it is in the SmZ$_A$ phase of DIO. The Aharony model is directed toward understand certain crystalline magnetic materials that have short-range ferromagnetic exchange forces, but where long-range dipolar interactions, which can favor mutual parallel or antiparallel alignment of dipoles depending on their relative position, are also important. In these systems, short-range interactions are included in a model Hamiltonian as nearest-neighbor Ising or Heisenberg-like, and the long-range interactions are calculated explicitly. Renormalization group analysis shows that the long-range interactions make the magnetic correlations dipolar-anisotropic near the transition in the high temperature phase [19,20,19,20,21, extending them along *z* by strongly suppressing longitudinal charge density $(\partial P_z / \partial z)$ fluctuations. The dipole-dipole (third) term produces extended correlations that grow as $\xi(\tau)$ along $x$ and $y$ but as $\xi(\tau)^2$ along $z$ [20], suppressing $\chi(\boldsymbol{q})$ for finite $q_z$ as is observed qualitatively from the visual appearance of the textures upon passing through the phase transition, and from their optical Fourier transforms. Because of this anisotropy, the correlation volume in this model grows in 3D as $V \sim \xi(\tau)^4$ rather that the isotropic $V \sim \xi(\tau)^3$, reducing the upper marginal dimensionality of the transition to 3D, and making the transition mean-field-like with logarithmic corrections, rather than fluctuation-dominated with 3D Ising universality [22]. LC cell thickness $d = 11$ $\mu$m. Scale bar = 200 $\mu$m (image corrected from the misassembled version shown in Ref. [7]).

***Section S5 – Landau model phase diagrams of NaNO₂, SC(NH₂)₂, and DIO***

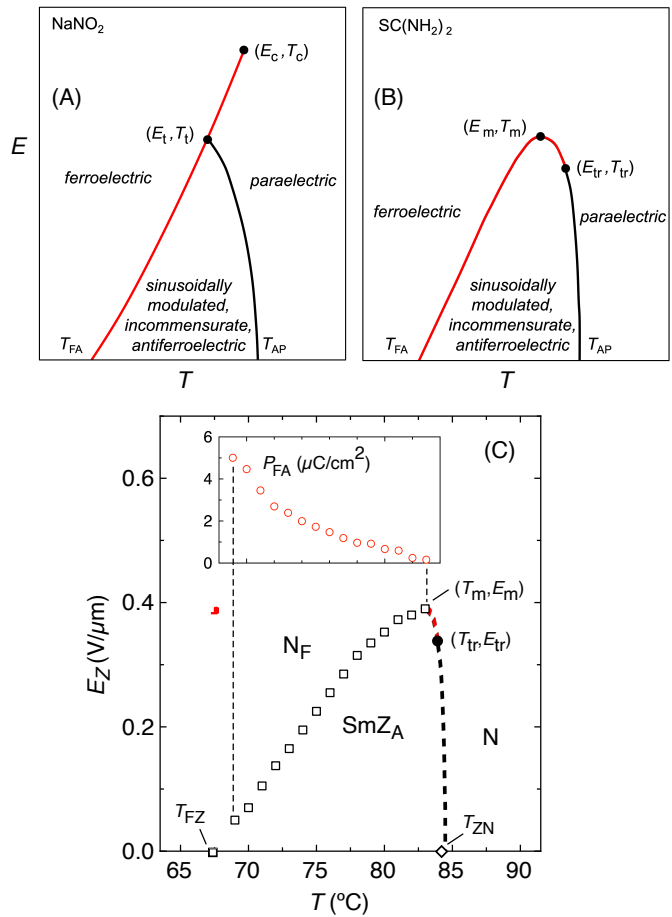

***Figure S12:*** Temperature-field $(T,E)$ phase diagrams for electric field applied along the polarization direction ($z$ in the case of DIO). (**A**) $(T,E)$ phase diagram computed using ***Eq. 1*** with $B < 0$ and $C > 0$, fitting the P – A – F phase behavior of NaNO₂ [23,24]. $(T_t, E_t)$ is a tricritical point, and $(T_c, E_c)$ the critical end point. (**B**) Temperature-field $(T, E)$ phase diagram computed using ***Eq. 1*** with $B > 0$ and $C = 0$, fitting the P – A – F phase behavior of SC(NH₂)₂ [25,26,27]. (**C**) Temperature-field $(T, E_z)$ phase diagram showing the P – A – F phase behavior of DIO. Data are from $V_{FA}$(□), $P_{FA}$(○) in ***Fig. 5B***, $T_{FZ}$ at zero field (□), and $T_{ZN}$ at zero field (◇). The corresponding field $E_{FA}(T)$ = $V_{FA}(T)/(100\ \mu m)$. The open squares give the first-order SmZ$_A$–N$_F$ phase boundary in the $(T, E_z)$ plane, and the open circles the polarization change $P_{FA}$ at this transition, which decreases with increasing field. According to the Clausius/Clapyron equation, the transition entropy $\Delta S$ also decreases along this line to zero at the point of maximum SmZ$_A$ stability, $(T_m, E_m)$, and then changes sign with increasing $T$. The solid red curve is a fit to $T_{ZF}$, the SmZ$_A$ – N$_F$ transition line using ***Eq. 1***, and the dashed line an estimate of the $T_{NZ}$ line, based on the data at its two ends and the $F_{IS}$ model described by ***Eq. 1***. In the model the P – A transition can be weakly first order or second order and have a tricritical point $(T_{tr}, E_{tr})$. The N$_F$ – SmZ$_A$ is observed to be weakly first order at zero field, but has not yet been studied vs. $E_z$.

*Supplementary Information References*

1    H. Nishikawa, K. Shiroshita, H. Higuchi, Y. Okumura, Y. Haseba, S. Yamamoto, K. Sago, H. Kikuchi, A fluid liquid-crystal material with highly polar order. *Adv. Mater.* **29**, 1702354 (2017). DOI: 10.1002/adma.201702354

2    R.J. Mandle, S.J. Cowling, and J.W. Goodby, A nematic to nematic transformation exhibited by a rod-like liquid crystal. *Phys. Chem. Chem. Phys.* **19**, 11429–11435 (2017). DOI: 10.1039/C7CP00456G

3    R.J. Mandle, S.J. Cowling, J.W. Goodby, Rational design of rod-like liquid crystals exhibiting two nematic phases. *Chemistry A European Journal* **23**, 14554–14562 (2017). DOI : 10.1002/chem.201702742

4    R.J. Mandle, N. Sebastián, J. Martinez-Perdiguero, A. Mertelj, On the molecular origins of the ferroelectric splay nematic phase. *Nature Communications* **12** (1), 4962 (2021). DOI: 10.1038/s41467-021-25231-0

5    A. Mertelj, L. Cmok, N. Sebastián, R.J. Mandle, R.R. Parker, A.C. Whitwood, J.W. Goodby, M. Čopič, Splay nematic phase. *Phys. Rev. X* **8**, 041025 (2018). DOI: 10.1103/PhysRevX.8.041025

6    V. Castelletto, A.M. Squires, I.W. Hamley, J. Stasiak, G.D. Moggridge, A SAXS study of flow alignment of thermotropic liquid crystal mixtures. *Liquid Crystals* **36**, 435–442 (2009). DOI: 10.1080/02678290902928542

7    F. Vita, M. Hegde, G. Portale, W. Bras, C. Ferrero, E.T. Samulski, O. Francescangeli, T. Dingemans, Molecular ordering in the high-temperature nematic phase of an all-aromatic liquid crystal. *Soft Matter* **12**, 2309–2314 (2016).  DOI: 10.1039/c5sm02738a

8    J. Engqvist, M. Wallin, S.A. Hall, M. Ristinmaa, T.S. Plivelic, Measurement of multi-scale deformation of polycarbonate using X-ray scattering with *in-situ* loading and digital image correlation. *Polymer* **82**, 190–197 (2016).  DOI: 10.1016/j.polymer.2015.11.028

9    A. Roviello, S. Santagata, A. Sirigu, Mesophasic Properties of linear copolymers, III: Nematogenic copolyesters containing non-mesogenic rigid groups. *Makromol. Chem., Rapid Commun.* **4**, 281–284 (1983). DOI: 10.1002/marc.1983.030040503

10   S. Nishitsuji, Y. Watanabe, T. Takebe, N. Fujii, M. Okano, M. Takenaka, X-ray scattering study on the changes in the morphology of low-modulus polypropylene under cyclic uniaxial elongation. *Polymer Journal* **52**, 279–287 (2020). DIO: 10.1038/s41428-019-0284-2

11   X. Chen, E. Korblova, D. Dong, X. Wei, R. Shao, L. Radzihovsky, M.A. Glaser, J.E. Maclennan, D. Bedrov, D.M. Walba, N.A. Clark, First-principles experimental demonstration of ferroelectricity in a thermotropic nematic liquid crystal: spontaneous polar domains and striking electro-optics. *Proceedings of the National Academy of Sciences of the United States of Americ*a **117**, 14021–14031 (2020).  DOI: 10.1073/pnas.2002290117

12   M.R. Tuchband, M. Shuai, K.A. Graber, D. Chen, C. Zhu, L. Radzihovsky, A. Klittnick, L.M. Foley, A. Scarbrough, J.H. Porada, M. Moran, J. Yelk, D. Bedrov, E. Korblova, D.M. Walba, A. Hexemer, J.E. Maclennan, M.A. Glaser, N.A. Clark, Double-helical tiled chain structure of the twist-bend liquid crystal phase in CB7CB. arXiv:1703.10787  (2017).

13   N. Sebastián, L. Cmok, R.J. Mandle, M. Rosario de la Fuente, I. Drevenšek Olenik, M. Čopič, A. Mertelj, Ferroelectric-ferroelastic phase transition in a nematic liquid crystal. *Phys. Rev. Lett.* **124**, 037801 (2020).  DOI: 10.1103/PhysRevLett.124.037801

14 S.T. Lagerwall, _Ferroelectric and antiferroelectric liquid crystals_ (Wiley VCH, Weinheim, 1999) ISBN 3-527-2983 1-2.

15 N.A. Clark, T.P. Rieker, J.E. Maclennan, Director and layer structure of SSFLC cells, _Ferroelectrics_ **85**, 79–97 (1988). DOI: 10.1080/00150198808007647

16 N.A. Clark, R.B. Meyer, Strain-induced instability of monodomain smectic A and cholesteric liquid-crystals. _Applied Physics Letters_ **22**, 494–493 (1973). DOI: 10.1063/1.1654481

17 A. Aharony, M.E. Fisher, Critical behavior of magnets with dipolar interactions. I. Renormalization group near four dimensions. _Physical Review B_ **8**, 3323–3341 (1973). DOI: 10.1103/PhysRevB.8.3323

18 A. Aharony, Critical behavior of magnets with dipolar interactions. V. Uniaxial magnets in _d_-dimensions. _Physical Review B_ **8**, 3363–3370 (1973). DOI: 10.1103/PhysRevB.8.3363

19 J. Kotzler, Critical phenomena in dipolar magnets. _Journal of Magnetism and Magnetic Materials_ **54-57**, 649–654 (1986). DOI: 10.1016/0304-8853(86)90197-6

20 J. Als-Nielsen, Experimental test of renormalization group theory on the uniaxial, dipolar coupled ferromagnet $LiTbF_4$. _Physical Review Letters_ **37**, 1161–1164 (1976). DOI: 10.1103/PhysRevLett.37.1161

21 J. Als-Nielsen, R.J. Birgeneau, Mean field theory, the Ginzburg criterion, and marginal dimensionality of phase transitions. _American Journal of Physics_ **45**, 554–560 (1977). DOI: 10.1119/1.11019

22 G. Ahlers, A. Kornblit, H.J. Guggenheim, Logarithmic corrections to the Landau specific heat near the Curie temperature of the dipolar Ising ferromagnet $LiTbF_4$. _Physical Review Letters_ **34**, 1227–1230 (1975). DOI: 10.1103/PhysRevLett.34.1227

23 D. Durand, F. Denoyer, D. Lefur, Neutron diffraction study of sodium nitrite in an applied electric field. _Journal de Physique_ **44** L207–L216 (1983). DOI : 10.1051/jphyslet:01983004405020700

24 D. Durand, F. Denoyer, R. Currat, M. Lambert Chapter 13 - Incommensurate phase in $NaNO_2$. _Incommensurate phases in dielectrics: 2 Materials_, R. Blinc and A.P. Levanyuk, Eds. (Elsevier Science Publishers, Amsterdam 1986). DOI: 10.1016/B978-0-444-86970-8.50010-X

25 P. Lederer, C.M. Chaves, Phase diagram of thiourea at atmospheric pressure under electric field: a theoretical analysis. _Journal de Physique Lettres_ **42**, L127–L130 (1981). DOI: 10.1051/jphyslet:01981004206012700

26 J. P. Jamet, Electric field phase diagram of thiourea determined by optical birefringence. _Journal de Physique Lettres_ **42**, L123–L125 (1981). DOI: 10.1051/jphyslet:01981004206012300

27 F. Denoyer, R. Currat, Chapter 14 – Modulated Phases in Thiourea - _Incommensurate phases in dielectrics: 2 Materials_, R. Blinc and A. P. Levanyuk, Eds. (Elsevier Science Publishers, Amsterdam 1986). DOI: 10.1016/B978-0-444-86970-8.50010-X